\documentclass{aa}  

\usepackage{graphicx}
\usepackage{txfonts}
\usepackage{multirow}
\usepackage{amssymb}
\usepackage{amsmath}
\usepackage{wasysym}
\usepackage{commath}
\usepackage{bigdelim}
\usepackage{subfigure}
\usepackage{lscape}
\usepackage{url}
\usepackage{caption}
\usepackage{gensymb}
\begin{document} 

\title{Results from a set of three-dimensional numerical experiments
  of a hot Jupiter atmosphere.}

\titlerunning{Hot Jupiters: numerical experiments.}

   \author{N. J. Mayne\inst{1}\fnmsep\thanks{E-mail:
       nathan@astro.ex.ac.uk},
          F. Debras\inst{1,2},
          I. Baraffe\inst{1,2}, 
          John Thuburn\inst{3},
          David S. Amundsen\inst{1,4,5},
          David M. Acreman\inst{1,6}, 
          Chris Smith\inst{7}, 
          Matthew K. Browning\inst{1},
          James Manners\inst{7},           
          \and
          Nigel Wood\inst{7}
          }

          \institute{Physics and Astronomy, College of Engineering,
            Mathematics and Physical Sciences, University of Exeter,
            EX4 4QL.  \and Ecole Normale Sup\'{e}rieure de Lyon, CRAL,
            UMR CNRS 5574, Universit\'{e} de Lyon, 69364 Lyon
            Cedex. \and Applied Mathematics Group, University of
            Exeter, Exeter, EX4 4QL, United Kingdom. \and Department
            of Applied Physics and Applied Mathematics, Columbia
            University, New York, NY 10025, USA. \and NASA Goddard
            Institute for Space Studies, New York, NY 10025, USA.
            \and Department of Computer Science, College of
            Engineering, Mathematics and Physical Sciences, University
            of Exeter, EX4 4QF. \and Met Office, FitzRoy Road, Exeter,
            Devon EX1 3PB, UK.  }

          \authorrunning{Mayne et al.}

   \date{Received September 15, 1996; accepted March 16, 1997}

% \abstract{}{}{}{}{} 
% 5 {} token are mandatory
 
   \abstract{We present highlights from a large set of simulations of
     a hot Jupiter atmosphere, nominally based on HD~209458b, aimed at
     exploring both the evolution of the deep atmosphere, and the
     acceleration of the zonal flow or jet. We find the occurrence of a
     super-rotating equatorial jet is robust to changes in various
     parameters, and over long timescales, even in the absence of
     strong inner or bottom boundary drag. This jet is diminished in one
     simulation only, where we strongly force the deep atmosphere
     equator--to--pole temperature gradient over long
     timescales. Finally, although the eddy momentum fluxes in our
     atmosphere show similarities with the proposed mechanism for
     accelerating jets on tidally-locked planets, the picture appears
     more complex. We present tentative evidence for a jet driven by a
     combination of eddy momentum transport and mean flow.}
 
   \keywords{Hydrodynamics -- Planets and satellites: atmospheres -- Methods: numerical}

   \maketitle
%
%________________________________________________________________
\section{Introduction}
\label{section:introduction}

Three-dimensional (3D) general circulation models (GCMs) have become
established tools used both to interpret current observations of
exoplanets, and to predict those to be made by future instruments
\citep[see][for recent examples]{kataria_2016,lee_2016}. Given the
complex, non--linear and interacting physical mechanisms acting within
a planetary atmosphere, combined with the incomplete nature of our
observational access to exoplanets, GCMs are likely to become
increasingly important in this field.

Currently, the most observationally constrained exoplanets are a
subset termed hot Jupiters, being Jovian in size and orbiting close to
their parent star \citep[see discussion and review
in][]{baraffe_2010}. Radial velocity and transit measurements provide
estimates of the masses and radii of these planets, and numerous
subsequent studies have inferred further atmospheric characteristics
such as chemical compositions, temperature structures and even wind
speeds \citep{kreidberg_2015,vidal_madjar_2011,louden_2015}. The
recent work of \citet{sing_2016} has also begun to classify these
objects spectrally, as has been done for stars. The proximity of hot
Jupiters to their parent star suggests that tidal forces are likely to
rapidly evolve hot Jupiters into a tidally-locked state, with
permanent day and night hemispheres \citep{baraffe_2010}. Therefore,
although much progress has been made (and much more can yet be made)
using 1D models, 3D models are required to truly unpick the
observations, and extract robust physical meaning.

Several GCMs (or similar models) with varying levels of sophistication
have been applied to hot Jupiters \citep[see for
example][]{cooper_2005,cho_2008,menou_2009,rauscher_2010,heng_2011,dobbs_dixon_2013,parmentier_2013,showman_2015,helling_2016,kataria_2016,lee_2016},
including our own adaptation of the Met Office GCM termed the
Unified Model (UM)
\citep{mayne_2014,mayne_2014b,amundsen_2014,helling_2016,amundsen_2016,amundsen_2017,boutle_2017}. However,
much of the progress has been driven by application of a single GCM,
the SPARC/MITgcm. In fact, a significant number of subsequent,
detailed analyses of hot Jupiter atmospheres have been performed using
output, or derivatives of the pioneering work of
\citet{showman_2009}. Studies based on the SPARC/MITgcm include
exploring the presence of TiO/VO, advection driven non--equilibrium
chemistry and changes in the dynamics over a range of hot Jupiters
\citep{parmentier_2013,agundez_2014,kataria_2016}. Despite this
progress lessons learned from the Earth and solar system communities
tell us that GCM results can be particularly model-dependent
\citep[e.g.][]{lebonnois_2011}. Simple model intercomparisons have
been done \citep{heng_2011}, and efforts have also started to compare
more complex models \citep{helling_2016}. In \citet{amundsen_2016} we
presented simulations of HD~209458b using our own adapted GCM which is
of commensurate sophistication to that of
\citet{showman_2009}. \citet{amundsen_2016} present qualitative and
quantitative differences between their results and those of
\citet{showman_2009}, emphasising the need for further intercomparison
of both the GCMs themselves and post-processing tools\footnote{We are
  working on a more direct comparison between the UM and SPARC/MITgcm,
  but such comparisons are difficult for models as complex as GCMs.}.

Previous results from GCM simulations of hot Jupiters hint at possible
initial condition sensitivity, if extensive damping is not used at the
bottom boundary \citep{liu_2013,cho_2015}, which could be caused by
the deep, radiatively unforced atmosphere (hereafter the `deep'
atmosphere refers to the region where the pressure is in excess of
$10^6$\,Pa or 10\,bar) dynamically evolving similarly to the oceans of
Earth \citep{mayne_2014}. \citet{amundsen_2016} also note an evolution
of the deep atmosphere temperature--pressure profile in their
simulations of HD~209458b. We have recently explored the evolution of
the deep atmosphere using a 2D model assuming a steady state solution,
the results of which will be discussed in an upcoming publication
(Tremblin, et al., submitted).

Despite the possible uncertainties, several features are
qualitatively, yet robustly reproduced, across most of the GCMs
applied to hot Jupiters, in particular the occurrence of a
super-rotating equatorial jet (coherent zonal flow). This jet is
seemingly confirmed by observations of a shift in the brightest part
of a hot Jupiter atmosphere away from the substellar point, observed
via phase curves and suggested to be caused by wind-driven advection
\citep{knutson_2007,zellem_2014} by a super-rotating equatorial
jet. Although, simulations suggest the cause is more subtle than this
and that the initial temperature structure is setup by fast moving
waves driven by the irradiation, subsequently acting to accelerate the
jet itself \citep{showman_2011}.

Although jets are commonplace amongst Solar system objects, and simple
2D arguments can be used to characterise their likely breadth in
latitude \citep[see reviews by][]{showman_2008b,showman_2011b},
analytical description of the mechanisms which accelerate the jet by
the convergence of prograde momentum (or pumping), is much more
challenging. The difficulty is that the diagnosis of such mechanisms
requires an understanding of the interactions of the mean flow with
eddies or perturbations. Despite this difficulty the mechanism for the
acceleration and support of Earth's mid--latitude jets is relatively
well understood. The result comes directly from the conservation of
wave activity \citep{vallis_2006}. Atmospheric Rossby waves, or
vortices, excited by baroclinic instability at mid--latitudes, travel
towards the pole (or equator). The characteristics of the Rossby wave
dispersion relation lead to the transport of eastward (or positive)
angular momentum into the excitation location and westward (negative)
angular momentum into the dissipation site, which in the case of Earth
is at low latitudes and at high latitudes. The total angular momentum
in the system is conserved, and this is why Earth's mid--latitude jets
are coupled to the circumpolar and equatorial retrograde flow
\citep{vallis_2006}. This effect can be diagnosed by exploring the
Eliassen-Palm flux, a vector quantity representing the relative
strengths of the eddy heat and momentum fluxes
\citep{eliassen_1960,eliassen_1961,vallis_2006}.

\citet{showman_2010} explore the mechanism for generating
super-rotating flows at the equator, extending the 2D analytical
results of \citet{matsuno_1966} (representing positive and negative
heating on opposing hemispheres) and \citet{gill_1980} (representing
positive and zero forcing on opposing
hemispheres). \citet{showman_2010} performed tests using a simple
shallow--water model, simulating superrotation at Earth's tropics and
with the addition of momentum exchange between the upper, `active'
atmosphere and lower, `quiescent' atmosphere super-rotating flows were
accelerated. This indicates strongly that the mechanism which
exchanges angular momentum vertically in an atmosphere is critical, in
balance with the horizontal interactions, to the generation of a
super-rotating jet at the equator, or in other words the process is
truly a 3D one. As discussed by \citet{showman_2011}, however, an
Earth--like mechanism is unlikely to operate in the atmospheres of hot
Jupiters \citep[see discussion
in][]{showman_2011}. \citet{showman_2011} develop an alternative
theory, still involving mean flow--eddy interaction, but reliant upon
planetary scale, equatorially trapped, standing Rossby and Kelvin
waves \citep[which are responses to the large scale forcing as
explained in][]{showman_2011}. This mechanism, as for the previous
more Earth-like case, relies on a balance between the vertical and
meridional eddy momentum fluxes in the atmosphere, with angular
momentum being extracted from a deep atmosphere reservoir
\citep{showman_2011}. \citet{showman_2011} suggest that the jet in
their simulations of HD~209458b, is accelerated in the first few tens
of days (all references within this work to days refer to Earth days
i.e. 86\,400 seconds) and reaches an equilibrium where the
acceleration terms balance to zero at the equator. The equations and
simulations of \citet{showman_2011} are, however, based on the
primitive equations of motion \citep[a simplified version of the
equations of motion for an atmosphere, see][for a full discussion in
relation to hot Jupiters]{mayne_2014}, incorporating the assumption of
vertical hydrostatic equilibrium, a shallow atmosphere and constant
gravity. Further testing of this mechanism has been performed using
results from a more dynamically complete model by
\citet{tsai_2014}. \citet{tsai_2014} adopt the $\beta$-plane
approximation and investigate the momentum fluxes in 3D, in an assumed
steady state. In the $\beta$-plane approximation the Coriolis force is
taken to vary linearly with latitude, $\phi$ (actual variation is
$\propto\cos\phi$). \citet{tsai_2014} found a picture consistent with
the ideas of \citet{showman_2011}.

In this work we explore the dynamical form and acceleration of the
super-rotating equatorial jet within various simulations based
nominally on HD~209458b. A key feature of this study is that our model
is a non-hydrostatic, deep atmosphere model, and we do not invoke a
drag at the bottom boundary. The layout of this paper is as follows:
in Section \ref{section:model_setup} we detail the model used and the
simulations we have performed, referring to previous works for the
details. Section \ref{section:results} then details our
results. Firstly, in Section \ref{sub_section:jet_robust} we present
the morphology of the dominant, zonal advection over long simulation
timescales, at various levels of approximation to the dynamical
equations, treatments of the heating and treatment of the deep
atmosphere thermal profile. Our results reveal a robust recurrence of
the equatorial super-rotating jet. Then in Section
\ref{sub_section:deep_atmosphere} we explore the evolution of the
deep, high pressure atmosphere, highlighting the gradual evolution of
the kinetic energy and the fact that it has not reached a steady state
in any of our simulations. In order to test the effect of this on the
flows in the upper atmosphere (i.e. at low pressures), we perform a
simple limiting numerical experiment, strongly forcing the deep
atmosphere over long timescales, revealing an eventual weakening of
the super rotating equatorial jet. Finally, in Section
\ref{sub_section:jet_pumping} we explore the acceleration and
maintenance of the zonal flows. Section \ref{subsub_section:rest}
presents the initial response (from rest) of the simulated atmosphere
to the heating, featuring similar eddy patterns as observed in
\citet{matsuno_1966,gill_1980,showman_2011,tsai_2014}. Then, in
Section \ref{subsub_section:steady_eddy} gradients of the momentum
fluxes in the simulated atmospheres are used to reveal the mechanisms
maintaining the zonal flows. We find similarities between momentum
transport, and the previous work of \citet{showman_2011} and
\citet{tsai_2014} but highlight additional complexities which merit
further study beyond the scope of this work (Debras et al., in
prep). In particular, the mean flow contributions to the momentum
transport are non-negligible. Our conclusions are stated in Section
\ref{section:conclusions}. Finally, the Appendix contains extended
detail of the flow structure (Appendix
\ref{app_section:flow_details}), results from a `shallow-hot Jupiter'
test (Appendix \ref{app_section:shj}), the derivation of the eddy-mean
flow interaction equation (Appendix
\ref{app_section:eddy_mean_derive}) and plots of additional terms
within this equation for our simulations (Appendix
\ref{app_section:eddy_terms}).

\section{Model setup}
\label{section:model_setup}

Results from this work are from simulations using the basic setups of
\citet{mayne_2014} and \citet{amundsen_2016}, which are differentiated
by the treatment of heating/cooling and we refer to these papers for
the basic numerical details. In \citet{mayne_2014}, the basic
equations solved by the UM are introduced and results from simulations
including a simple Newtonian relaxation, or radiative forcing scheme
are presented, where the temperature is simply relaxed to a
pre-calculated radiative equilibrium profile (hereafter termed TF
simulations referring to `Temperature Forced'). This work also defines
the various levels of simplification to the dynamical equations termed
`primitive', `shallow', `deep' and `full' in order of the most
simplified to the most complete. Essentially, the most approximated
equations are the `'primitive', assuming vertical hydrostatic
equilibrium, gravity constant with height and a
shallow--atmosphere. The `shallow' equations are formed by relaxing
the hydrostatic approximation, and the `deep' by further relaxing the
shallow--atmosphere approximation. Finally, the `full' equations
invoke none of these approximations. \citep[see][for
details]{mayne_2014}. \citet{amundsen_2016} features models using only
the `full' equations but incorporating a two-stream, dual band,
correlated-\textit{k} radiative transfer scheme, which has previously
been tested in \citet{amundsen_2014,amundsen_2017}, alongside minor
updates in the treatment of diffusion and minor parameter changes
(hereafter termed RT, referring to `Radiative Transfer',
simulation). The model uses SI units, and these are adopted throughout
this work.

Although TF simulations are computationally much cheaper than their RT
counterparts allowing us to reach extensive total integration times,
they have several key disadvantages. As TF simulations linearly relax
the temperature to a prescribed profile, they lack the inclusion of
atmospheric interactions (i.e. thermal emission and absorption), and
do not model the non--linear response of the atmosphere to significant
levels of heating and cooling both of which are captured in the RT
simulations. Additionally, the TF setups rely on initial calculation
of equilibrium profiles, often from simple and physically incomplete
1D models, further reducing their accuracy and also flexibility. Our
results exhibit clear differences between the TF and RT simulations
(see Section \ref{section:results}), some elements of which will be
caused by such issues.

\subsection{Damping}
\label{sub_section:model_damp}

Of particular note is the treatment of damping in the model. In a
physical fluid eddies and turbulence will act to cascade energy to
smaller scales, and ultimately convert kinetic energy into thermal
energy \citep[see discussion in][]{li_2010b}. Hot Jupiter GCM
simulations do not correctly capture such small scale
turbulence. Additionally, physical processes which are not explicitly
modelled can also lead to further dissipation, for example magnetic
braking in hot Jupiters \citep{menou_2012}.  Explicit damping is often
applied to account for these effects. However, we currently have no
way of constraining the strength of the dissipation
\citep{li_2010b,heng_2011}, except perhaps by matching simulated
offsets in hot Jupiter hotspots from the substellar point with
observed phase curve offsets, although processes not explicitly
modeled complicate this \citep{zellem_2014}. Therefore, such damping
and dissipation parametrisation are primarily used to achieve
numerical stability in a physically plausible way, but are not
robustly constrained. In particular the low pressure atmosphere, near
the outer boundary is sensitive to perturbations which grow as they
travel from higher pressure regions, and become unstable. For the
inner, high pressure, boundary dissipation from the numerical scheme
(i.e. discretisation) is more significant, and the fluid more
stable. However, as the pressure (and therefore density) increases
exponentially with depth into the hot Jupiter atmosphere, the angular
momentum for a given flow rate will increase. Therefore, even very
slow fluid flows at the inner boundary can significantly affect the
angular momentum budget \citep{mayne_2014}, and minor inaccuracies of
the numerical solver can drastically affect the simulated dynamical
structure \citep{cho_2015}.

In our model we include several forms of artificial
dissipation. Firstly, as detailed in \citet{mayne_2014} and updated in
\citet{amundsen_2016} we include an explicit diffusion of the zonal
flow (only). However, we incorrectly reported the coefficient
for this scheme previously, in \citet{amundsen_2016}, stating a single
coefficient for the zonal diffusion operator ( $K_\lambda\sim
0.16$). Actually, due to simplification in the routine performing the
diffusion several terms are implicitly included in the resulting
applied diffusion coefficient, $K^{\rm \prime}_\lambda$, such that it
is given by
\begin{multline}
K_\lambda^\prime=\frac{\Delta t K_\lambda}{(r^2\cos^2\phi(\Delta\lambda)^2)}.\\
\label{k_lambda}
\end{multline}
Where $\Delta t$ and $\Delta \lambda$ are the timestep and
longitudinal grid spacing, respectively \footnote{In practice the
  diffusion is set using a characteristic e--folding timescale,
  $K_\lambda=0.25\left(1-e^{\frac{-1}{t_K}}\right)$, where $t_{K}$ is
  the set number of simulation timesteps.}. We also include a vertical
`sponge layer' as detailed in \citet{mayne_2014} to damp vertical
velocities close to the outer, low pressure boundary, and represent
vertical waves propagating out of our modelled domain. For this work,
to allow us more flexibility, we have also included a vector laplacian
damping (to mimic an explicit viscosity) in the momentum equation. The
vector Laplacian of the vector wind field is calculated,
$\nabla^2(\mathbf{u}) \equiv
\nabla\left(\nabla\cdot\mathbf{u}\right)-\nabla\times\left(\nabla\times\mathbf{u}\right)$.
The resulting components of this provide additional terms to equations
(1), (2) and (3) of \citet{mayne_2014}, governing the $u$, $v$ and $w$
wind components, which are in the longitude ($\lambda$), latitude
($\phi$) and vertical directions ($r$), respectively. A separate
multiplicative coefficient is then prescribed for the horizontal and
vertical directions as $\frac{\nu_{\lambda,\phi}}{\rho}$ and
$\frac{\nu_{\mathrm r}}{\rho}$, respectively, where $\rho$ is the
density.

In effect the diffusion and vector Laplacian schemes apply the same,
physical equation, but differ via their calculation of the
coefficient.  In practice neither of these schemes operate on the
vertical component of velocity. Given our spatial resolution
($\sim10^5$ \& $\sim10^6$\,m in the vertical and horizontal
directions, respectively), and maximum windspeeds ($\sim 10^2$,
$\sim 10^3$\,ms$^{-1}$, in the vertical and horizontal directions,
respectively), the vertical flow is much more accurately
resolved. However, the vector laplacian scheme is applied to both the
zonal and meridional directions, whereas the diffusion scheme is only
applied in the zonal direction \citep[see discussion
in][]{amundsen_2016}. Given typical simulation parameters
($\Delta t\sim 120$\,s, $r\sim 10^{8}$\,m, $\Delta \lambda\sim 0.04$\,
radians, see Table \ref{basic_par}) and densities throughout our
domain which range from $\sim$1 to 2$\times 10^{-7}$\,kg\,m$^{-3}$,
$K_\lambda^\prime\sim\frac{8\times 10^{-12} K_\lambda}{\cos^2\phi}$
and the applied vector laplacian coefficient,
$\nu^{\prime}_{\lambda,\phi}\sim (10^{-7}\rightarrow
1)\,\nu_{\lambda,\phi}$. Therefore, the maximum applied diffusion
coefficient (at the equator) is typically between 12 and five orders
of magnitude smaller than the vector laplacian, for the same input
value. The spatial variation of the coefficient also leads to the two
schemes behaving differently. For the vector laplacian, the variation
of the coefficient with density effectively means this process mimics
kinematic visocity and will act most strongly on the horizontal wind
in the upper, low pressure regions of the atmosphere. However, the
diffusion coefficient varies strongly with latitude such that it
reduces towards the pole, meaning this scheme effectively acts to
suppress grid--scale noise in the dominant zonal flow. In Appendix
\ref{app_section:shj}, we demonstrate the effect of varying the
`strength' of the diffusion on a shallow--hot Jupiter test case
\citep[introduced by ][]{menou_2009}. It is important to reiterate
that although generally physically motivated, these
damping/dissipation mechanisms are poorly constrained for hot
Jupiters, and are primarily used to achieve numerical stability.

Additional horizontal damping at the inner or high pressure boundary
is also often employed \citep[see discussion in][]{cho_2015}. This
`bottom drag' is employed similar to the Rayleigh friction schemes
used to model the frictional damping of the Earth's surface on
horizontal winds, and is motivated by the potential for magnetic drag
in the deep atmosphere, \citep{rogers_2014}. However, the strength of
this drag, and vertical profile are very poorly constrained, and it is
not clear if a Rayleigh type drag is appropriate at all. Of course,
when employing such a bottom drag GCM simulations are much more robust
to conservation of total angular momentum, and less sensitive to
initial conditions \citep{liu_2013,cho_2015}. In this work \emph{we do
  not include a bottom drag damping} as we are, in part, interested in
the evolution of the deep atmosphere, and its interaction with the
lower pressure regions.

\subsection{Model variations}
\label{sub_section:model_var}

In Table \ref{basic_par} we state the main parameters for both the TF
and RT simulations. These differ slightly as the TF simulations were
setup to match the work of \citet{heng_2011} and the RT simulations to
be compared with \citet{showman_2009} but these differences are
unlikely to affect our main conclusions. We have run a large set of
simulations to explore various scenarios, but only directly report
results here for a subset. We have also performed additional
simulations of the `shallow-hot Jupiter' setup of \citet{menou_2009},
the results of which are discussed in Appendix
\ref{app_section:shj}. Table \ref{model_names} details the elapsed
simulation time, parameters adjusted from the default (shown in Table
\ref{basic_par}), dynamical equation set \citep[following the
nomenclature of][]{mayne_2014} and the `short name' adopted throughout
this manuscript.

Only a single RT simulation is included as the TF simulations allow
greater, direct, control of the heating and run significantly
faster. Although TF simulations are less physically accurate than
their RT counterparts \citep[see discussion
in][]{showman_2009,amundsen_2016}, they have still proved useful in
the study of hot Jupiter (and other) atmospheres. Recent examples of
the use of TF simulations include exploration of the efficiency of
dynamical redistribution of heat \citep{komacek_2016} and the
dependence on the atmospheric flow on bulk composition
\citep{zhang_2017}. For the RT simulation we do not include TiO and VO
formation as evidence for its presence has only been suggested for
Wasp-121b \citep{evans_2016} so far, and not HD~209458B.

\begin{table*}
  \caption{Value of the standard parameters for the temperature forced (TF) and radiative transfer (RT) simulations.}
\label{basic_par}
\centering
\begin{tabular}{lcc}
  \hline\hline
  Quantity&TF&RT\\
  \hline
  Horizontal resolution&\multicolumn{2}{c}{$144_\lambda$, $90_\phi$}\\
  Vertical resolution, $N_z$&\multicolumn{2}{c}{66}\\
  Dynamical Timestep (s)&1200&30\\
  Radiative Timestep (s)&-&150\\
  Initial inner boundary pressure, $p_{\rm max}$ (Pascals, Pa)&$220\times 10^5$&$200\times 10^5$\\
  Rotation rate, $\Omega$ (s$^{-1}$)&\multicolumn{2}{c}{$2.06\times 10^{-5}$}\\
  Radius, $R_{\rm p}$ (m)&$9.44\times10^7$&$9.0\times 10^7$\\
  Radius to outer boundary, $R_{\rm top}$ (m)&$1.1\times 10^7$&$9.0\times 10^6$\\
  Surface gravity, $g_{\rm p}$ (ms$^{-2}$)&9.42&10.79\\
  Specific heat capacity (constant pressure), $c_{\rm p}$ (Jkg$^{-1}$K$^{-1}$)&14\,308.4&13\,000.0\\
  Ideal gas constant, $R$ (Jkg$^{-1}$K$^{-1}$)$^{(1)}$&4593&3556.8\\
  Diffusion setting, $K_\lambda$ (see Section \ref{sub_section:model_damp} for applied coefficient, $K^\prime_\lambda$) &\multicolumn{2}{c}{0.158}\\
  Horizontal vector laplacian damping coefficient, $\nu_{\lambda, \phi}$&0.1&0.0\\
  Vertical vector laplacian damping coefficient, $\nu_{\rm r}$&\multicolumn{2}{c}{0.0}\\
  Vertical, `sponge', damping coefficient $R_{w}$&\multicolumn{2}{c}{0.15}\\
  Temperature characterising intrinsic luminosity of the planet (K), $T_{\rm int}$&-&100\\
  \hline
\end{tabular}
\end{table*}

\begin{table*}
  \caption{Details, alongside short names, of the key simulations used in this
    work (a more extensive set of simulations were performed, but we only
    highlight those where results are directly used in this work). The
    heating/cooling scheme (temperature forced, TF, or radiative transfer,
    RT), total elapsed simulation time (rounded down to the nearest
    100\,days), variables or settings which are different from the
    standard models (see Table \ref{basic_par} for the basic parameters),
    and the sophistication of the dynamical equations is shown
    \citep[see][for definition]{mayne_2014}. The time sampling in all
    cases was ten days, with an additional run outputting at 1\,day
    performed only over the first 100\,days for each simulation. $p$ is
    the pressure, $\tau_{\rm rad}$ the radiative timescale and $T_{\rm
      eq}$ the equilibrium temperature.}
\label{model_names}
\centering
\begin{tabular}{lcclll}
  \hline\hline
  Short Name&Heating/Cooling&Length (Earth days)&\multicolumn{2}{l}{Adjusted parameters}&Equation set\\
  \hline
  Std Prim&TF&10\,200&\multicolumn{2}{l}{$R_{w}=0.20$}&``Primitive''\\
  Std Full&TF&13\,300&\multicolumn{2}{l}{-}&``Full''\\
  Std RT&RT&1\,600&\multicolumn{2}{l}{-}&``Full''\\
  Reduced $p_{\rm max}$&TF&15\,400&\multicolumn{2}{l}{$p_{\rm max}=10^6$\,Pa, $R_{w}=0.20$}&``Full''\\
  \multirow{2}{*}{Deep $\Delta T_{\rm eq\rightarrow pole}$}&\multirow{2}{*}{TF}&\multirow{2}{*}{10\,800}&\multirow{2}{*}{$p>10^6 {\rm Pa}$:}&$\tau_{\rm rad}(p)=\tau_{\rm rad}(10^6 {\rm Pa})$&\multirow{2}{*}{``Full''}\\
  &&&&$T_{\rm eq}=T_{\rm eq}+1\,000.0\sin^2(\phi)$&\\
  \hline
\end{tabular}
\end{table*}

Our TF simulations have been evolved for much longer than those of
\citet{mayne_2014}, allowing us to explore the long term evolution for
simulations adopting various levels of simplification to the dynamical
equations (e.g., Std Prim \& Std Full)\footnote{Note we have also
  performed simulations using the ``shallow'' and ``deep''
  versions.}. The profiles for the heating in the TF simulations are
those of \citet{mayne_2014}, which are adjusted from \citet{heng_2011}
which, in turn, are based on those of \citet{iro_2005}. The radiative
timescale is an (approximately) exponentially increasing function of
pressure, ranging from $\sim 10^3-10^8$ over pressures of
$\sim10^2-10^6$\,Pa, and is infinite for pressures of $10^6$\,Pa or
higher.

As discussed in \citet{mayne_2014} evolution of the deep atmosphere is
very gradual and is likely to require extremely long simulation
integrations to reach a steady state. \citet{mayne_2014} note a
gradual increase in the difference between the equatorial temperature
and that of the pole for the deep atmosphere. In several of their
simulations the poles are warmed, and the equator cooled, which given
the absence of forcing for the standard models can only be caused by
compression/expansion or advection potentially by material lifting
over the hotspot and falling towards the poles. A similar temperature
evolution of the deep atmosphere can be seen in the RT simulations of
\citet{amundsen_2016}, Figure 7. However, for the models where the
initial profile is hotter, presented in \citet{amundsen_2016}, the
latitudinal temperature gradient is reduced. This behaviour hints that
the 3D model is evolving to a hotter equilibrium state than the one
with which it is initialised, however this is beyond the scope of this
work, where we are more interested in the effects of different
scenarios for the unconstrained deep atmosphere, and will be discussed
in an upcoming publication (Tremblin, et al., submitted). Therefore,
we have simply performed a numerical experiment to explore the limits
of this effect, and determine whether it will be significant for the
flows in the upper atmosphere.  In the simulation Deep $\Delta T_{\rm
  eq\rightarrow pole}$ we impose an additional, latitudinal
temperature gradient in the deep atmosphere using a constant radiative
timescale given as the value at $10^6$\,Pa (10\,bar). For this setup
although the equilibrium temperature is increased by
$(1\,000.0\,K)\sin^2(\phi)$ (where $\phi$ is latitude) the final
temperature difference between the equator and pole will be much less
as this must first compensate for the existing contrast enforced by
the original equations of $\sim-500$\,K
\citep{heng_2011,mayne_2014}. We have also run a simulation with the
bottom boundary at $10^6$\,Pa to explore the effect of omitting the
deep region entirely (Reduced $p_{\rm max}$). It is important to note
that these simulation setups are explorative and performed to
investigate the interaction of the deep and low pressure
atmosphere. We are unlikely to obtain observational constraints on the
state of the deep atmosphere of hot Jupiters in the near future and so
are forced to make choices regarding how to model this region.

All of our models are initialised with zero winds, and in solid body
rotation with a hydrostatically balanced atmosphere. The initial
temperature pressure profile used is either a midway profile between
the hottest and coldest profiles for the TF simulations, or a globally
averaged radiative equilibrium profile for the RT simulations
\citep[for further details refer to][]{mayne_2014,amundsen_2016},
unless otherwise stated.

\section{Results}
\label{section:results}

In this section we present results for the dynamical state of our
simulated atmospheres. In Section \ref{sub_section:jet_robust} we
demonstrate that the occurrence of a super rotating equatorial jet is
robust in our simulations, and present the flow regime for some
examples. In Section \ref{sub_section:deep_atmosphere} we demonstrate
that the deep atmosphere is still evolving in our simulations, and
show that this effect may disrupt the flow regime in the upper
atmosphere, but only if strongly forced over long timescales. Finally,
in Section \ref{sub_section:jet_pumping} we present the mean flow and
eddy interaction, demonstrating that the jet in our simulations is
driven by a mix of eddy and mean flow momentum transport.

\subsection{Zonal flow}
\label{sub_section:jet_robust}

\begin{figure*}
\begin{center}
  \subfigure[Std Prim: 200-1\,200\,days]{\includegraphics[width=9.0cm,angle=0.0,origin=c]{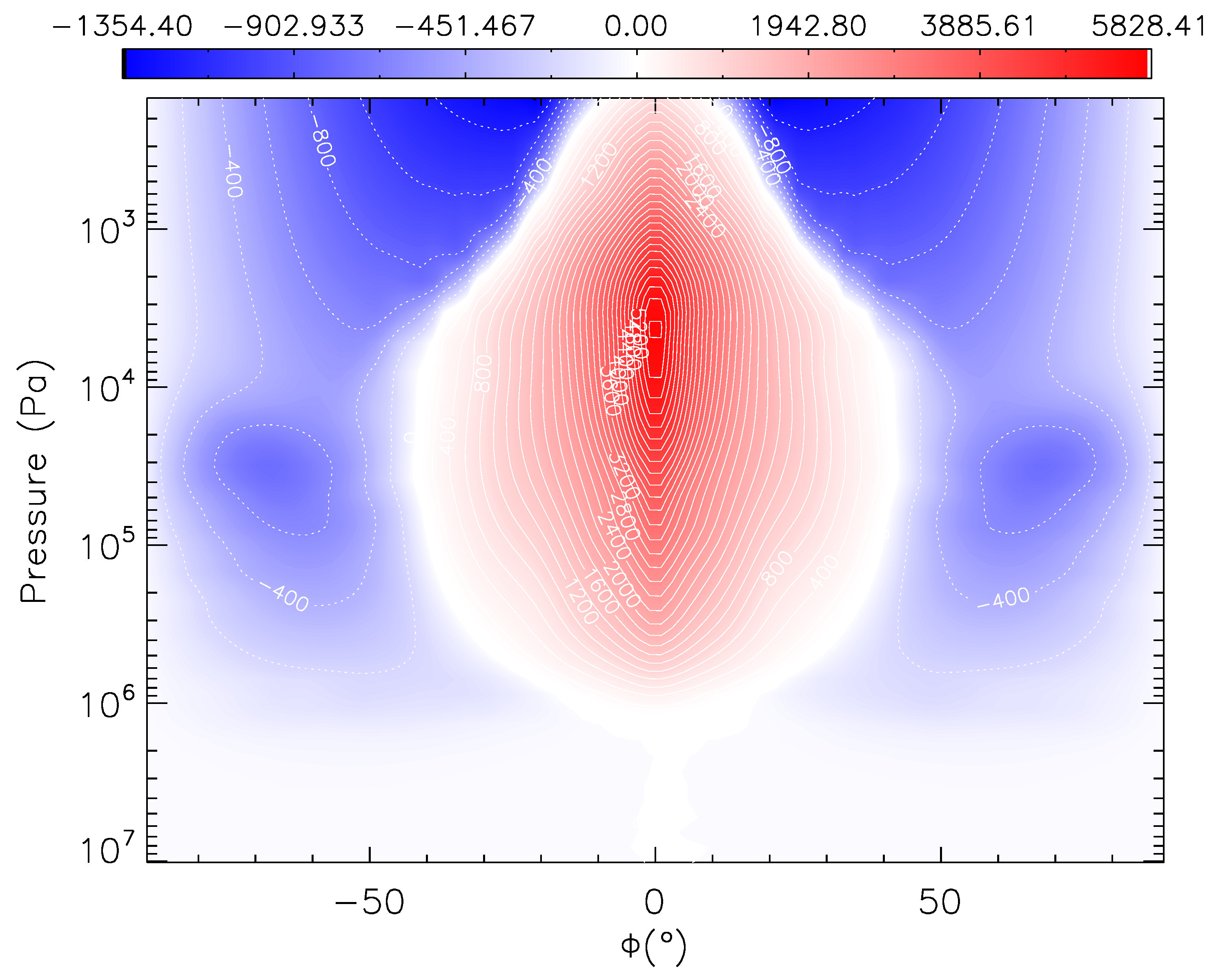}\label{std_prim_200_1200_uvel_bar}}
  \subfigure[Std Full: 200-1\,200\,days]{\includegraphics[width=9.0cm,angle=0.0,origin=c]{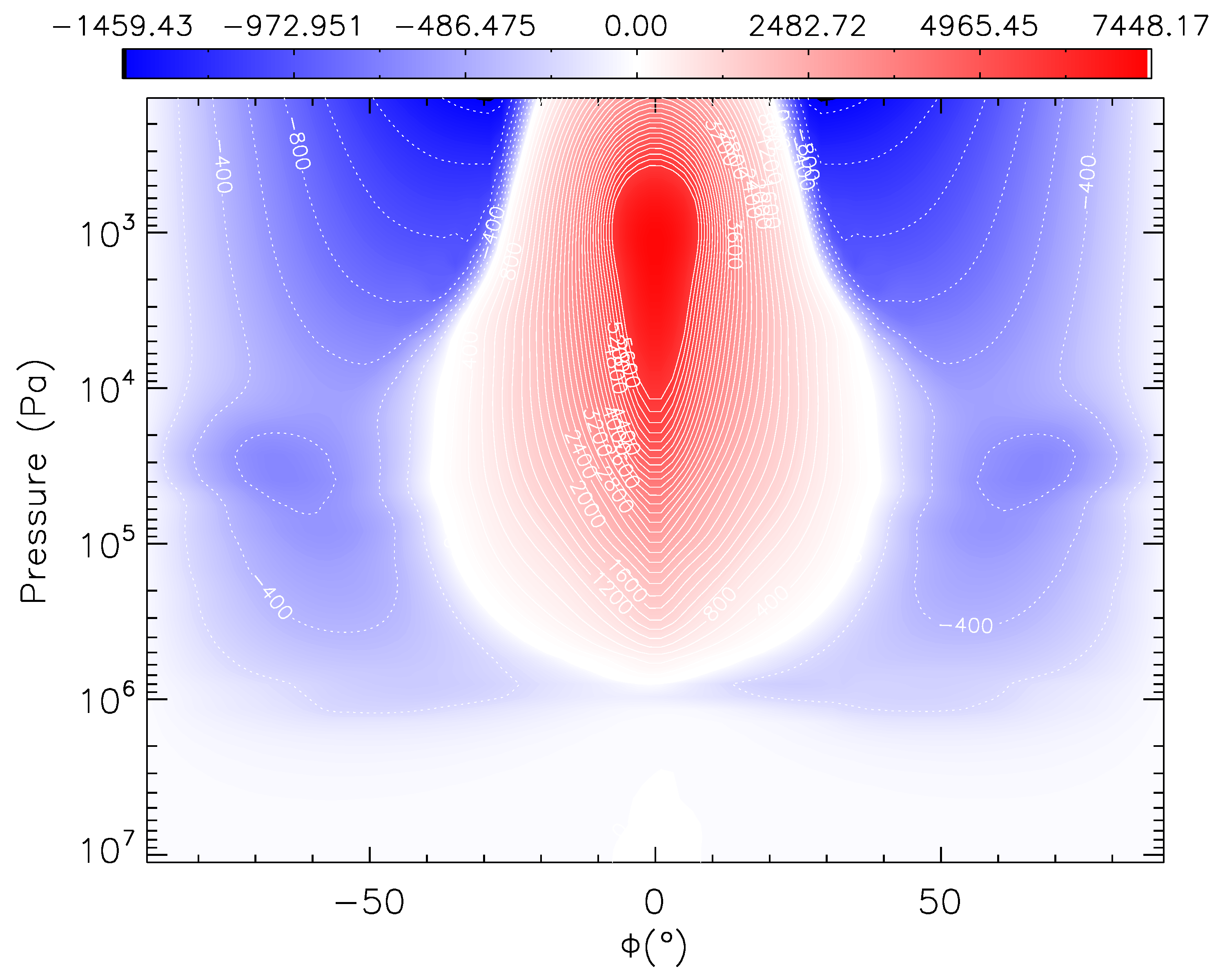}\label{std_full_200_1200_uvel_bar}}
  \subfigure[Std Prim: 9\,000-10\,000\,days]{\includegraphics[width=9.0cm,angle=0.0,origin=c]{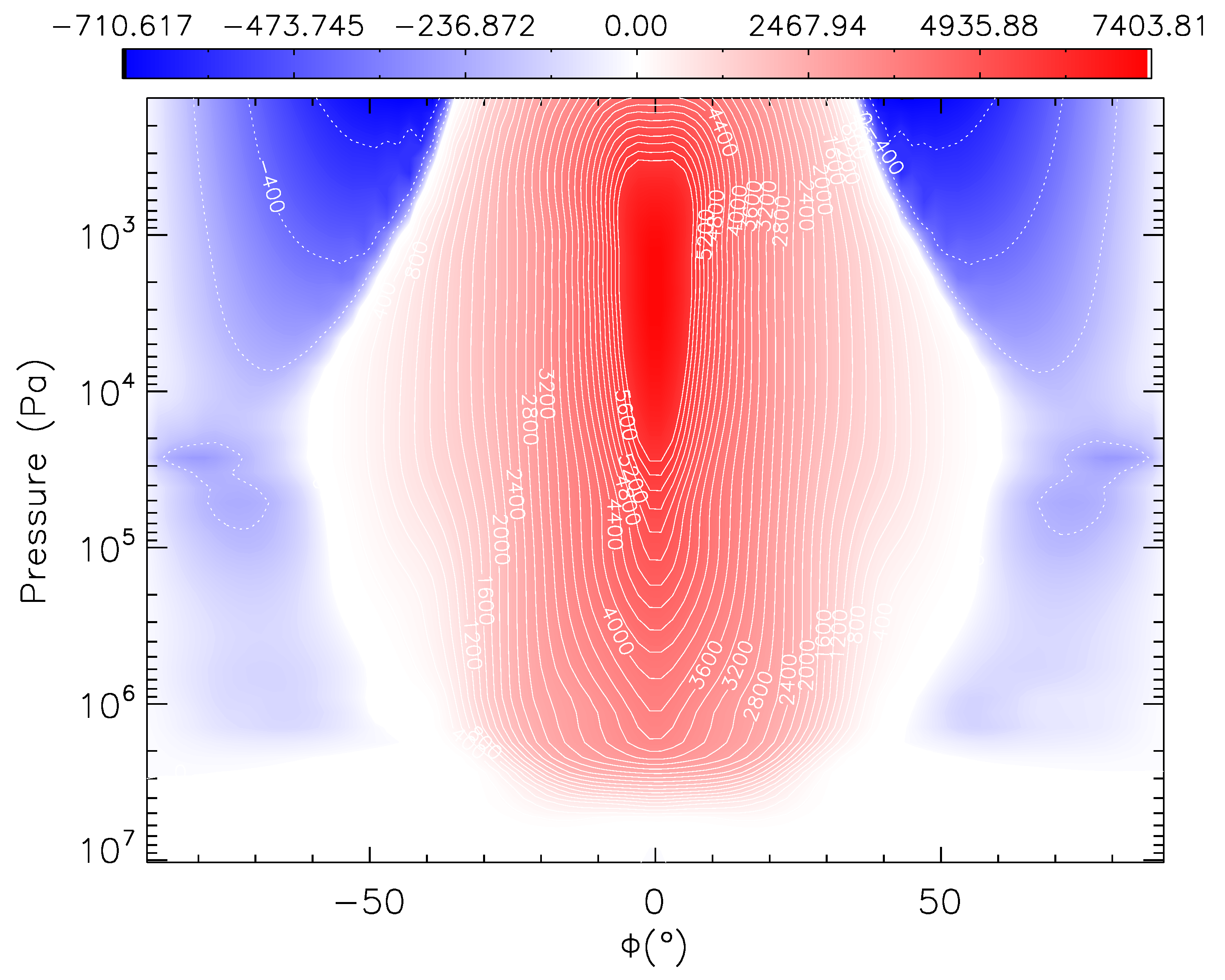}\label{std_prim_9000_10000_uvel_bar}}
  \subfigure[Std Full: 9\,000-10\,000\,days]{\includegraphics[width=9.0cm,angle=0.0,origin=c]{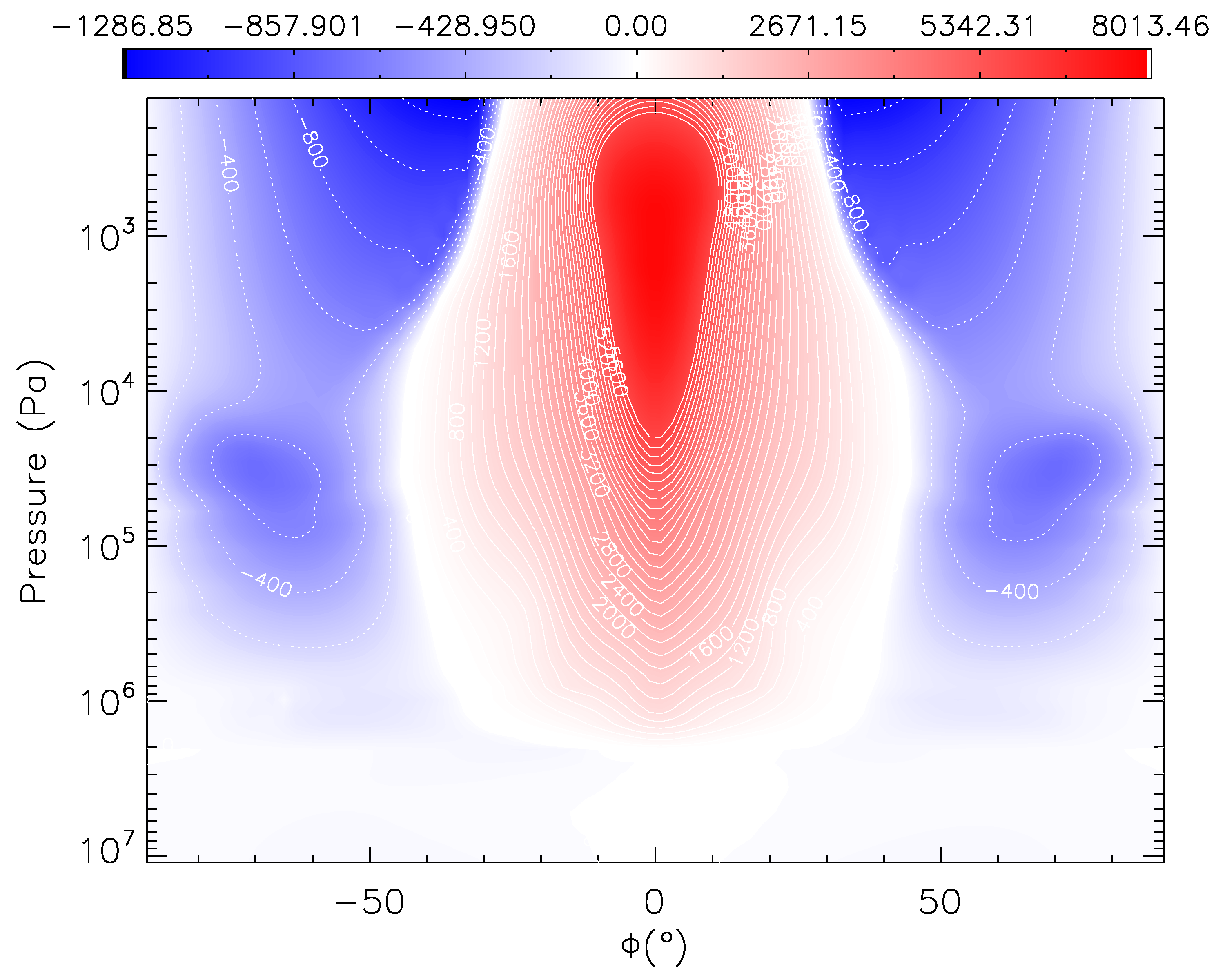}\label{std_full_9000_10000_uvel_bar}}
\end{center}
\caption{Figure showing the zonal and temporal mean of the zonal wind
  (ms$^{-1}$) as a function of latitude ($\phi^{\degree}$) and
  pressure ($\log_{10}(p\,[{\rm Pa}])$), for the Std Prim and Std Full
  simulations (see Table \ref{model_names} for explanation of
  simulation names), \textit{left} and \textit{right columns},
  respectively. The temporal averaging periods are 200--1\,200 and
  9\,000--10\,000\,days, shown as the \textit{top}, and \textit{bottom
    rows}, respectively. \label{std_uvel_bar}}
\end{figure*}

Figure \ref{std_uvel_bar} shows the zonal and temporally averaged
(mean) zonal wind, where red is prograde and blue is retrograde for
the TF simulations where the completeness of the dynamical equations
solved is varied (Std Prim and Std Full, \textit{left} and
\textit{right columns}, respectively), at two different epochs
(200--1\,200 \& 9\,000--10\,000\,days as the \textit{top} and
\textit{bottom rows}, respectively). It is clear that the jet evolves
over the period of about 10\,000 Earth days ($\sim$3\,000 rotation
periods). The breadth, in latitude, and depth range, as well as the
peak prograde velocity all increase. Additionally, the morphology of
the jet changes with the level of assumptions used in the dynamical
equations. For the more simplified case the jet is generally slower
and covers a narrower range of pressures, whilst being slightly
broader in latitude\footnote{Results from the simulations of
  intermediate complexity (i.e. `shallow' and `deep' are consistent
  with this trend.}. The difference found when moving from `full' to
`primitive' is similar to what one might expect when reducing the
rotation rate. However, despite these changes it is clear that,
broadly and qualitatively speaking, the same jet structure is
apparent; a prograde equatorial jet of a few kms$^{-1}$ flanked by
retrograde jets, covering a broad range of pressures down to about
$10^6$\,Pa.

\begin{figure}
\begin{center}
  \subfigure[Std RT: 600-1\,600\,days]{\includegraphics[width=9.0cm,angle=0.0,origin=c]{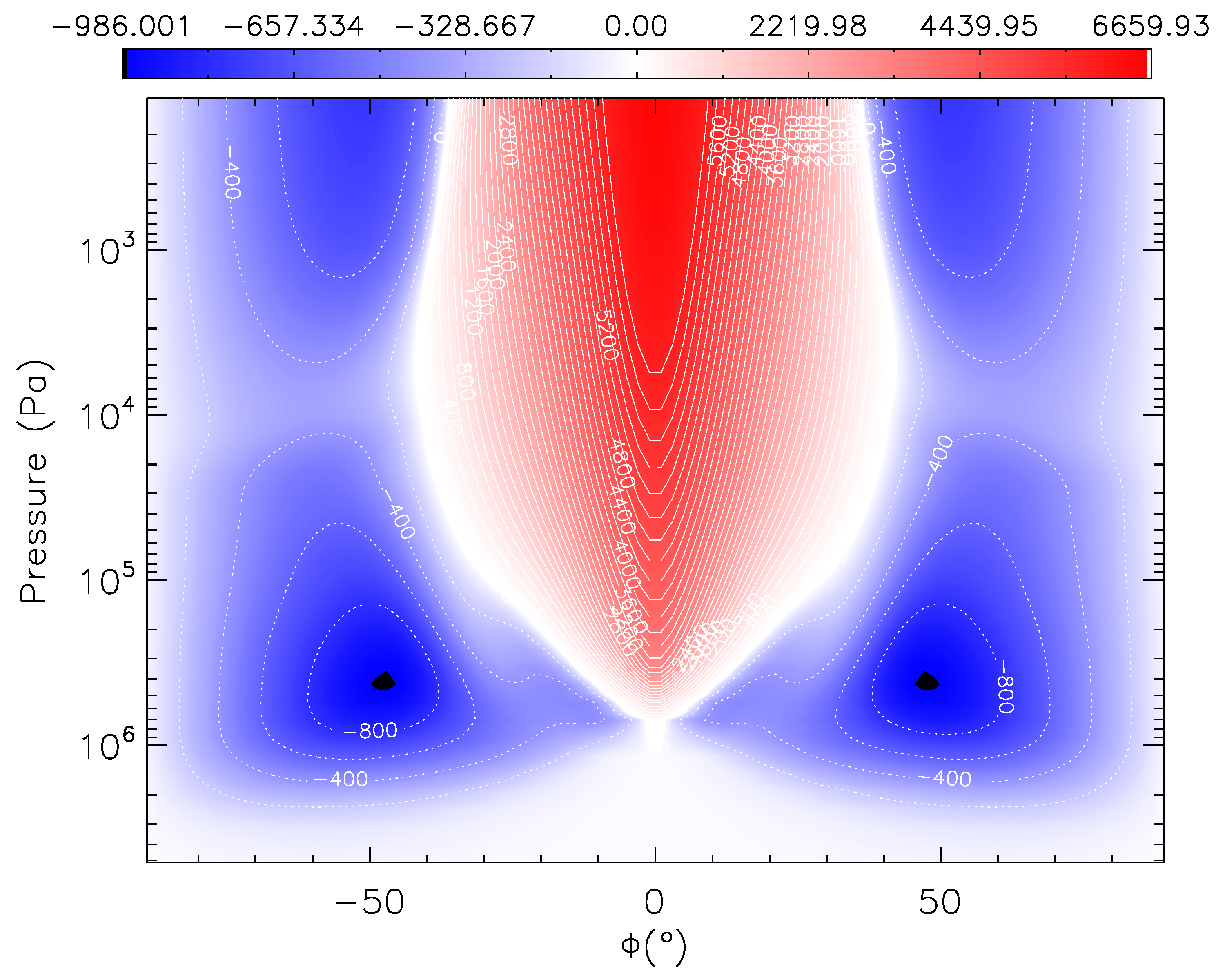}\label{rt_notiovo_full_600_1600_uvel_bar}}
\end{center}
\caption{Figure showing the zonal and temporal mean of the zonal wind
  (ms$^{-1}$) for the Std RT simulation (see Table \ref{model_names}
  for explanation of simulation names) as a function of latitude
  ($\phi^{\degree}$) and pressure ($\log_{10}(p\,[{\rm Pa}])$). The
  temporal averaging periods of 600--1\,600\,days was chosen as the
  latest time available. Note the reduced extent in pressure range due
  to lower achieved pressures at the base of the simulated atmosphere,
  compared to the Std Prim and Std Full
  simulations. \label{rt_uvel_bar}}
\end{figure}

Figure \ref{rt_uvel_bar} shows the results for the zonal flow for the
Std RT simulation, in the same format as Figure \ref{std_uvel_bar},
but for 600-1\,600\,days. However, the Std RT simulation bottom
boundary is placed at lower pressures, and this combined with the
subsequent evolution means the highest available pressure is slightly
lower than that of the Std Prim or Std Full cases. Additionally, the
much shorter total elapsed simulation time, compared to the TF
simulations, means that the deeper atmosphere is likely to still be
evolving. In fact, as we demonstrate in Section
\ref{sub_section:deep_atmosphere}, and present in Figure
\ref{conservation_layer}, the kinetic energy of the atmosphere appears
to still be evolving at pressure higher than $\sim 10^6$\, Pa. As
discussed in \citet{showman_2009} and \citet{amundsen_2016}, moving
from a temperature forcing to more complete radiative transfer scheme
alters the simulated dynamics of the atmosphere. For the RT simulation
the jet is slower and maintains a broader latitude profile at lower
pressures when compared to the most comparable standard TF simulation
(from simulations not presented here a similar result is found when
reducing the gravity, or slowing the rotation speed of the
planet). However, the qualitative result is still consistent with that
of the Std Prim and Std Full simulations.

The set of results presented here, alongside further simulations (not
presented), show broad qualitative agreement and are consistent with
previously published results
\citep[e.g.][]{showman_2009,heng_2011,dobbs_dixon_2013}, in that all
produce a prograde equatorial jet. However, it is also clear that
various parameter choices can affect the flow, and create differences
which can be seen even in a relatively coarse measure such as the
zonally and temporally averaged zonal wind plots

\begin{figure*}
\begin{center}
 \subfigure[Std Prim: 4.69$\times 10^5$\,Pa, 1\,200\,days]{\includegraphics[width=9.0cm,angle=0.0,origin=c]{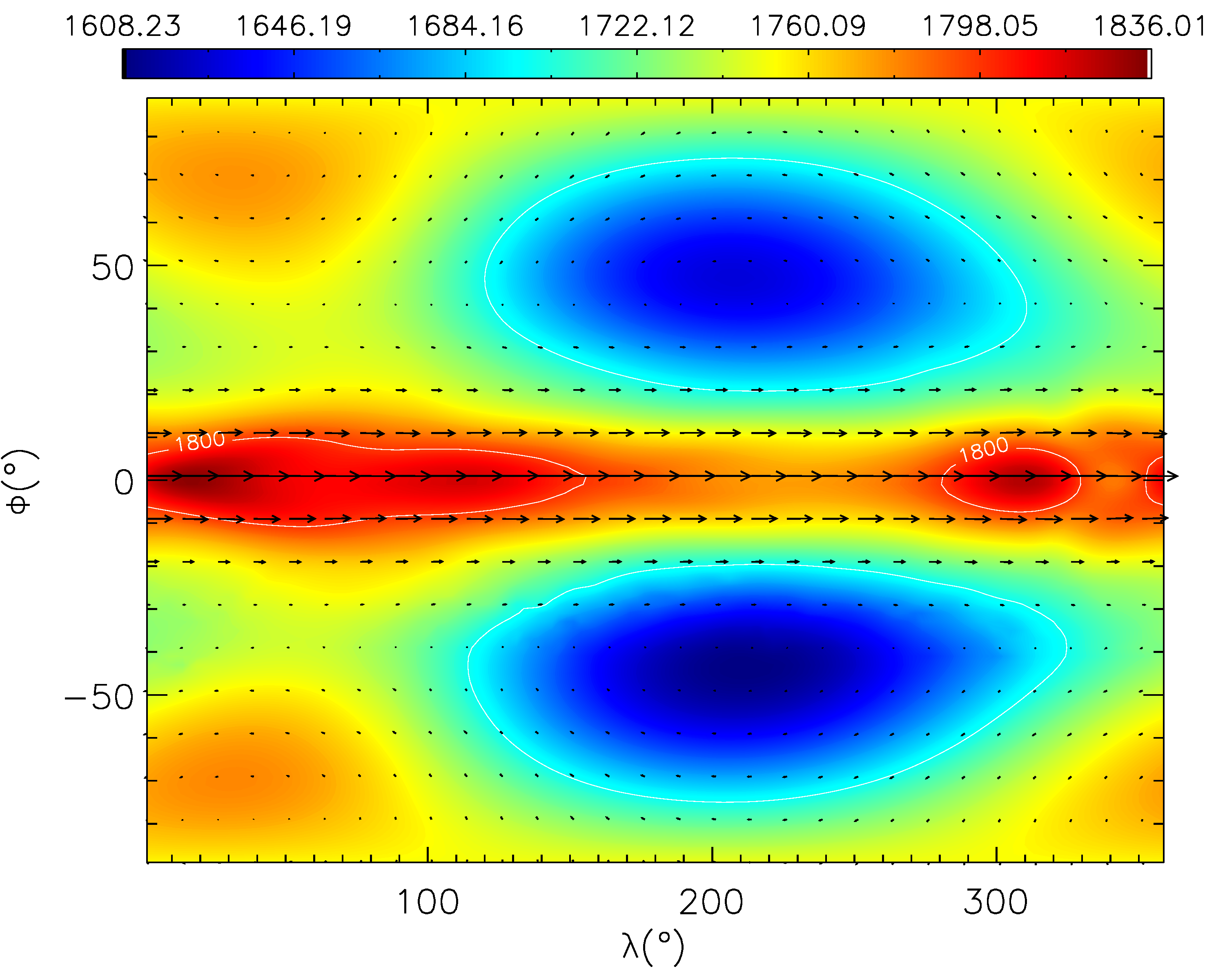}\label{std_prim_4_69e5_1200_slice}}
  \subfigure[Std Full: 4.69$\times 10^5$\,Pa, 1\,200\,days]{\includegraphics[width=9.0cm,angle=0.0,origin=c]{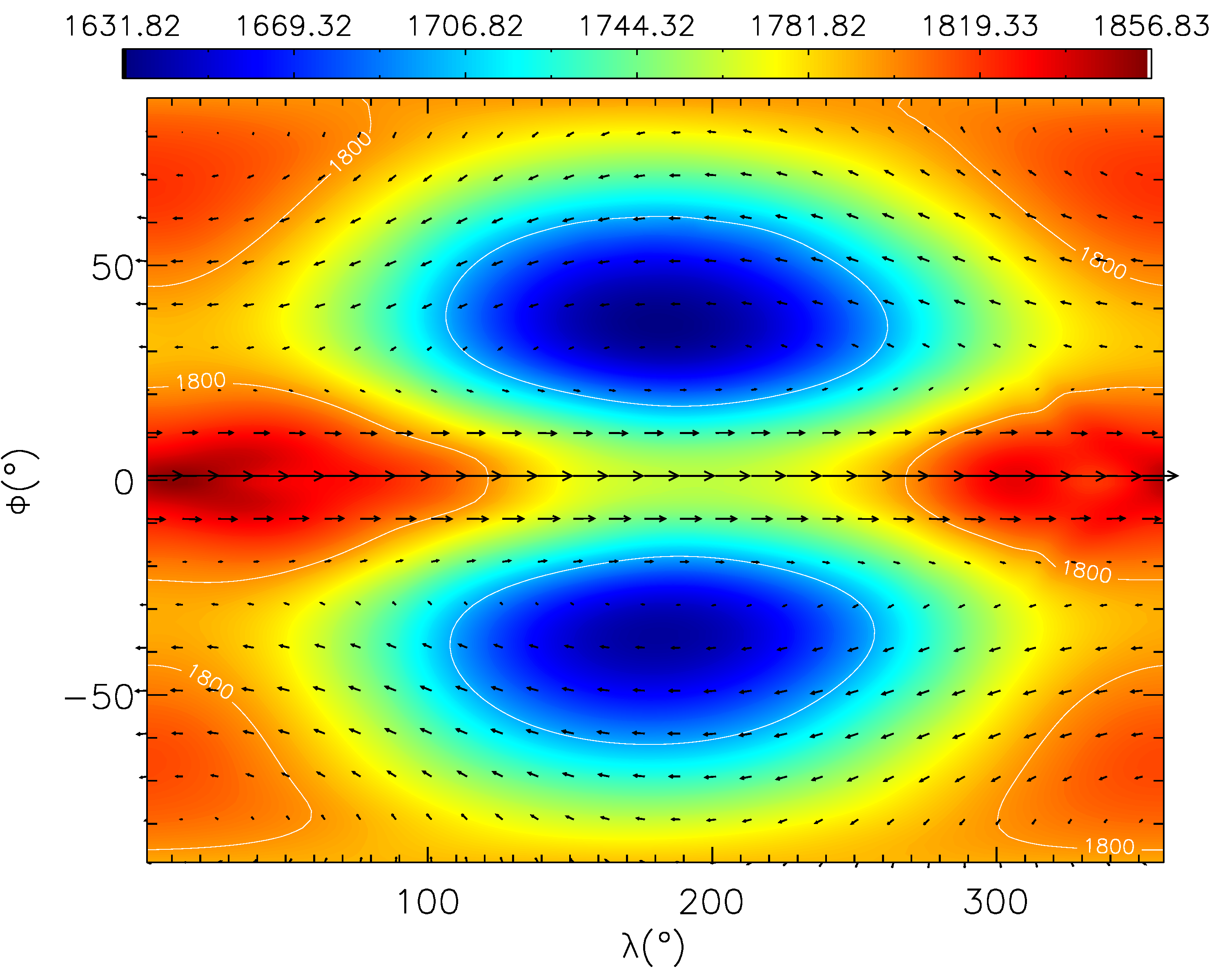}\label{std_full_4_69e5_1200_slice}}
 \subfigure[Std Prim: 21.9$\times 10^5$\,Pa, 1\,200\,days]{\includegraphics[width=9.0cm,angle=0.0,origin=c]{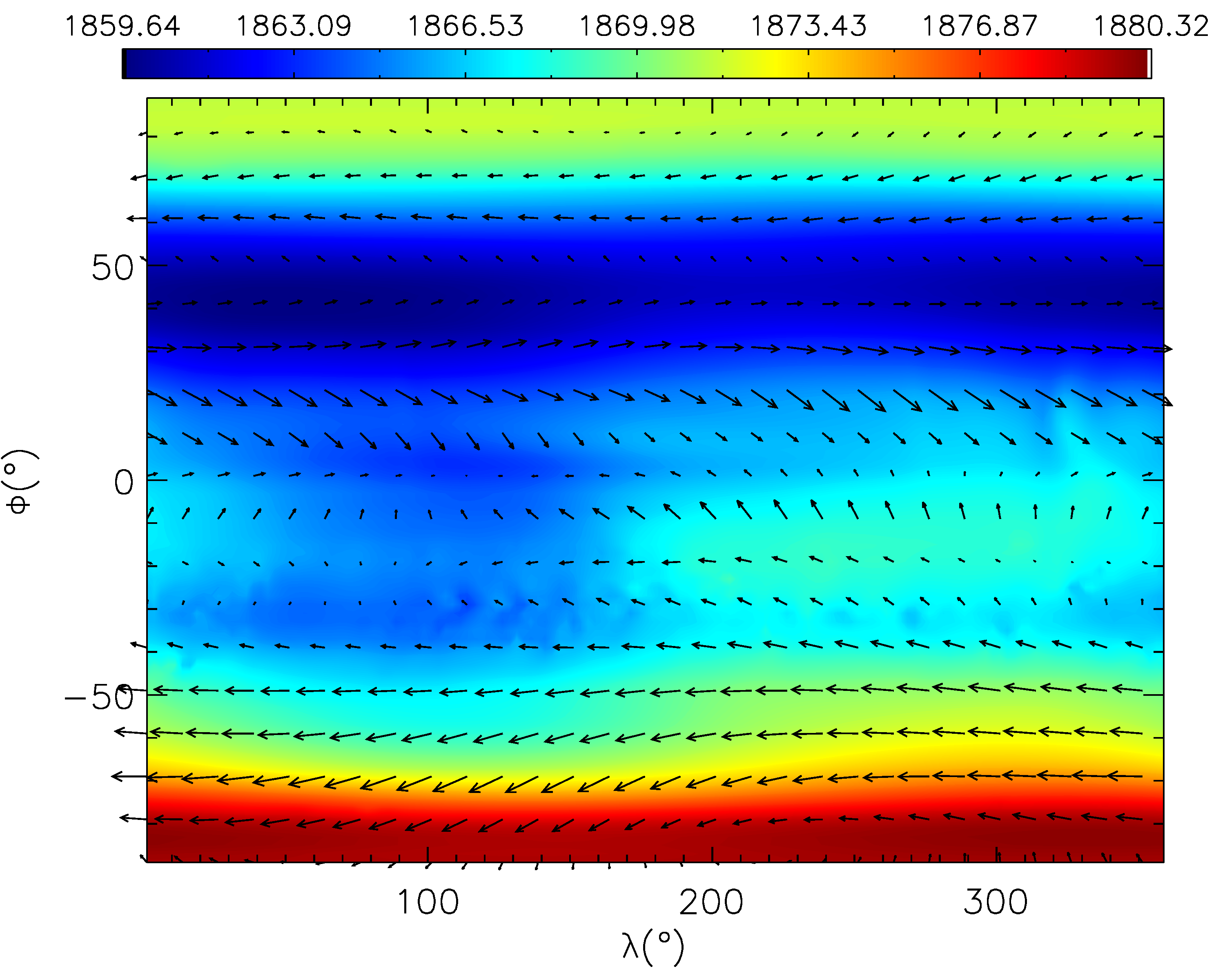}\label{std_prim_21_9e5_1200_slice}}
 \subfigure[Std Full: 21.9$\times 10^5$\,Pa, 1\,200\,days]{\includegraphics[width=9.0cm,angle=0.0,origin=c]{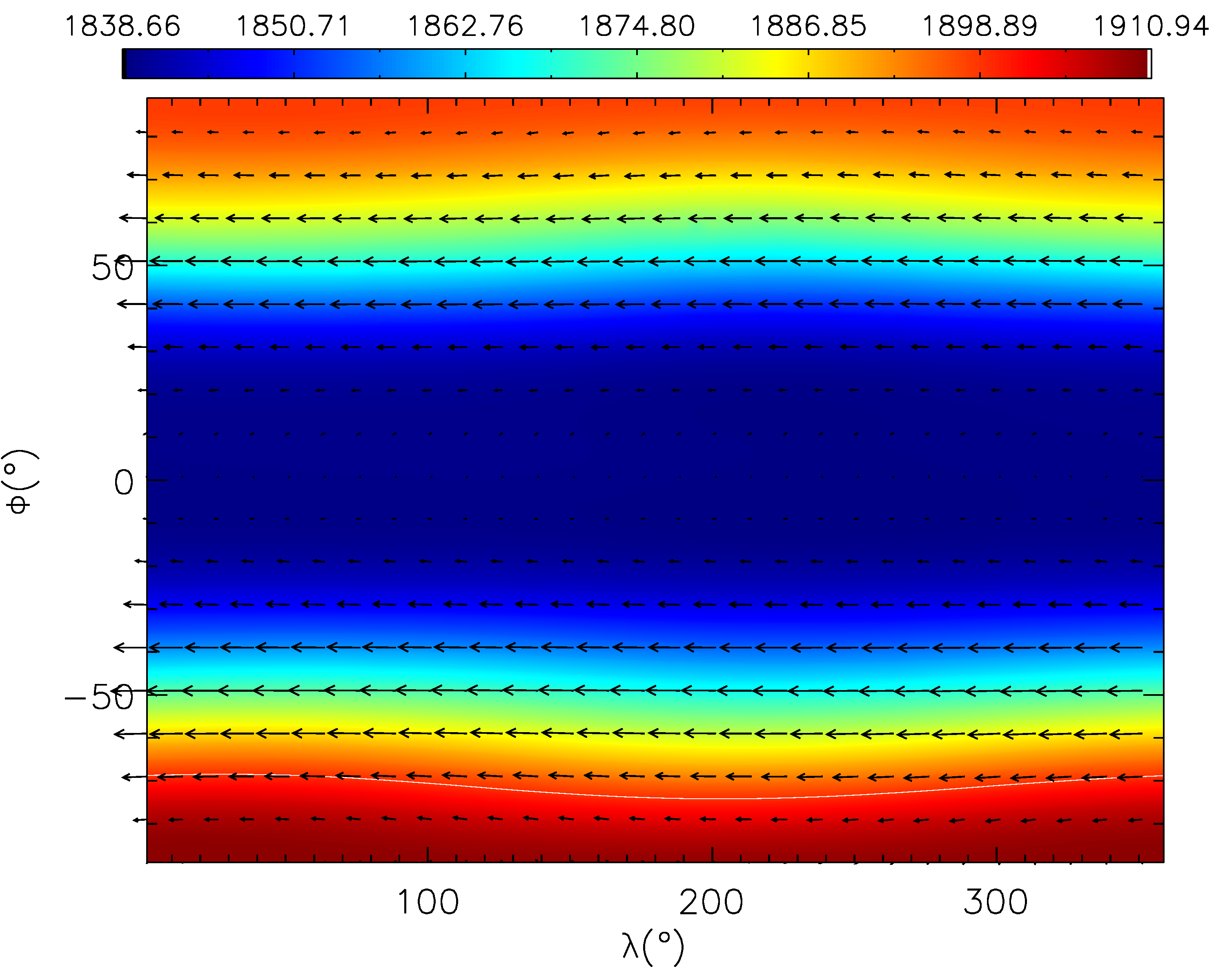}\label{std_full_21_9e5_1200_slice}}
\end{center}
\caption{Figure showing temperature (K, contours) and horizontal wind
  velocity (ms$^{-1}$, vector arrows, note: the number of arrows has
  been reduced from the simulation resolution in order to aid
  interpretation) at isobaric surfaces (against latitude,
  $\phi^{\degree}$ and longitude, $\lambda^{\degree}$) of
  4.69$\times 10^5$ and 21.9$\times 10^5$\,Pa (\textit{top} and
  \textit{bottom rows}, respectively) after 1\,200\,days for the Std
  Prim and Std Full simulations as the \textit{left} and \textit{right
    columns}, respectively (see Table \ref{model_names} for
  explanation of simulation names). The maximum magnitudes of the
  horizontal velocities are $\sim$150, $\sim$200, $\sim$10 \&
  3\,ms$^{-1}$ for the \textit{top left}, \textit{top right},
  \textit{bottom left} \& \textit{bottom right panels},
  respectively. \label{slice_std_1200_bot}}
\end{figure*}

\begin{figure*}
\begin{center}
  \subfigure[Std Prim: 4.69$\times 10^5$\,Pa, 10\,000\,days]{\includegraphics[width=9.0cm,angle=0.0,origin=c]{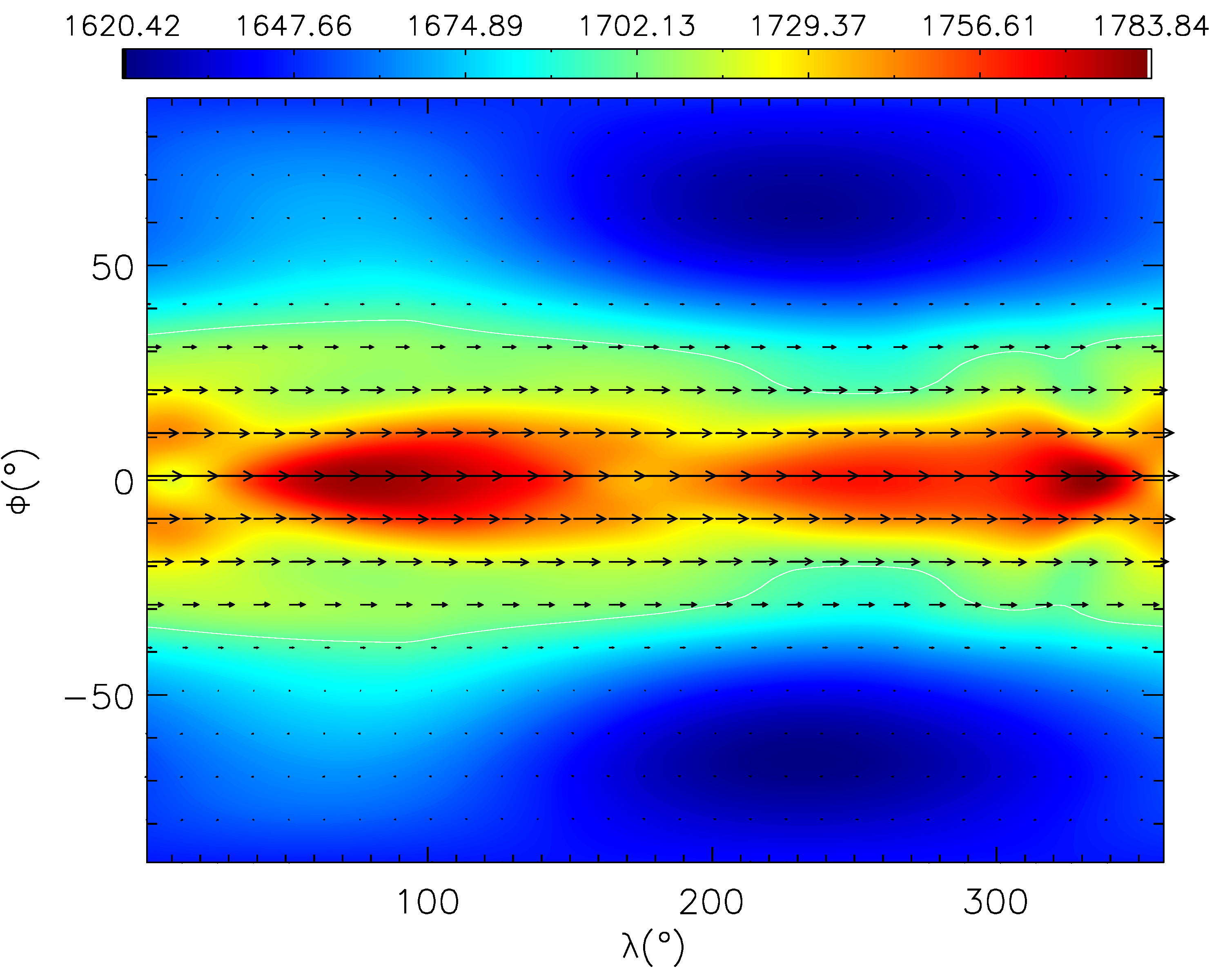}\label{std_prim_4_69e5_10000_slice}}
  \subfigure[Std Full: 4.69$\times 10^5$\,Pa, 10\,000\,days]{\includegraphics[width=9.0cm,angle=0.0,origin=c]{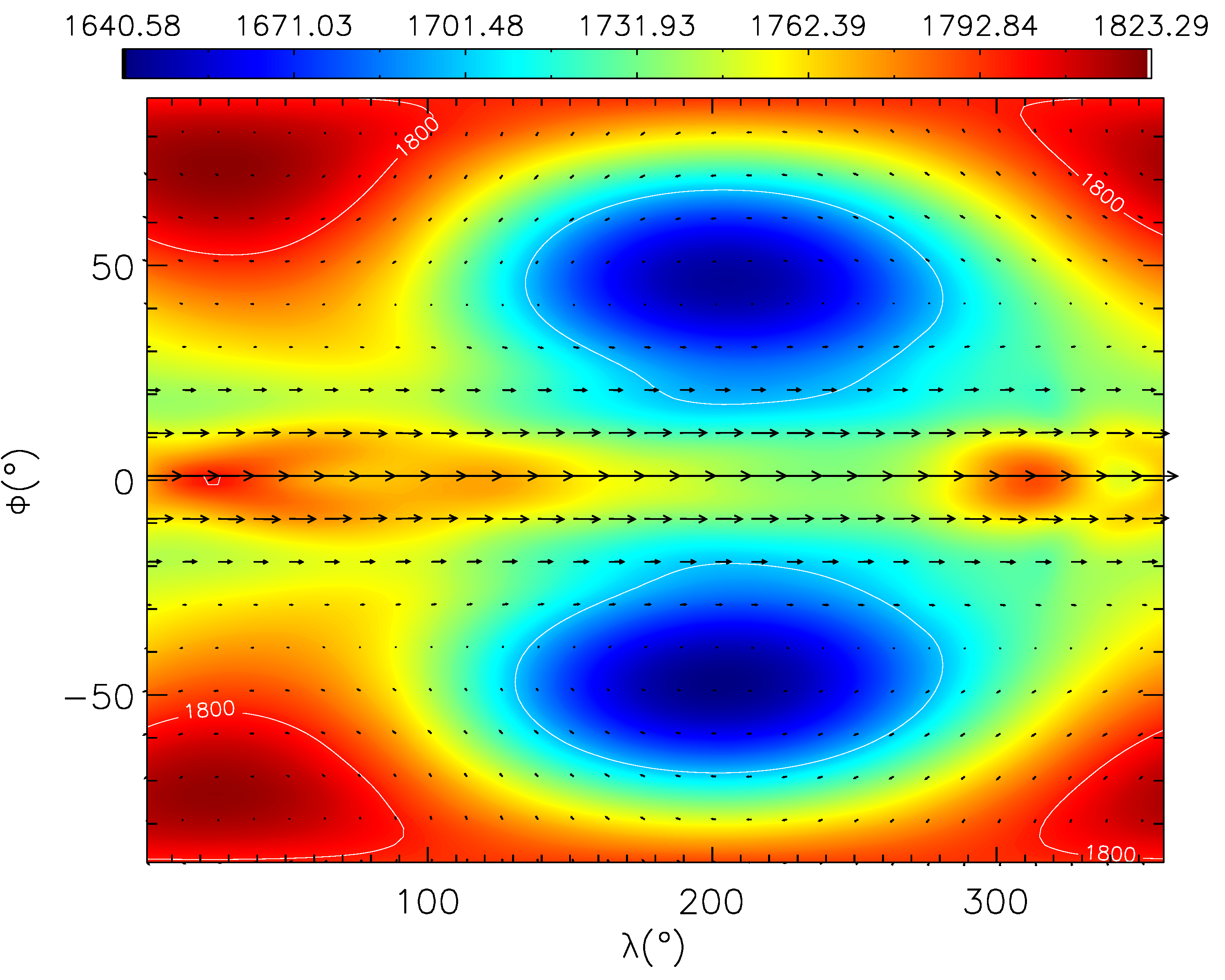}\label{std_full_4_69e5_10000_slice}}
 \subfigure[Std Prim: 21.9$\times 10^5$\,Pa, 10\,000\,days]{\includegraphics[width=9.0cm,angle=0.0,origin=c]{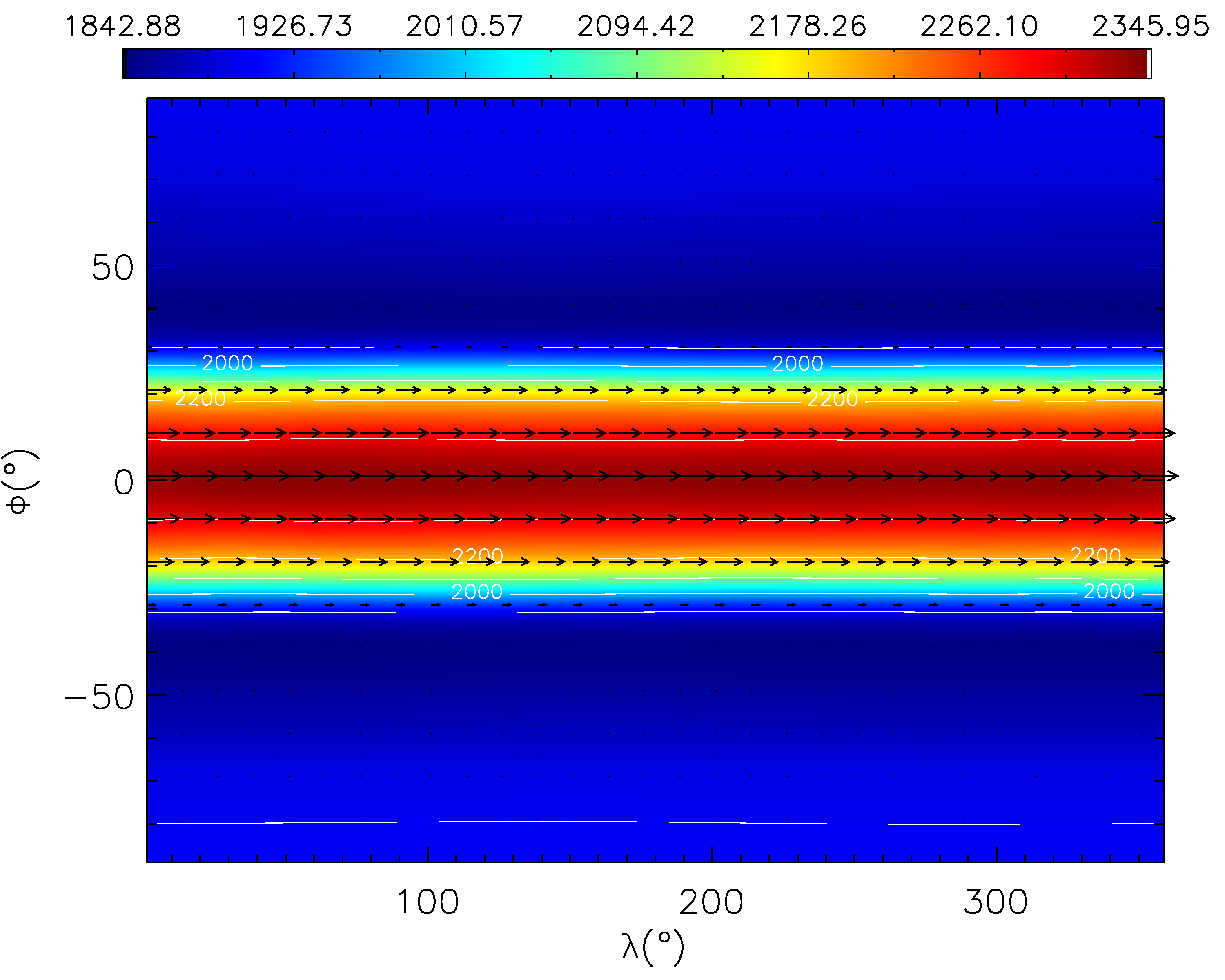}\label{std_prim_21_9e5_10000_slice}}
 \subfigure[Std Full: 21.9$\times 10^5$\,Pa, 10\,000\,days]{\includegraphics[width=9.0cm,angle=0.0,origin=c]{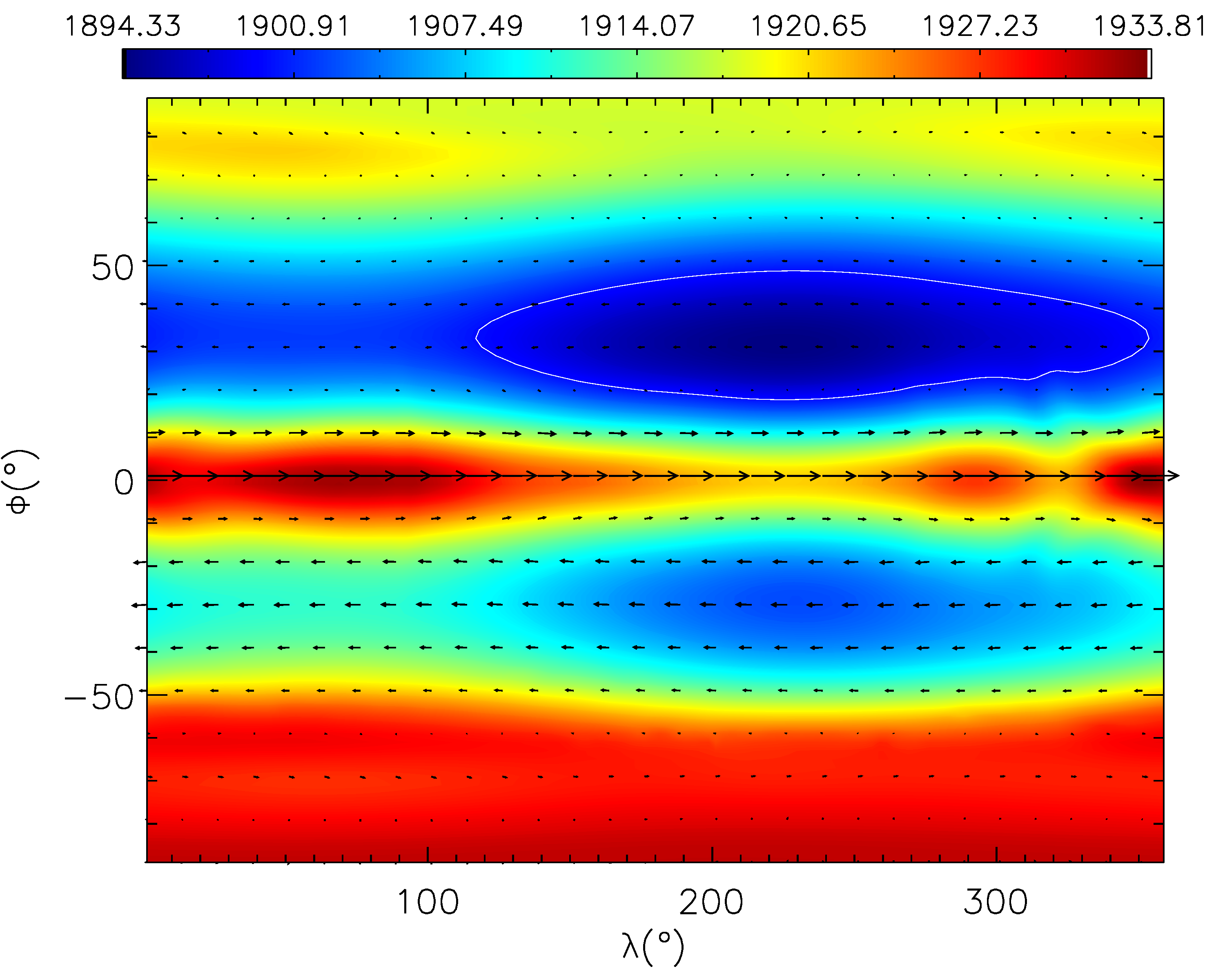}\label{std_full_21_9e5_10000_slice}}
\end{center}
\caption{As Figure \ref{slice_std_1200_bot} but after 10,000
  \,days. The maximum magnitudes of the horizontal velocities are
  $\sim$70, $\sim$180, $\sim$30 \& 3\,ms$^{-1}$ for the \textit{top
    left}, \textit{top right}, \textit{bottom left} \& \textit{bottom
    right panels}, respectively. \label{slice_std_10000_bot}}
\end{figure*}

\begin{figure*}
\begin{center}
  \subfigure[Std Full: 21.9$\times 10^5$\,Pa, 8\,000\,days]{\includegraphics[width=9.0cm,angle=0.0,origin=c]{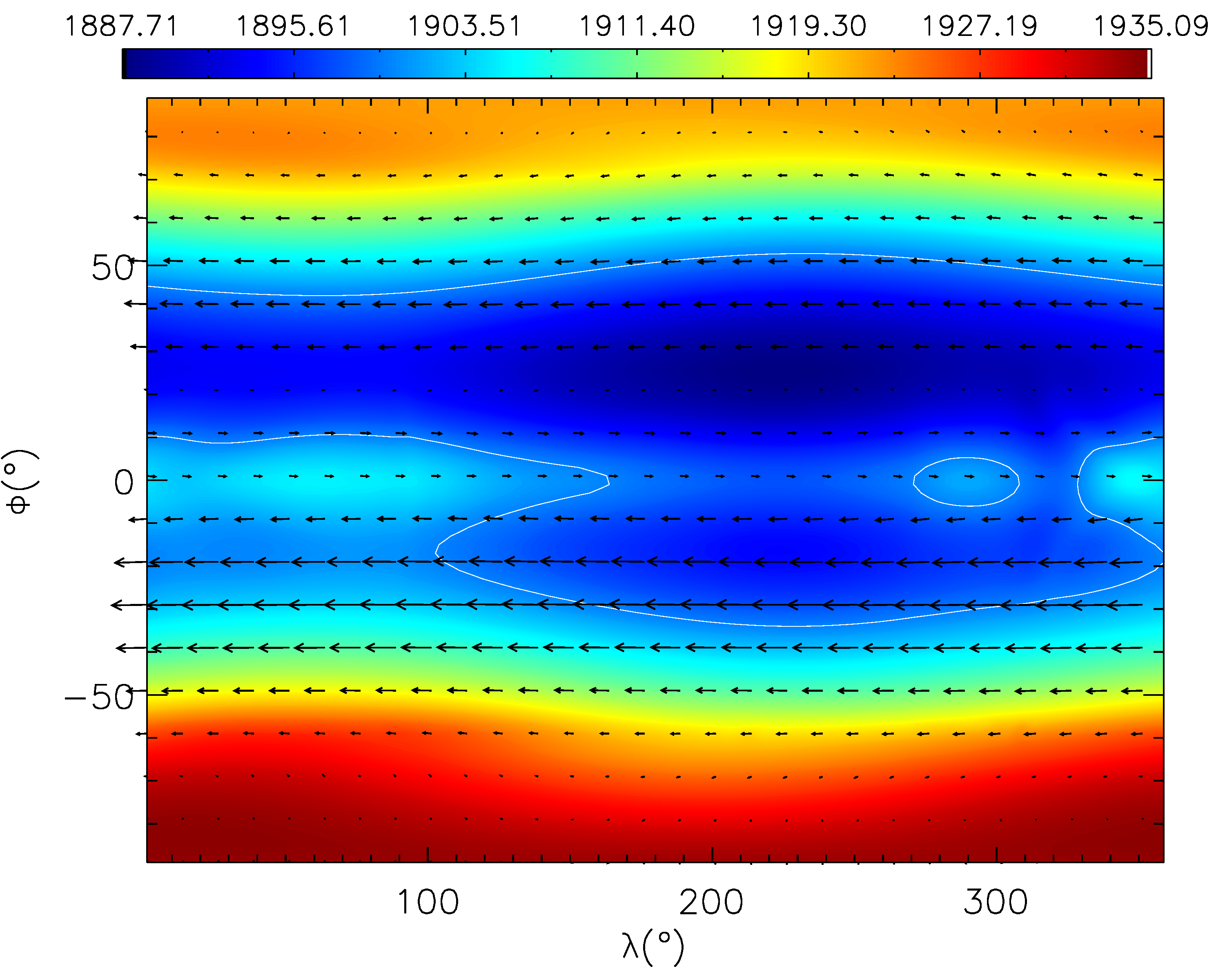}\label{std_full_21_9e5_8000_slice}}
  \subfigure[Std Full: 21.9$\times 10^5$\,Pa, 13\,000\,days]{\includegraphics[width=9.0cm,angle=0.0,origin=c]{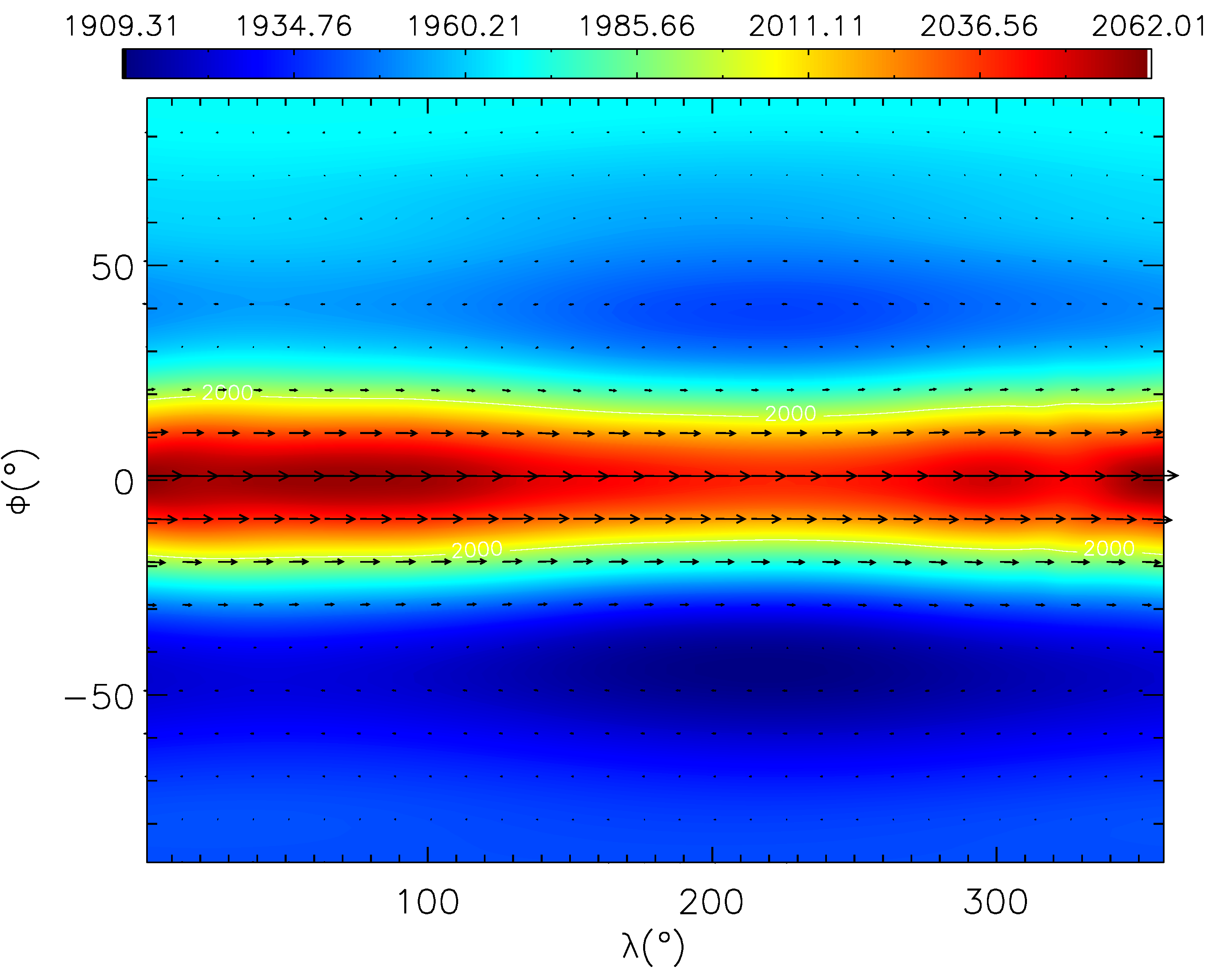}\label{std_full_21_9e5_13000_slice}}
\end{center}
\caption{Figure showing deep atmosphere isobaric slice, for the Std
  Full simulation, at a pressure of 21.9$\times 10^5$\,Pa after 8\,000
  and 13\,000\,days, \textit{left} and \textit{right panels},
  respectively. The temperature (K, contours) and horizontal winds
  (ms$^{-1}$, vector arrows) are still evolving (see Table
  \ref{model_names} for explanation of simulation names). The maximum
  magnitudes of the horizontal velocities are $\sim$2 \& 8\,ms$^{-1}$
  for the \textit{left} \& \textit{right panels},
  respectively. \label{slice_std_full_deep}}
\end{figure*}

Figures \ref{slice_std_1200_bot} and \ref{slice_std_10000_bot} show
the temperature (colour scale and contours) and horizontal wind
(vector arrows) structure, after 1\,200 and 10\,000\,days
respectively. The flow is depicted on the two highest pressure
isobaric surfaces as used in \citet{heng_2011} and subsequently
\citet{mayne_2014}, namely 4.69$\times 10^5$Pa, and
21.9$\times 10^5$\,Pa, as the \textit{top} and \textit{bottom rows},
respectively. The corresponding Figures for the lower pressure
surfaces (213 and 21\,600\,Pa) are presented in Appendix
\ref{app_section:flow_details} (and reveal little variation across
time, or simulation setup). After 10\,000\,days the Std Prim
simulation has a smooth, and relatively time invariant structure
dominated by the homogenisation of temperature around the
equator. However, the Std Full simulation, whilst still supporting a
similar equatorial flow maintains vortices at 4.69$\times 10^5$Pa, and
a more complex flow structure at 21.9$\times 10^5$\,Pa. For the Std
Full simulation the deepest isobaric slice shows time evolution, even
after 10\,000\,days, as demonstrated by Figure
\ref{slice_std_full_deep}, where this isobaric surface is shown at two
further times, 8\,000 and 13\,000 days. The behaviour in this region
is purely dynamical, driven by both circulations and adiabatic
compressions/expansions from the upper atmosphere since the forcing
timescale is infinite below $10^6$\,Pa
\citep{heng_2011,mayne_2014}. The deepest layer of the Std Full
simulation, 21.9$\times 10^5$\,Pa after 10\,000\,days shows some
asymmetry about the equator (see Figure
\ref{slice_std_10000_bot}). However, this is due to fluctuations of
the still evolving fluid, as can be seen in Figure
\ref{slice_std_full_deep}.

\begin{figure*}
\begin{center}
  \subfigure[Std RT: 4.69$\times 10^5$\,Pa, 1\,600\,days]{\includegraphics[width=9.0cm,angle=0.0,origin=c]{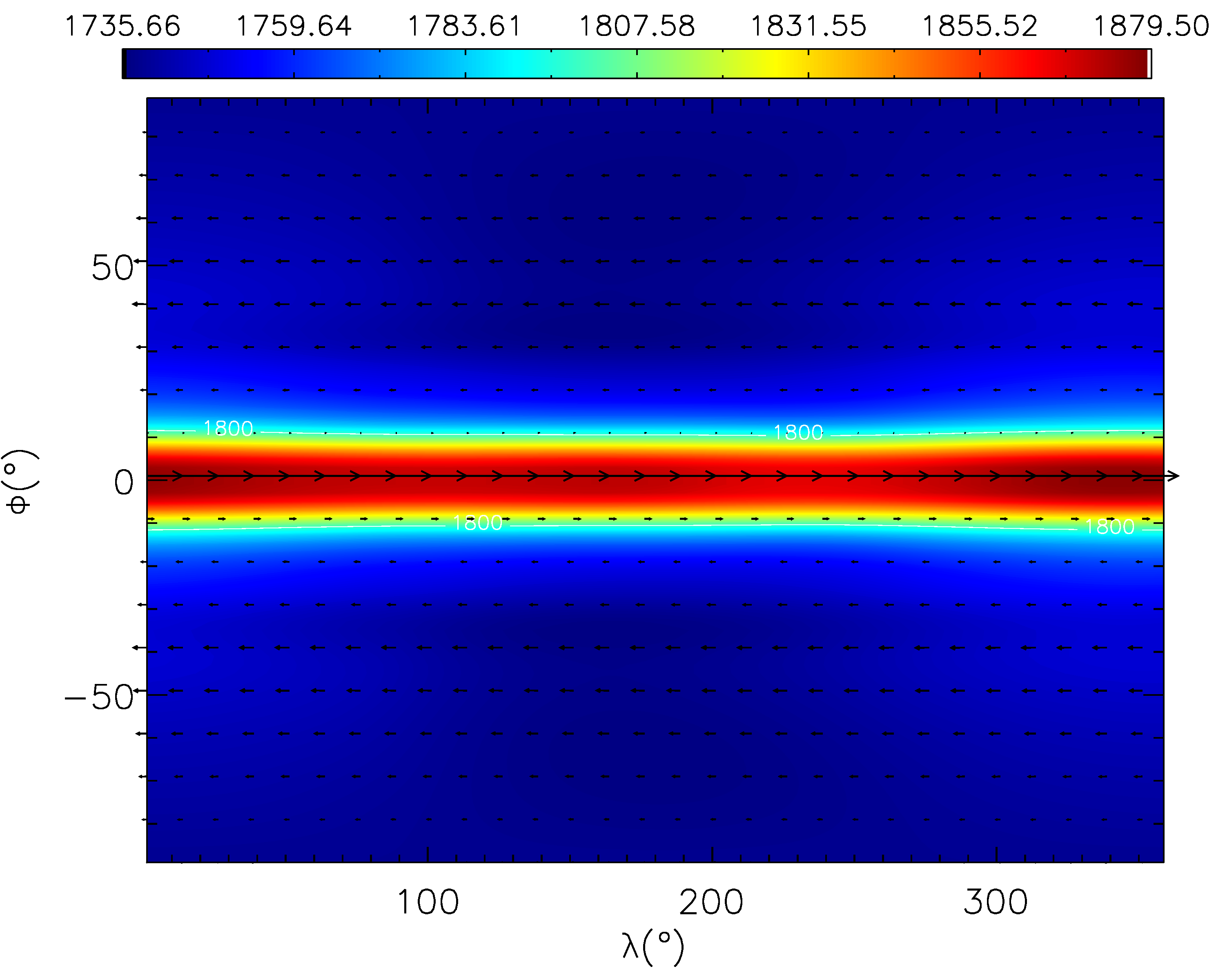}\label{rt_notiovo_full_4_69e5_1600_slice}}
 \subfigure[Std RT: 21.9$\times 10^5$\,Pa, 1\,600\,days]{\includegraphics[width=9.0cm,angle=0.0,origin=c]{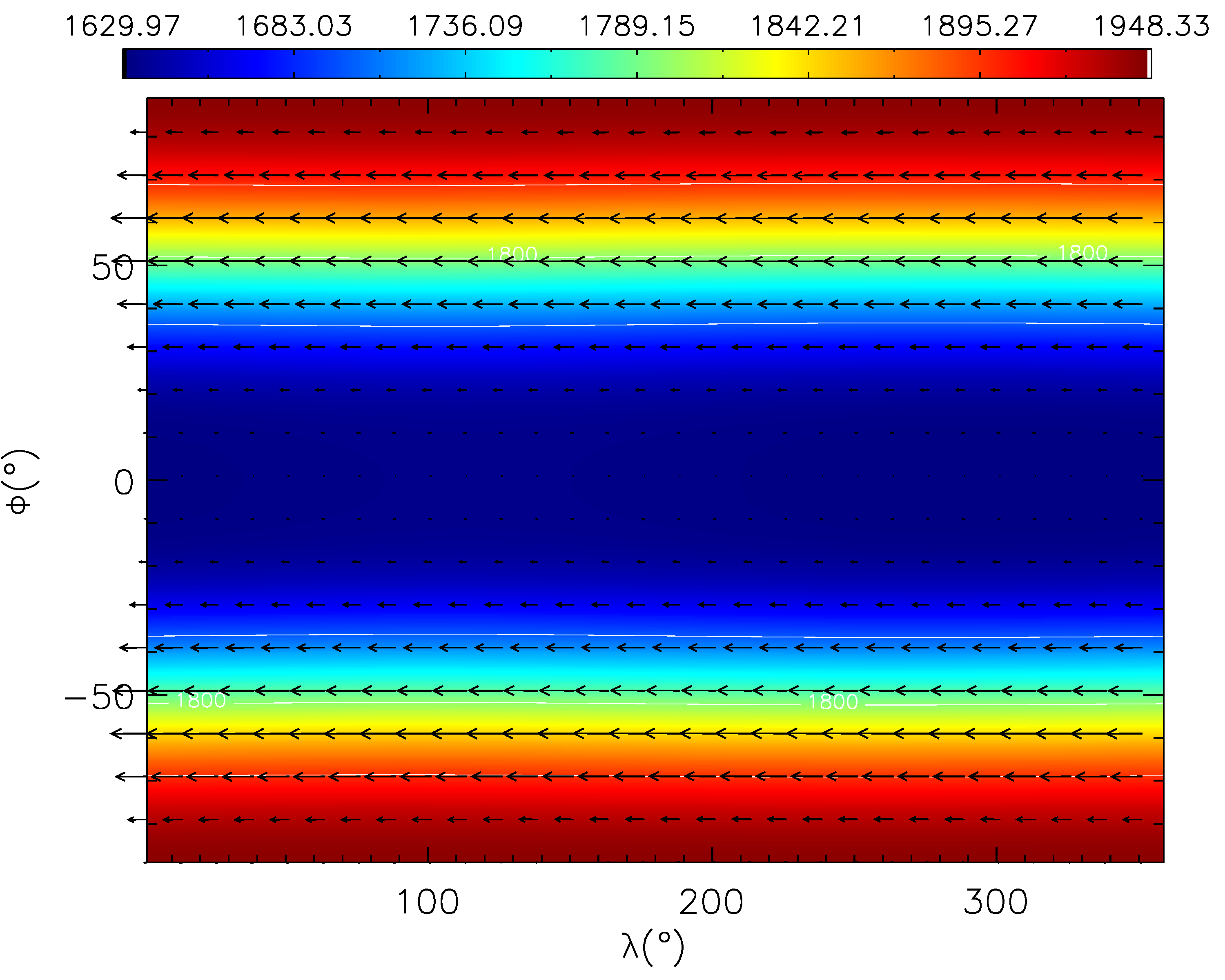}\label{rt_notiovo_full_21_9e5_1600_slice}}
\end{center}
\caption{As for Figure \ref{slice_std_1200_bot} but for the Std RT
  simulation after 1\,600\,days. The maximum magnitudes of the
  horizontal velocities are $\sim$16 \& 4\,ms$^{-1}$ for the
  \textit{left} \& \textit{right panels},
  respectively. \label{slice_rt_bot}}
\end{figure*}

Figure \ref{slice_rt_bot}, shows the same information as Figure
\ref{slice_std_1200_bot}, but for the Std RT simulation, after
1\,600\,days.  Again the lower pressure surfaces are shown, and
discussed in Appendix \ref{app_section:flow_details}. The Std RT
`snapshots' have been taken at the latest simulation time,
1\,600\,days, to capture the atmosphere in its most evolved state, as
opposed to matching the 1\,200\,days of the first `snapshot' of the
Std Prim and Std Full simulations. The main differences between the
thermodynamic and dynamical structure between RT and TF simulations
have been discussed in the previous works of \citet{showman_2009} and
\citet{amundsen_2016}. Here we note that the deep atmosphere in this
case is qualitatively different to the TF simulations. In particular,
the slice at 21.9$\times 10^5$\,Pa shows an equator to pole
temperature gradient of $\sim+300$\,K. As discussed in
\citet{amundsen_2016} and Section \ref{sub_section:model_var} this may
also be due to the 3D model adjusting from an incorrect, or
inconsistent initial profile to a steady state (discussed further in,
Tremblin et al., submitted).
 
\subsection{Deep atmosphere evolution}
\label{sub_section:deep_atmosphere}

As discussed in the previous section, as well as \citet{mayne_2014}
and \citet{amundsen_2016}, the dynamical (and thermodynamic) state of
the deep atmosphere is slowly evolving throughout the simulations
(even out to several thousand rotation periods). Additionally,
although the flows in upper, lower pressure atmosphere are broadly
consistent across our simulations, as one would expect due to the
strong forcing, the flows do vary for the deeper, higher pressure
regions. Clearly, the evolutionary timescale for the deep atmosphere
is likely to be very long. The thermal timescale, $\tau_{\rm th}$, can
be estimated using \citep[see for example][]{geroux_2016},
\begin{equation}
\tau_{\rm th}\sim\frac{C_{\rm p}T\Delta m}{L},
\end{equation}
where $\Delta m$ is the mass in the layer, and $L$ the luminosity
(estimated from $L=4\pi r^2\sigma T_{\rm int}^4$, where $\sigma$ is
the Stefan-Boltzmann constant). Assuming a characteristic temperature
of $\sim 100$\,K, the timescale is $\sim 10^4$ years, at pressures of
$10^7$\,Pa. As the pressure, and therefore density, increase
exponentially moving to higher pressures, this timescale will also
increase exponentially (as the mass is linearly dependent on the
density). Therefore, we have performed two numerical experiments to
explore the possible impact of the deep atmosphere on the lower
pressure regions accessible to observations, by artificially forcing,
or omitting the deep regions entirely.

\begin{figure*}
\begin{center}
  \subfigure[Deep $\Delta T_{\rm eq\rightarrow pole}$: 9\,000-10\,000\,days]{\includegraphics[width=9.0cm,angle=0.0,origin=c]{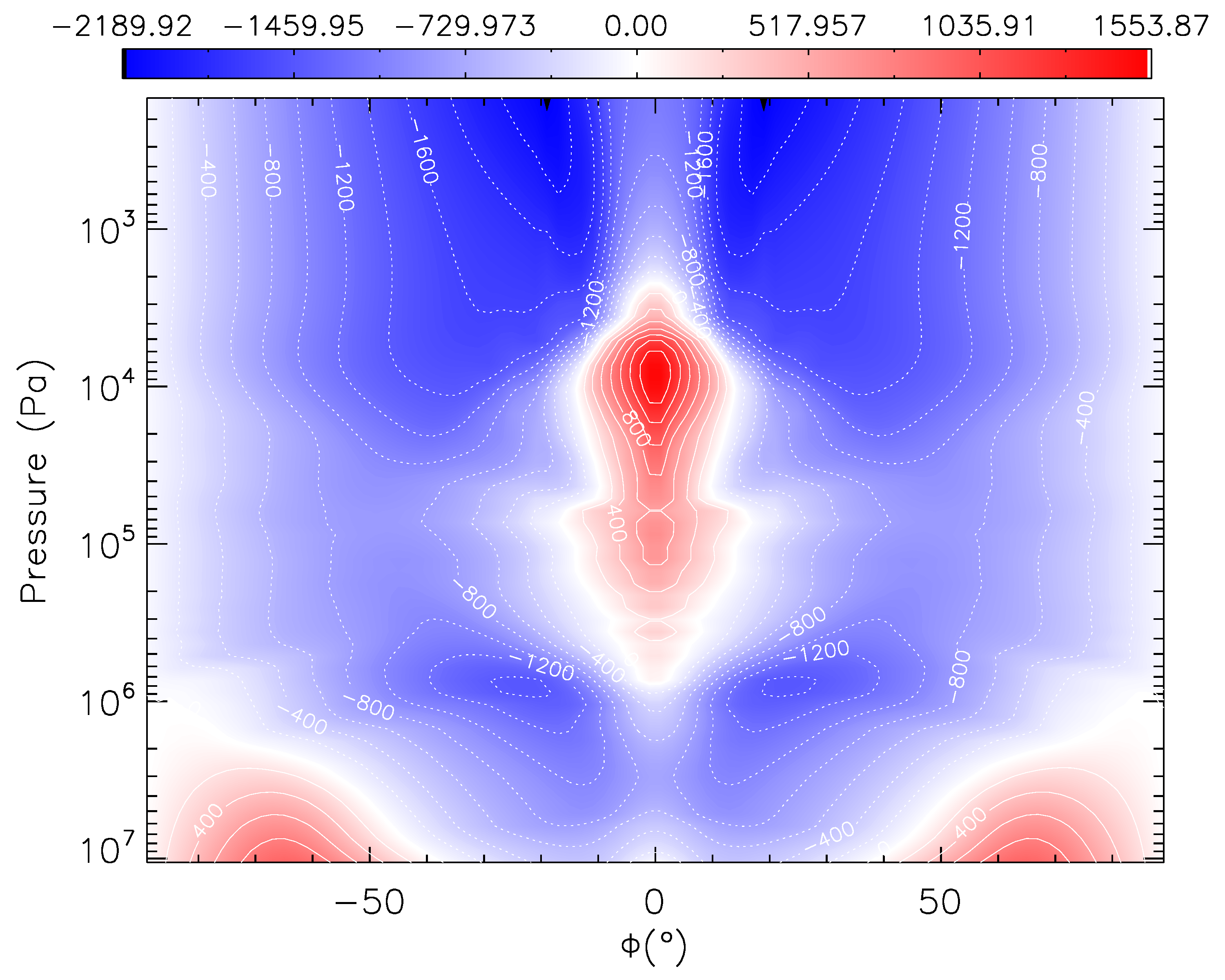}\label{teq_pole_deep_atm_full_9000_10000_uvel_bar}}
 \subfigure[Reduced $p_{\rm max}$: 9\,000-10\,000\,days]{\includegraphics[width=9.0cm,angle=0.0,origin=c]{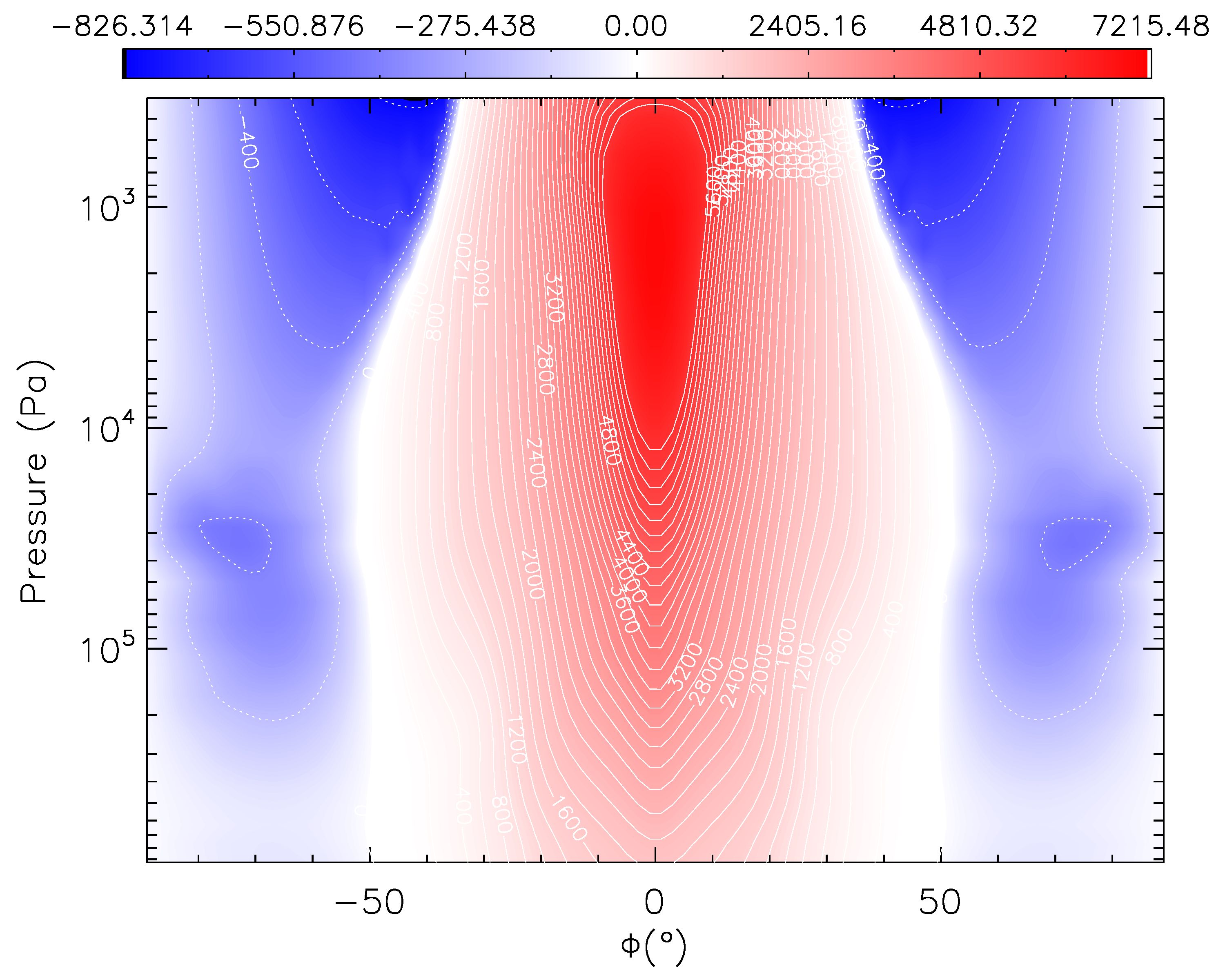}\label{p0_10bar_full_9000_10000_uvel_bar}}
\end{center}
\caption{Figure showing the zonal and temporally
  (9\,000--10\,000\,days) averaged zonal wind (ms$^{-1}$) as a function
  of latitude ($\phi^{\degree}$) and pressure
  ($\log_{10}(p\,[{\rm Pa}])$), in the same format as Figure
  \ref{std_uvel_bar}, for simulations exploring the treatment of the
  deep atmosphere. The $\Delta T_{\rm eq\rightarrow pole}$ and Reduced
  $p_{\rm max}$ (see Table \ref{model_names} for explanation of
  simulation names) simulations are shown as the \textit{left} and
  \textit{right columns}, respectively. Note, of course, the Reduced
  $p_{\rm max}$ simulation only extends to a pressure $\sim 10^6$\,Pa
  (10\,bar). \label{deep_uvel_bar}}
\end{figure*}

As demonstrated by \citet{showman_2011} equatorial superrotation can
be achieved, in the context of a tidally-locked hot Jupiters, by the
liberation of angular momentum from a deep atmosphere
`reservoir'. Additionally, \citet{mayne_2014} highlight an evolution
in the equator-to-pole temperature difference of the high pressure
atmosphere, in their simulations. Therefore, we have performed
simulations omitting the radiatively inactive deep atmosphere, termed
Reduced $p_{\rm max}$, or enforcing a strong equator-to-pole
temperature gradient, termed Deep $\Delta T_{\rm eq\rightarrow
  pole}$. Figure \ref{deep_uvel_bar} shows the zonal flow, in the same
format as Figure \ref{std_uvel_bar}, for the Deep $\Delta T_{\rm
  eq\rightarrow pole}$ and Reduced $p_{\rm max}$ (note the y-axis only
extends to pressures of $\sim 10^6$\,Pa or 10\,bar) simulations, as
the \textit{left} and \textit{right columns}, respectively.
Interestingly, the absence of the deep atmosphere section does not
disrupt the formation of a prograde equatorial jet. However, for the
simulation forcing the equator-to-pole temperature gradient after
10\,000\,days the jet has significantly slowed and covers a much
smaller region in both pressure and latitude. The angular momentum is
balanced in this model by the prograde jets around the poles at high
pressures, which are much faster than in the Std Full case (see Figure
\ref{std_uvel_bar}). Almost all of the simulations we have performed
conserve angular momentum to better than $\sim$5\% over periods of
around 10,000 Earth days (most are better than 1\%). The few
exceptions include the simulation with the bottom boundary at
$10^6$\,Pa (10 bar) where the angular momentum almost doubles, and all
are discussed later in this section.

\begin{figure*}
\begin{center}
  \subfigure[Deep $\Delta T_{\rm eq\rightarrow pole}$ Full: 4.69$\times 10^5$\,Pa, 10\,000\,days]{\includegraphics[width=9.0cm,angle=0.0,origin=c]{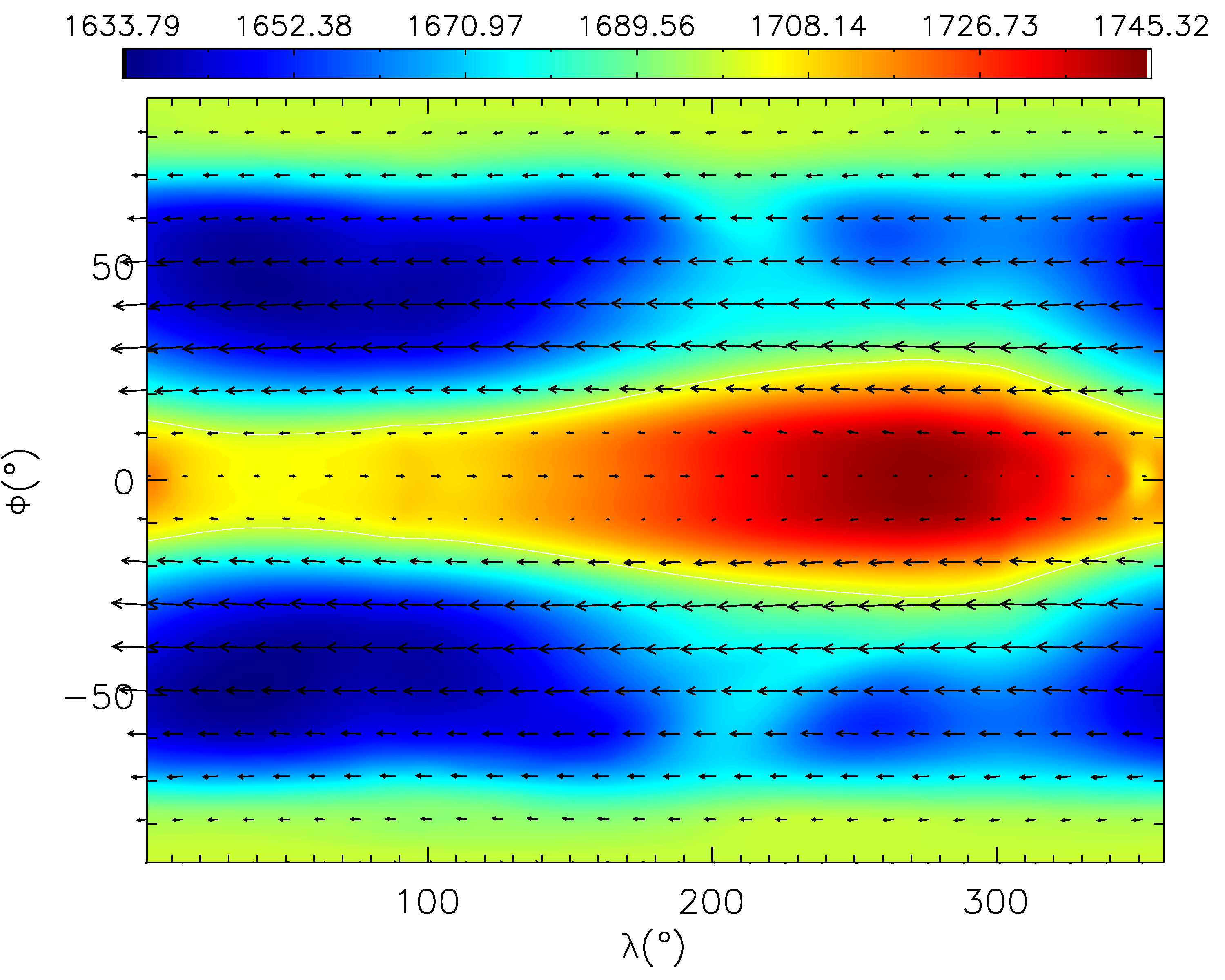}\label{teq_pole_deep_atm_full_4_69e5_10000_slice}}
  \subfigure[Deep $\Delta T_{\rm eq\rightarrow pole}$: 21.9$\times 10^5$\,Pa, 10\,000\,days]{\includegraphics[width=9.0cm,angle=0.0,origin=c]{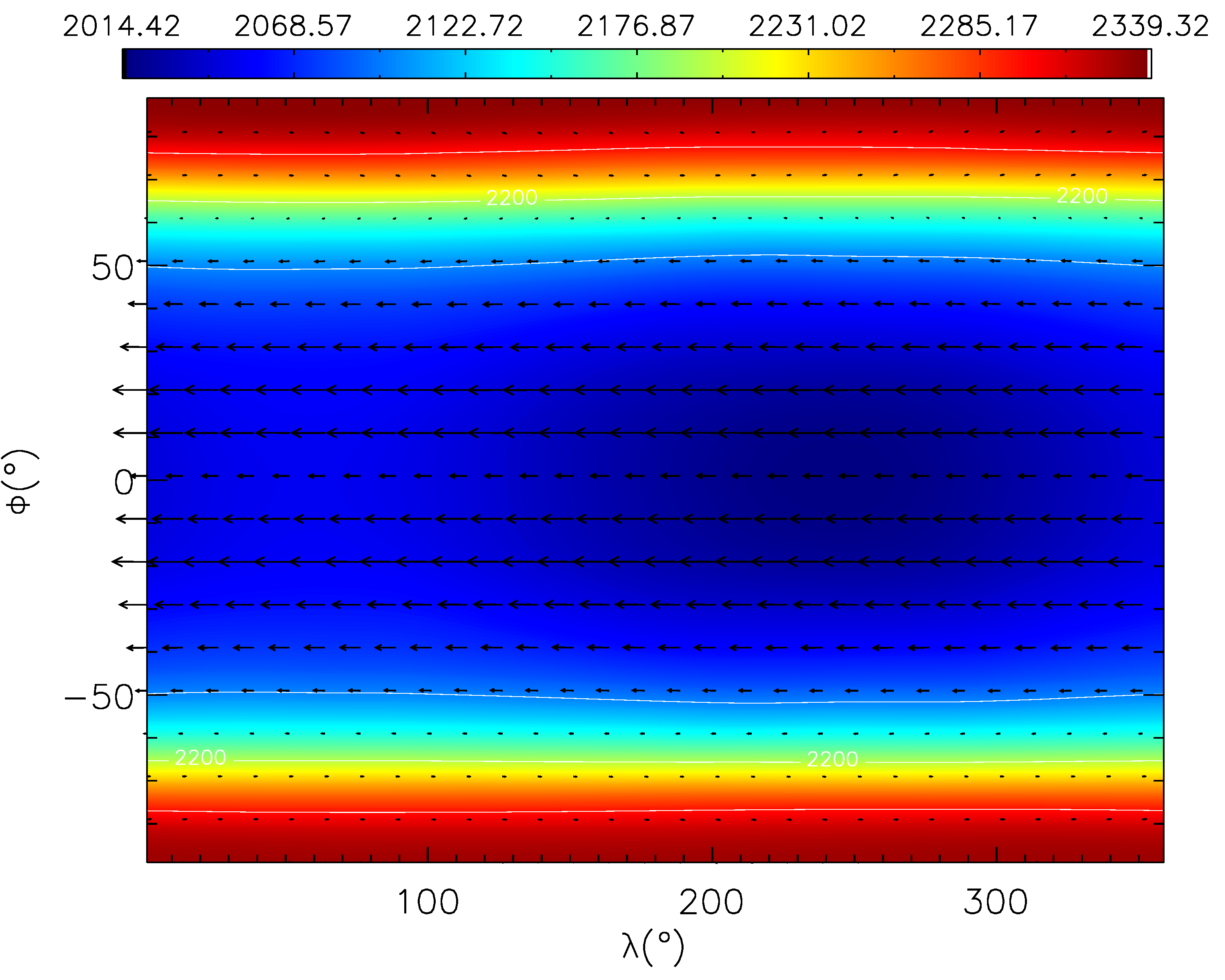}\label{teq_pole_deep_atm_full_21_9e5_10000_slice}}
\end{center}
\caption{As for Figure \ref{slice_std_10000_bot} but for the Deep
  $\Delta T_{\rm eq\rightarrow pole}$ simulation (see Table
  \ref{model_names} for explanation of simulation names). The maximum
  magnitudes of the horizontal velocities are $\sim$70 \&
  40\,ms$^{-1}$ for the \textit{left} \& \textit{right panels},
  respectively. \label{slice_deep_bot}}
\end{figure*}

Figure \ref{slice_deep_bot} shows the isobaric slices for the same
pressures, and in the same format as Figure \ref{slice_std_10000_bot},
but for the Deep $\Delta T_{\rm eq\rightarrow pole}$ simulation. The
lower pressure slices are again presented in Appendix
\ref{app_section:flow_details}, as Figure \ref{slice_deep_top}. The
corresponding figures for the Reduced $p_{\rm max}$ are omitted as for
the available pressures the results closely match those of the Std
Full simulation in terms of morphology. For the Deep
$\Delta T_{\rm eq\rightarrow pole}$ simulation the flow is strongly
retrograde at mid to high latitudes, with a weak equatorial prograde
flow, and subsequently weaker homogenisation of the temperature about
the equator compared to the standard simulations (compare the
21\,600\,Pa slice from Figures \ref{slice_std_10000_top} and
\ref{slice_deep_top}, as well as the 4.69$\times 10^5$\,Pa slice from
Figures \ref{slice_std_10000_bot} and \ref{slice_deep_bot}). Deeper in
the atmosphere a temperature difference of $\sim+300$\,K is achieved
between the equator and the pole, similar to that seen in the Std RT
model (compare Figures \ref{slice_deep_bot} and \ref{slice_rt_bot}),
and we see an extension of the retrograde flow.

\begin{figure*}
\begin{center}
  \subfigure[Std Prim: KE]{\includegraphics[width=9.0cm,angle=0.0,origin=c]{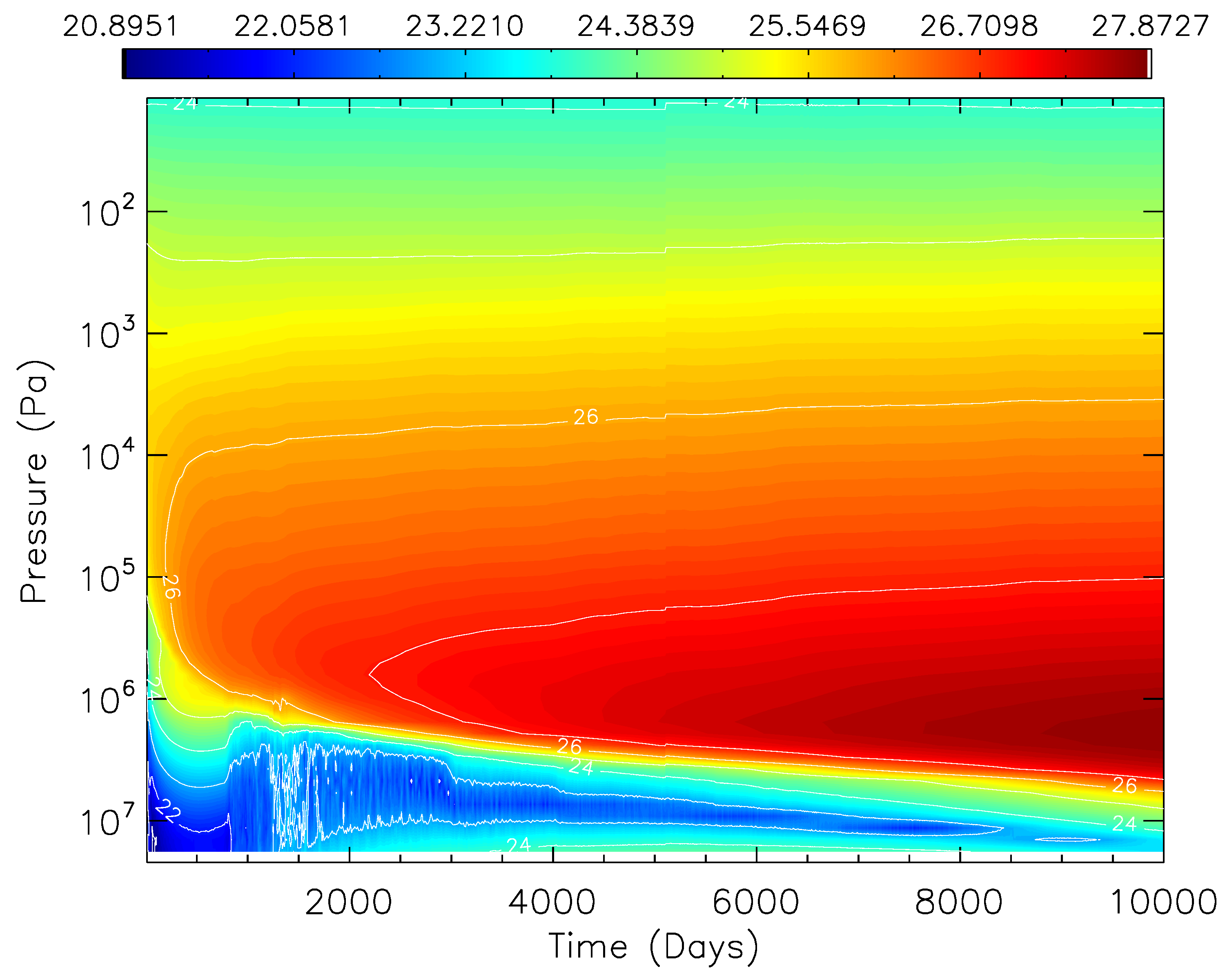}\label{std_prim_layer_ke}}
  \subfigure[Std Full: KE]{\includegraphics[width=9.0cm,angle=0.0,origin=c]{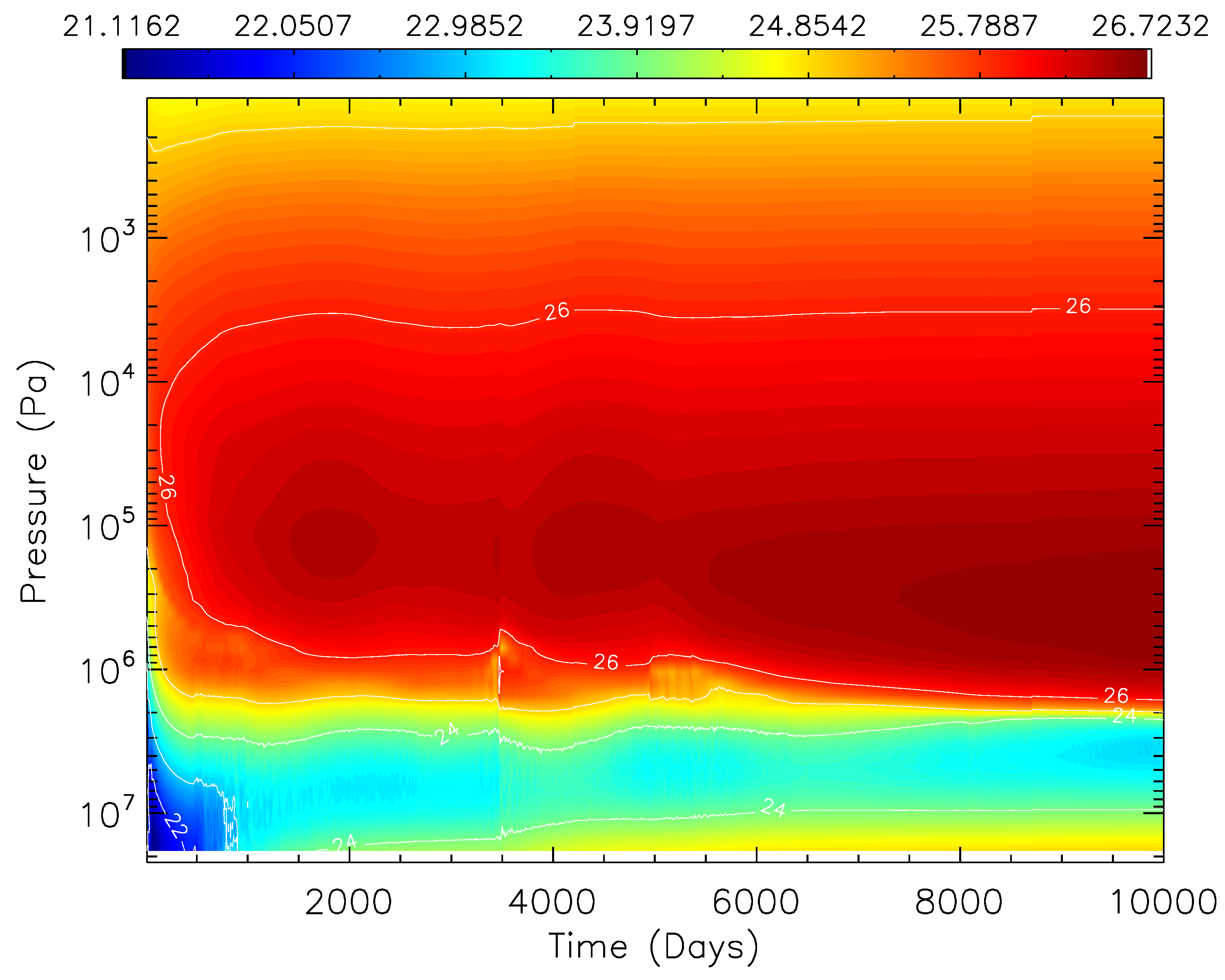}\label{std_full_layer_ke}}
  \subfigure[Deep $\Delta T_{\rm eq\rightarrow pole}$: KE]{\includegraphics[width=9.0cm,angle=0.0,origin=c]{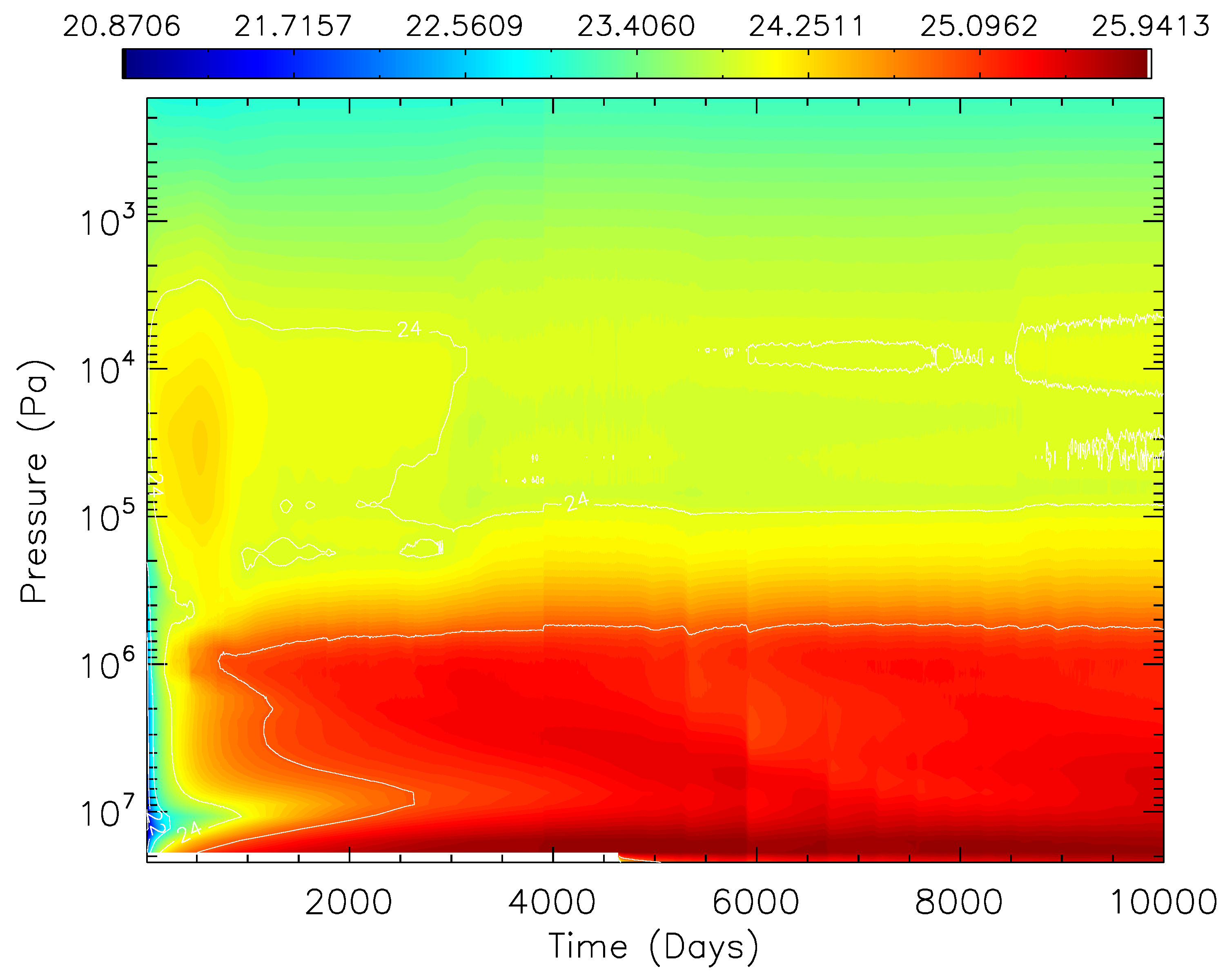}\label{teq_pole_rev_layer_ke}}
  \subfigure[Std RT: KE]{\includegraphics[width=9.0cm,angle=0.0,origin=c]{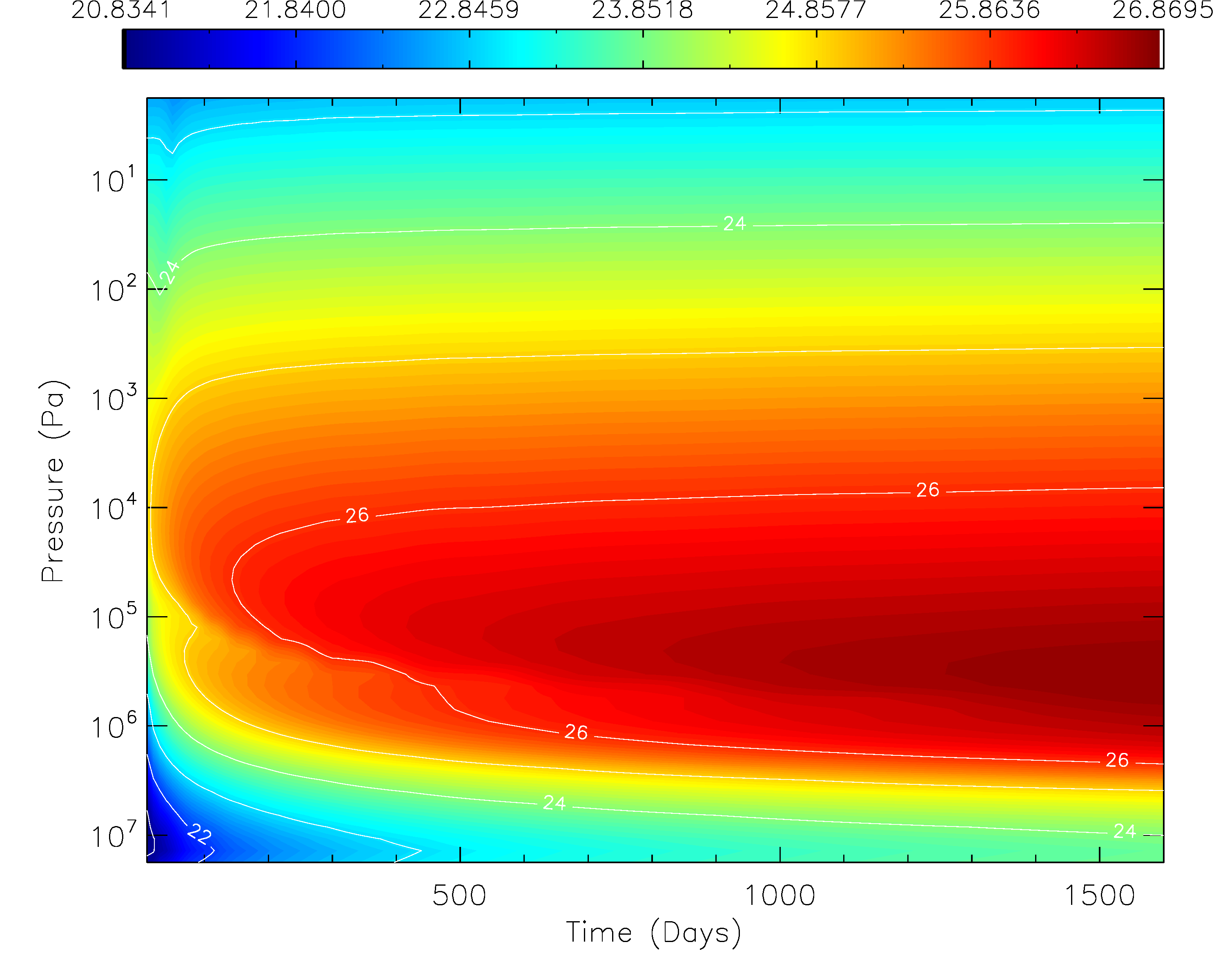}\label{rt_notiovo_layer_ke}}
 \end{center}
 \caption{Figure showing the logarithm of the kinetic energy
   ($\log_{10}$(KE\,[J])) as a function of pressure and time for the
   Std Prim, Std Full, Deep $\Delta T_{\rm eq\rightarrow pole}$ and ST
   RT simulations (note the Std RT simulation has only run for
   $\sim$1\,600\,days as opposed to $\sim$10\,000\,days in the other
   cases) as the \textit{top left}, \textit{top right}, \textit{bottom
     left} and \textit{bottom right panels},
   respectively. \label{conservation_layer}}
\end{figure*}

To explore the evolution of the deep atmosphere in our simulations,
Figure \ref{conservation_layer} shows the logarithm of the total
kinetic energy ($\log_{10}$(KE\,[J]), including the atmospheric mass)
as a function of time and pressure for the Std Prim, Std Full, Deep
$\Delta T_{\rm eq\rightarrow pole}$ and Std RT simulations as the
\textit{top left}, \textit{top right}, \textit{bottom left} and
\textit{bottom right panels}, respectively, following the figure
presented previously in \citet{rauscher_2010}. Note that the Std RT
simulation only extends to 1\,600\,days, whereas the others are
presented to 10\,000\,days. It is clear that the upper or lower
pressure atmosphere accelerates very rapidly, with the KE growing
rapidly and then reaches an almost steady value. The deeper atmosphere
seems to evolve much more slowly in the Std Prim case than that of the
Std Full simulation (although both are clearly far from completing
their evolution after 10\,000\,days). The deep atmosphere of the Std
RT simulation also appears to evolve very gradually, but of course the
total simulation time here is much shorter. \citet{mayne_2014}
concluded that the deep atmosphere had quickly finished evolving in a
simulation assuming a `shallow' atmosphere (i.e. similar to our Std
Prim case, but the assumption of vertical hydrostatic balance is
relaxed), using the deep atmosphere equator to pole temperature
contrast. However, Figure \ref{conservation_layer} actually shows that
this might just be caused by an evolution which is much slower than
the Std Full case. Additionally, it is clear that the KE peaks at
around 1-10 bar, $1$--$10\times 10^5$\,Pa except for the case of the
Deep $\Delta T_{\rm eq\rightarrow pole}$, where the energy budget is
dominated by the material at higher pressures, as one might expect
given the flow morphology. Although this experiment of artificially
forcing the deep atmosphere is somewhat arbitrary, it does illustrate
the possible effects of deep atmosphere flows on the upper
atmosphere. The final critical point is that the evolution of the deep
atmosphere appears to require very lengthy elapsed simulation times,
using Figure \ref{conservation_layer} as a guide. Therefore, it is
quite likely that the final solutions presented by GCMs after current
published simulation times may well be dependent on the initial state
of the deep atmosphere, both dynamically, and thermodynamically
\citep[see discussion in][]{mayne_2014,amundsen_2016}. However, it
must be noted, our artificial forcing of the deep atmospheres
thermodynamic state is quite strong (adopting the timescale assumed
for $10^6$\,Pa), and over 10\,000\,days has still not completely
destroyed the prograde equatorial jet.

\begin{figure*}
\begin{center}
  \subfigure[$u_{\rm max}$]{\includegraphics[width=9.0cm,angle=0.0,origin=c]{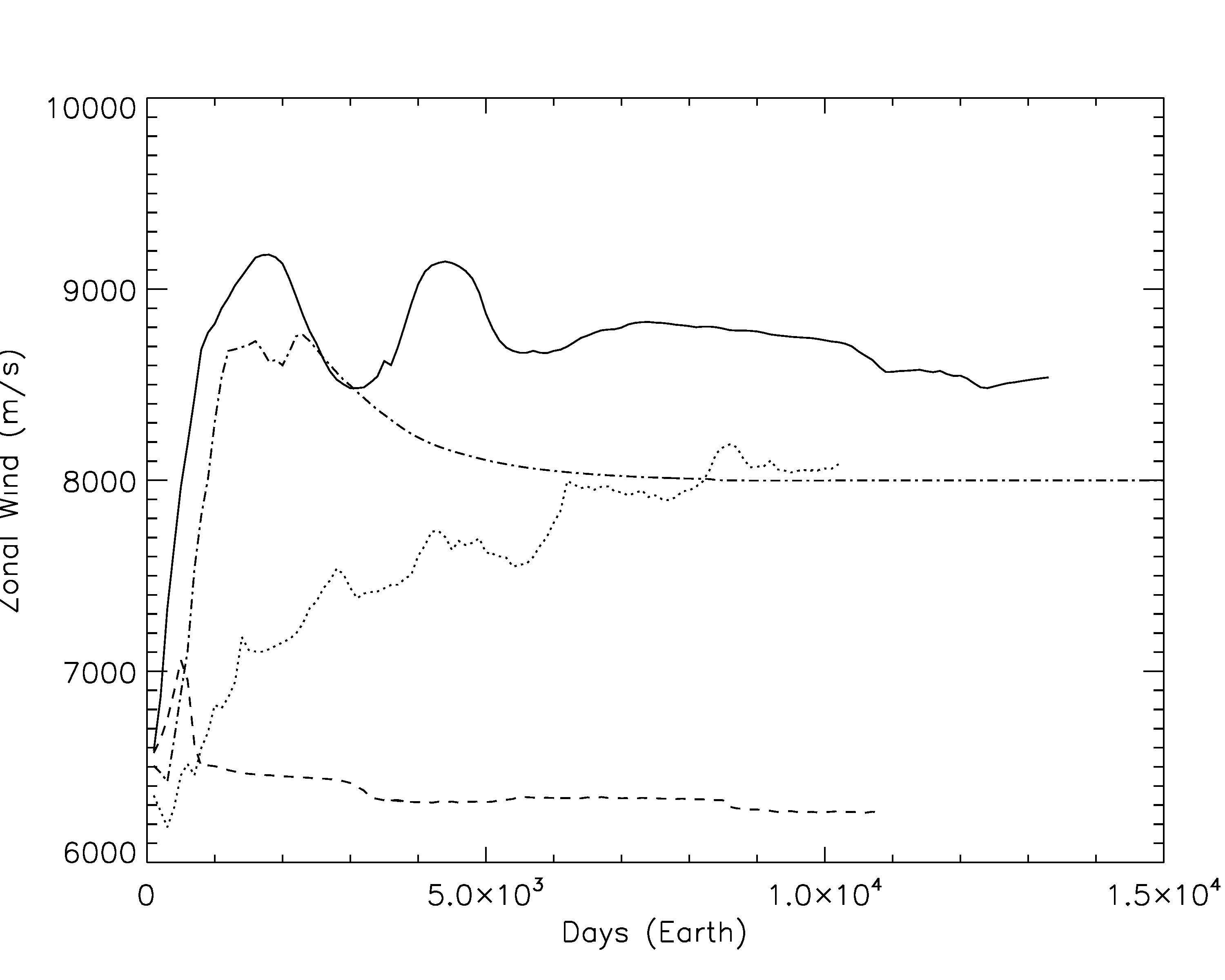}\label{uvel_max}}
  \subfigure[AAM/AAM(t=0)]{\includegraphics[width=9.0cm,angle=0.0,origin=c]{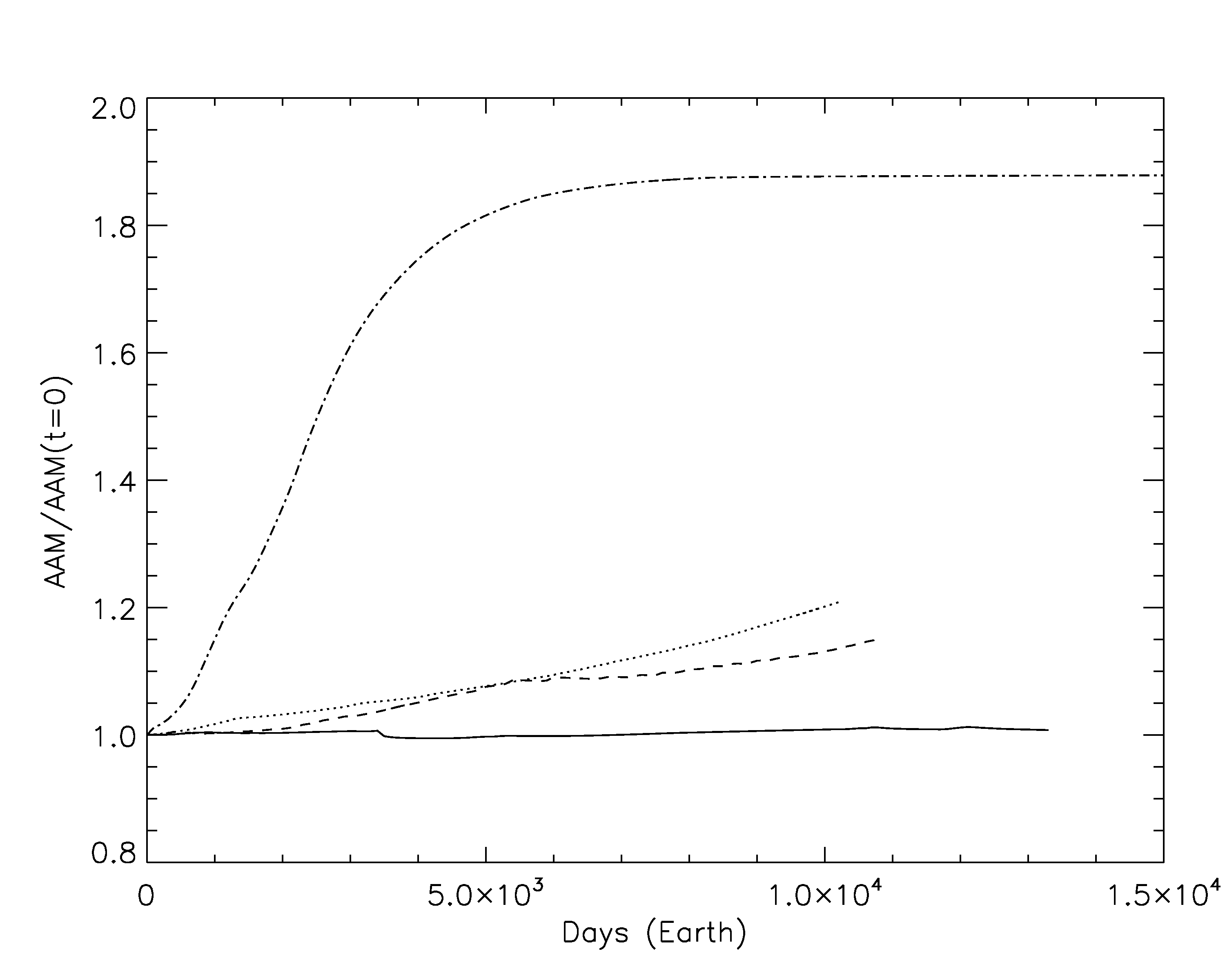}\label{conserve_AAM}}
\end{center}
\caption{Figures showing the total normalised axial angular momentum
  (AAM) (kg\,ms$^{-1}$) and maximum zonal velocity ($u_{\rm max}$,
  ms$^{-1}$) for several simulations as the \textit{left}, and
  \textit{right panels}, respectively. The solid line shows the Std
  Full, the dotted line the Std Prim, the dashed line the Deep $\Delta
  T_{\rm eq\rightarrow pole}$ and the dashed-dotted line the Reduced
  $p_{\rm max}$ simulations (see Table \ref{model_names} for
  explanation of simulation names). In terms of AAM conservation the
  Std Full simulation is representative of our simulation set, and the
  remaining three included simulations (Std Prim, Deep $\Delta T_{\rm
    eq\rightarrow pole}$ and Reduced $p_{\rm max}$) the only examples
  where AAM conservation is much poorer. Note the total simulation
  times of the individual runs (i.e. Std Prim, Std Full, Deep $\Delta
  T_{\rm eq\rightarrow pole}$ and Reduced $p_{\rm max}$) are
  different. \label{conservation}}
\end{figure*}

For the simulations presented in Figure \ref{conservation_layer} the
total KE is still increasing towards the end of the simulation time,
as one might expect given the increase in the deep atmosphere
component. As the pressure, and therefore density, increases
exponentially as one moves closer to the inner simulation boundary, a
significant KE contribution can be achieved with very slow
velocities. Generally, the maximum zonal velocity ($u_{\rm max}$) is
used to determine a quasi-steady state in the literature for hot
Jupiter simulations. Figure \ref{conservation}, \textit{left panel},
shows $u_{\rm max}$ as a function of time for the Std Full, Std Prim,
Deep $\Delta T_{\rm eq\rightarrow pole}$ and Reduced $p_{\rm max}$
simulations. As Figure \ref{conservation} shows, the fact that the KE
has not equilibrated is not detected in the maximum zonal velocity,
with all of these simulations appearing to reach a quasi-steady
solution. This is, of course driven by the fact that the deep
atmosphere velocities are much slower than those in the upper, low
pressure, regions. This suggests that assuming a steady state based on
the maximum zonal velocity will not apply to the high pressure
atmosphere.

The deep atmosphere also represents a huge reservoir of axial angular
momentum (AAM), which can be used to accelerate zonal flows. Updrafts
of material can effectively convert significant amounts of
\textit{planetary angular momentum} (simply due to the solid body
rotation of the atmosphere) into \textit{atmospheric angular momentum}
(from the winds) \citep[see discussion in][]{mayne_2014}, as the total
(i.e. the sum of the \textit{planetary} and \textit{atmospheric}
components) is conserved. Additionally, as with the KE even very small
velocities in the high pressure regions can make significant
contributions to the total AAM budget. The \textit{right panel} of
Figure \ref{conservation} shows the total normalised AAM, relative to
the initial value, for the same simulations presented in the
\textit{left panel} of Figure \ref{conservation}, namely the Std Full,
Std Prim, Deep $\Delta T_{\rm eq\rightarrow pole}$ and Reduced
$p_{\rm max}$ simulations. The Reduced $p_{\rm max}$ shows a near
doubling of AAM in the first $\sim$4\,000\,days. The Std Prim and Deep
$\Delta T_{\rm eq\rightarrow pole}$ simulations both gain around 15\%
AAM after about 10\,000 days, which is still a reasonable level of
accuracy for such low resolution simulations run over such long
periods. All of our remaining simulations conserve AAM to better than
5\%, and in many cases 1\%, and the Std Full model is presented as
indicative of this. If angular momentum conservation issues were
dominated solely by the fast dynamics in the upper atmosphere, one
would expect the Reduced $p_{\rm max}$ simulation to show much poorer
absolute AAM conservation. This is as this simulation does not include
the vast AAM contribution from the deep atmosphere. The absolute
change in AAM for the Reduced $p_{\rm max}$ simulation is
$\sim 13.5\times10^{32}$kg\,ms$^{-1}$ (from an initial total AAM of
$\sim 15.4\times 10^{32}$kg\,ms$^{-1}$), whereas the absolute change
for the Std Full simulation is a factor $\sim$5 lower,
$\sim 2.6\times10^{32}$kg\,ms$^{-1}$ (from an initial total AAM of
$\sim 33.9\times 10^{33}$kg\,ms$^{-1}$) indicating that indeed the
inaccuracies in the AAM conservation are dominated by the fast winds
in the low pressure atmosphere. However, \citet{cho_2015} noted
significant changes in AAM using models without bottom boundary drag,
implying that velocities close to the bottom boundary lead to
numerical losses. Here we do not employ a bottom boundary drag, and
allow our deep atmosphere to slowly accelerate. For our Reduced
$p_{\rm max}$ simulations fast winds are accelerated much closer to
the bottom, or inner, boundary than in the other simulations. We are
still investigating the reasons for this issue.

\subsection{Jet pumping}
\label{sub_section:jet_pumping}

Two of the main mechanisms for accelerating jets in planetary
atmospheres are either the exchange of momentum via mean flows, or the
convergence of momentum from the creation and destruction of eddies
\citep[see][for a review in the terrestrial planet
case]{showman_2013}. For hot Jupiter atmospheres the expected large
scale of the atmospheric eddies (i.e. Rossby radius of deformation),
and zonally asymmetric forcing (i.e. day and night side) led
\citet{showman_2011} to propose an eddy momentum acceleration
mechanism for the jets relying on the interaction of standing Kelvin
and Rossby waves. \citet{tsai_2014} explored the validity of this
mechanism in their simulations by presenting, in 3D adopting the
$\beta$-plane (where the Coriolis force is simplified as varying
linearly with latitude, instead of $\propto\cos\phi$), the eddy
momentum fluxes as a function of pressure and a proxy for
latitude. The simulations, and analysis of \citet{showman_2011} are
performed in a pressure--based (hydrostatic) framework. Although in
previous sections we have interpolated our prognostic variables onto
pressure surfaces, the eddy momentum fluxes can not easily be
translated in such a fashion. This is as the terms involve gradients,
in the height--based framework, at constant $r$ in the longitude,
$\lambda$, direction, which cancel to zero when zonally
averaged. Simply interpolating the variables onto pressure surfaces
invalidates this step, and the terms no longer vanish, introducing
extra terms into the eddy--mean interaction equation. The derivation
of this equation is explicitly stated in Appendix
\ref{app_section:eddy_mean_derive}, and this point
highlighted. Therefore, to analyse the transport of momentum in the
atmosphere we shift to using height, $z$, as a vertical
coordinate. The simulations of \citet{tsai_2014}, however, are
performed using a height--based model, and the results interpolated
onto a pressure grid, meaning they have omitted the extra
contributions to the equation derived from the, resulting non--zero,
zonally averaged vertical gradient terms.

\subsubsection{Initial Response from Rest}
\label{subsub_section:rest}

\begin{figure*}
\begin{center}
\subfigure[Std Full: $\Sigma_{\lambda}\left(\rho u\right)$, 1\,day]{\includegraphics[width=8.6cm,angle=0.0,origin=c]{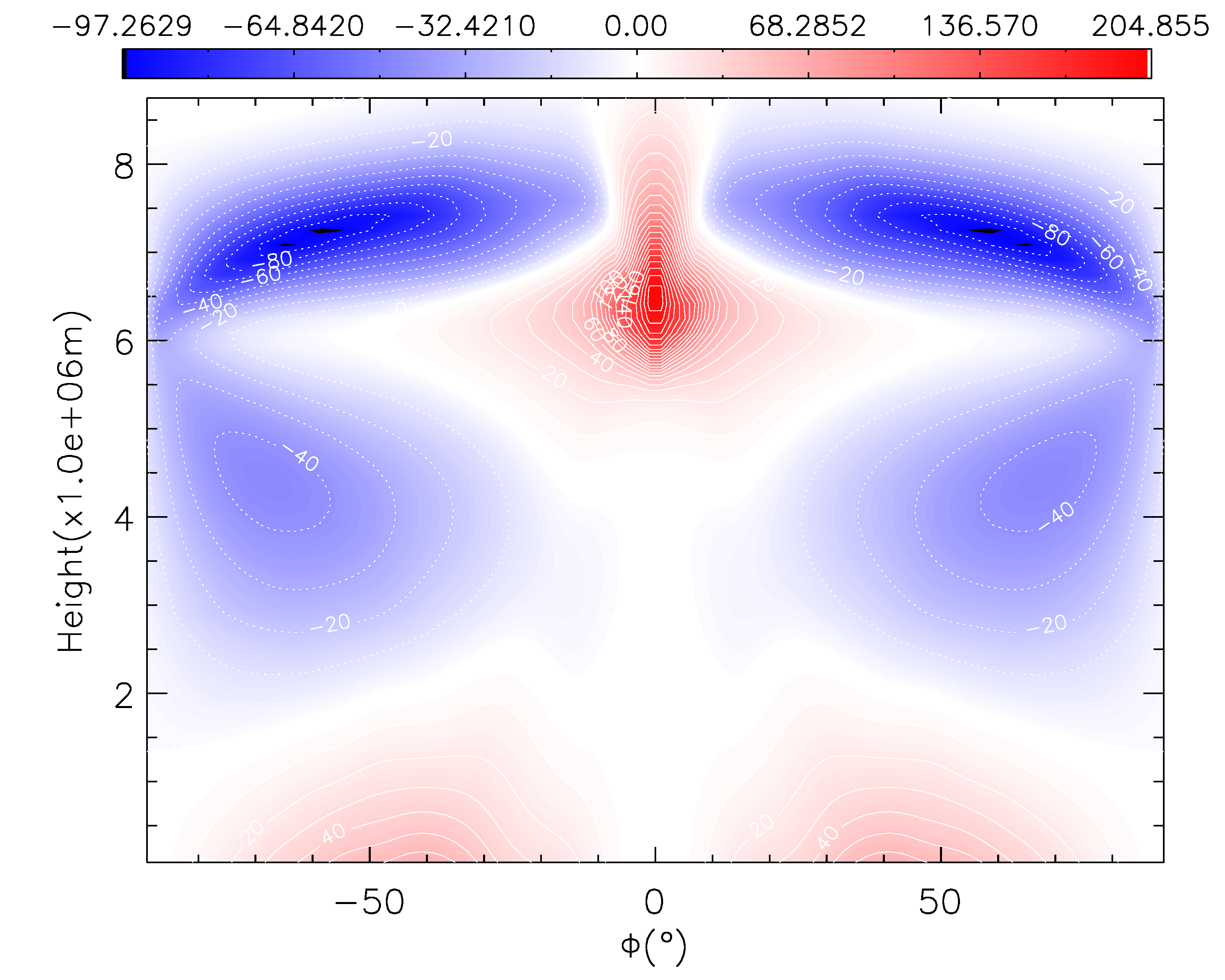}\label{std_full_lin_mmtm_slice_ylim_1}}
\subfigure[Std RT: $\Sigma_{\lambda}\left(\rho u\right)$, 1\,day]{\includegraphics[width=8.6cm,angle=0.0,origin=c]{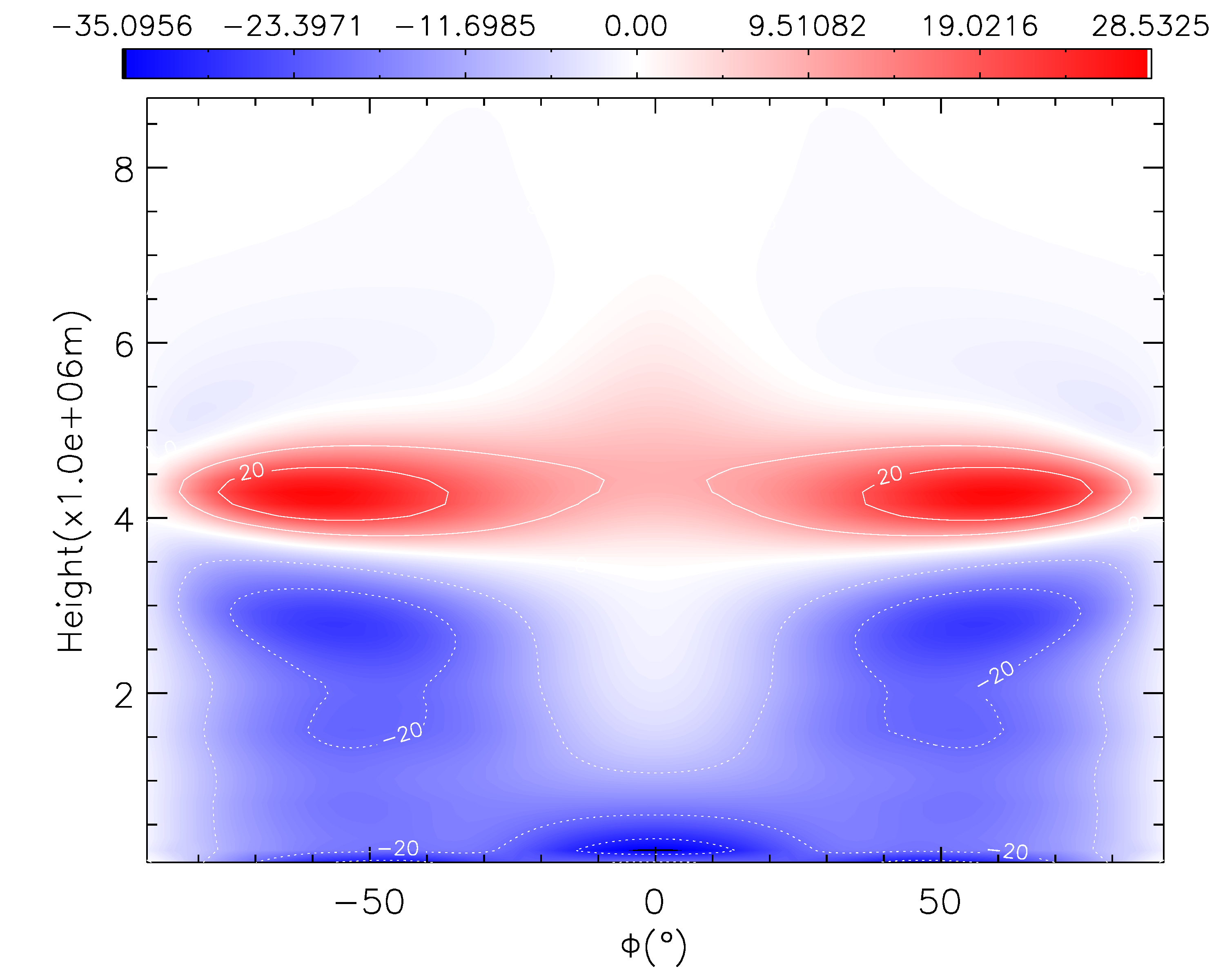}\label{rt_notiovo_full_lin_mmtm_slice_ylim_1}}
\subfigure[Std Full: $\Sigma_{\lambda}\left(\rho u\right)$, 10\,days]{\includegraphics[width=8.6cm,angle=0.0,origin=c]{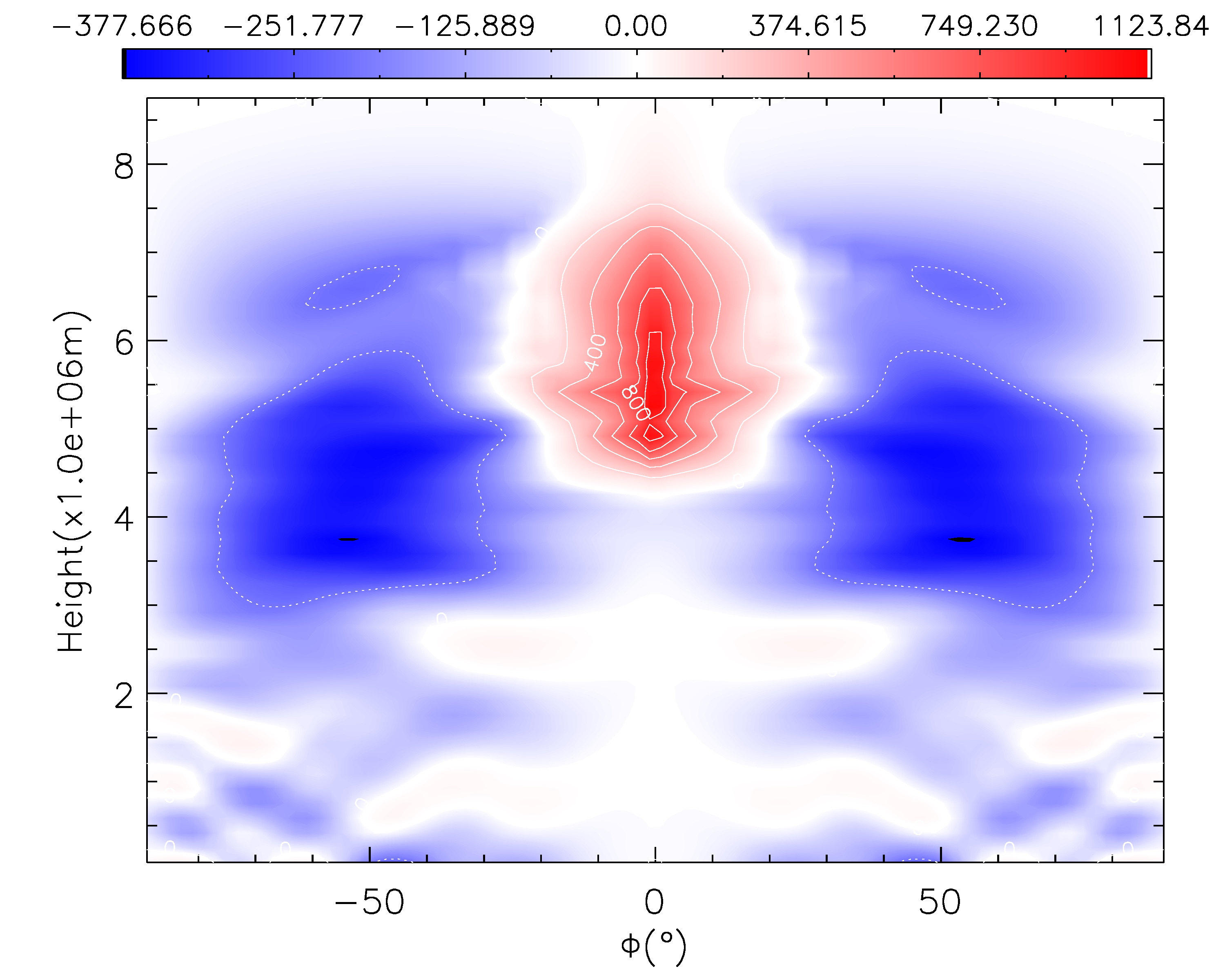}\label{std_full_lin_mmtm_slice_ylim_10}}
\subfigure[Std RT: $\Sigma_{\lambda}\left(\rho u\right)$, 10\,days]{\includegraphics[width=8.6cm,angle=0.0,origin=c]{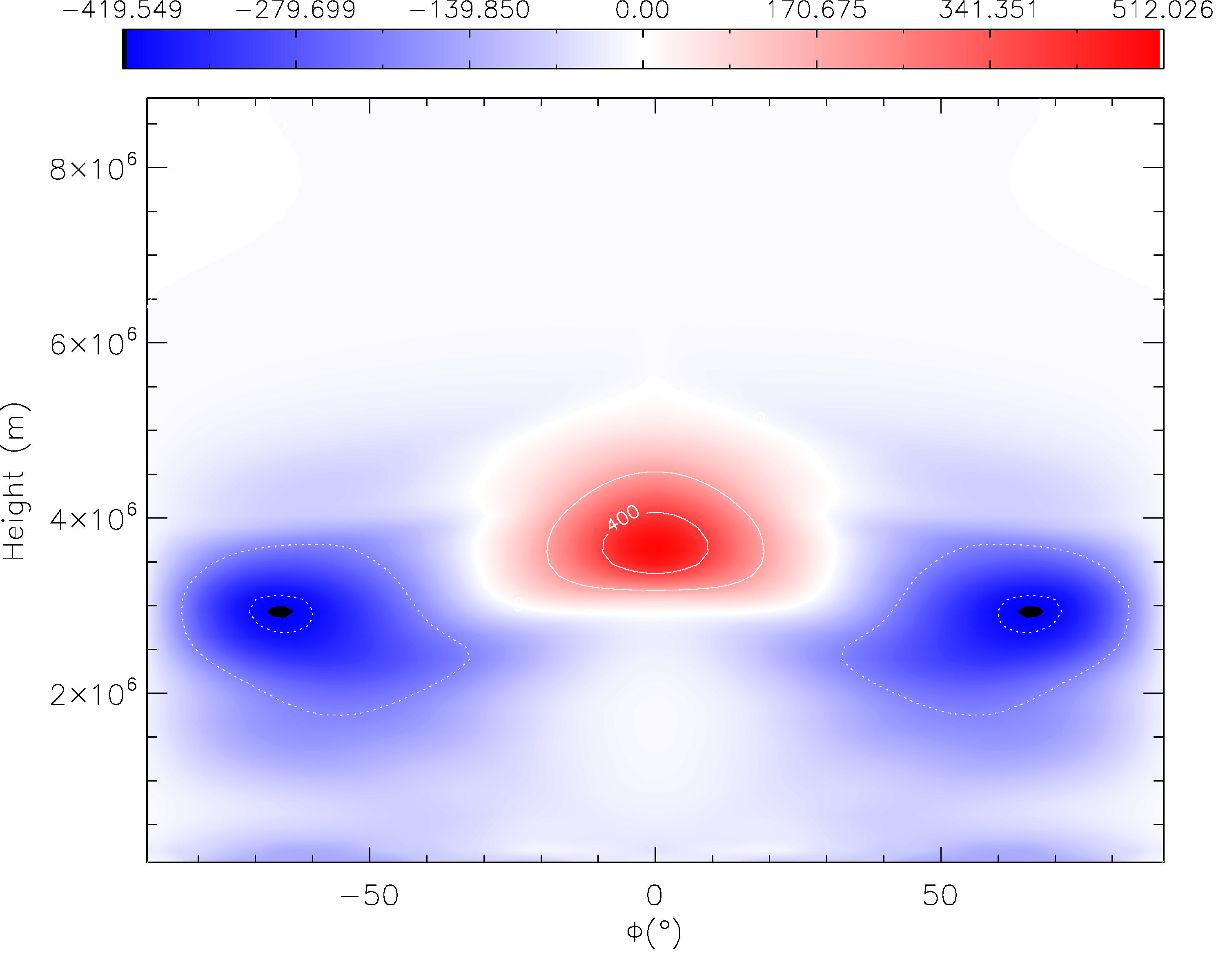}\label{rt_notiovo_full_lin_mmtm_slice_ylim_10}}
\subfigure[Std Full: $\Sigma_{\lambda}\left(\rho u\right)$, 1\,000\,days]{\includegraphics[width=8.6cm,angle=0.0,origin=c]{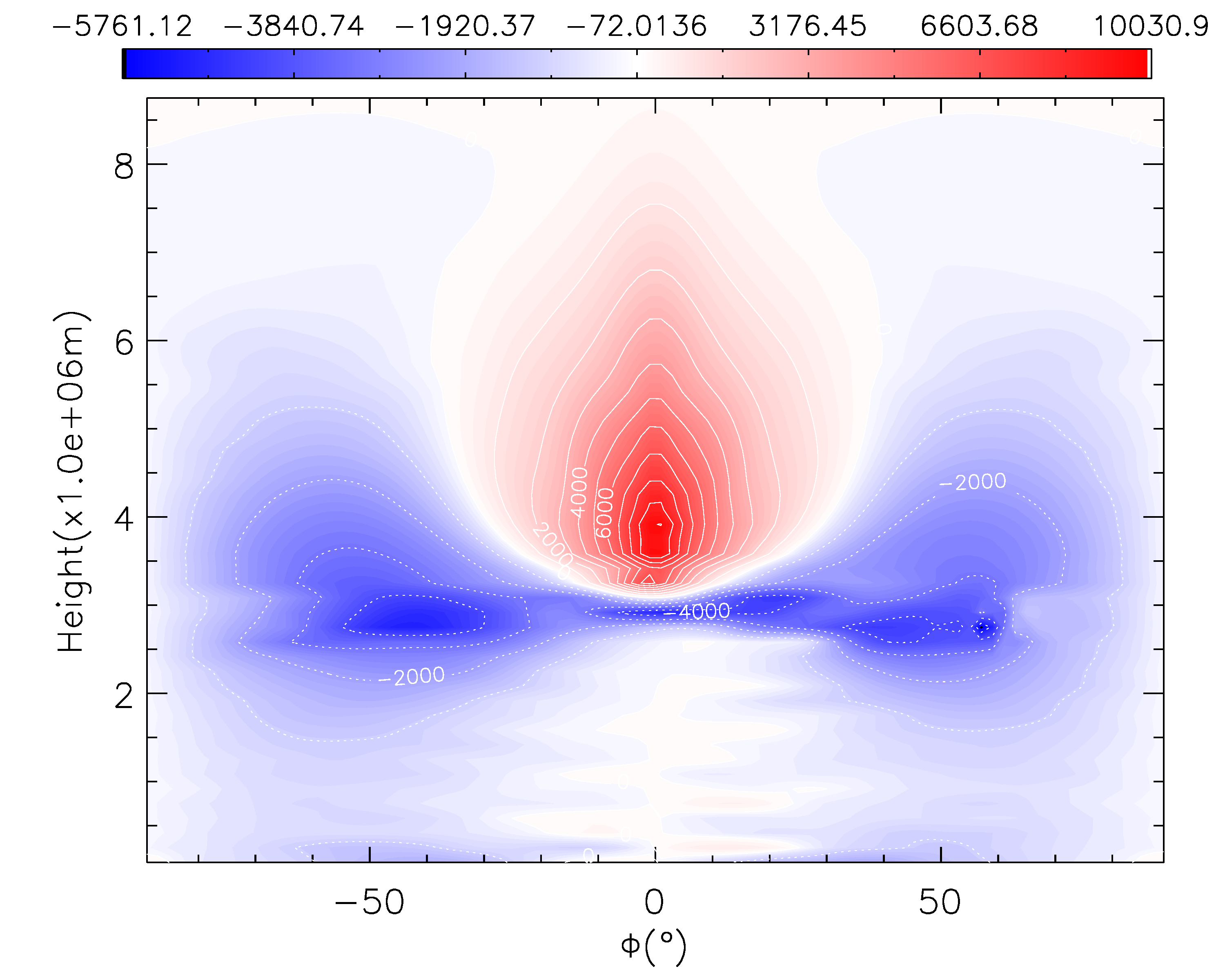}\label{std_full_lin_mmtm_slice_ylim_1000}}
\subfigure[Std RT: $\Sigma_{\lambda}\left(\rho u\right)$, 1\,000\,days]{\includegraphics[width=8.6cm,angle=0.0,origin=c]{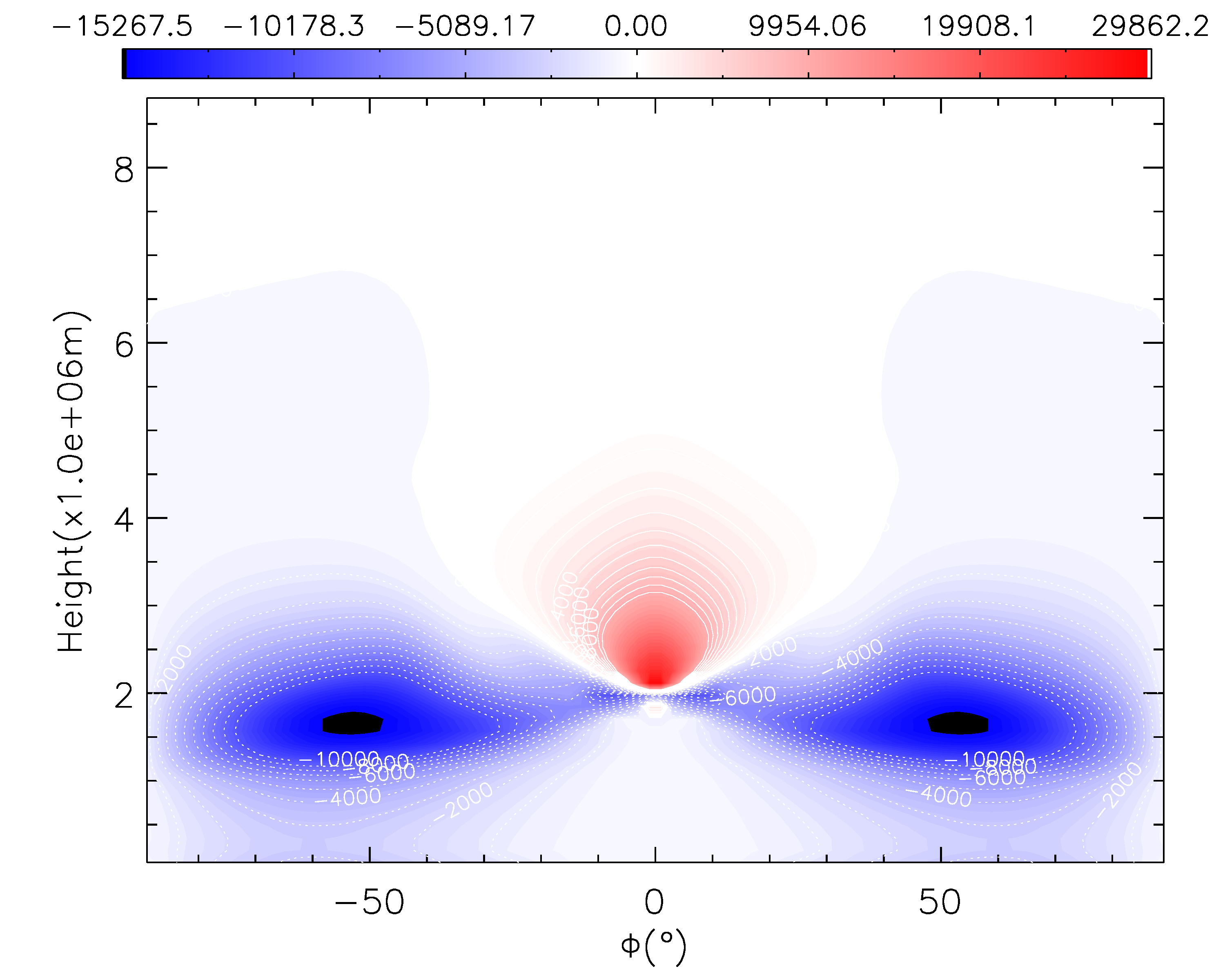}\label{rt_notiovo_full_lin_mmtm_slice_ylim_1000}}
 \end{center}
\vspace{-0.6cm}
 \caption{Figure showing the zonally summed linear zonal momentum
   (contours, kgm$^{-3}$ms$^{-1}$), as a function of latitude ($\phi$)
   and height ($z$, m). of the Std Full and Std RT simulations, as the
   \textit{left} and \textit{right columns}, respectively (see Table
   \ref{model_names} for explanation of simulation names). Red is
   positive or prograde and blue negative or retrograde. The
   quantities are zonally summed and presented at instantaneous times
   of 1, 10 and 1\,000\,days as the \textit{top}, \textit{middle} and
   \textit{bottom rows}, respectively. \label{lin_mmtm_comp}}
\end{figure*}

Firstly, Figure \ref{lin_mmtm_comp} presents the evolution of the
zonally summed linear zonal momentum ($\rho u$) for the Std Full and
Std RT simulations, as the \textit{left} and \textit{right columns},
respectively. The \textit{top}, \textit{middle} and \textit{bottom
  rows} show the zonal momentum after 1, 10 and 1\,000\,days,
respectively.

\begin{figure}
\begin{center}
\subfigure[Std Full: $\Sigma_{\lambda}\left(\rho u\right)$, 10\,000\,days]{\includegraphics[width=9.0cm,angle=0.0,origin=c]{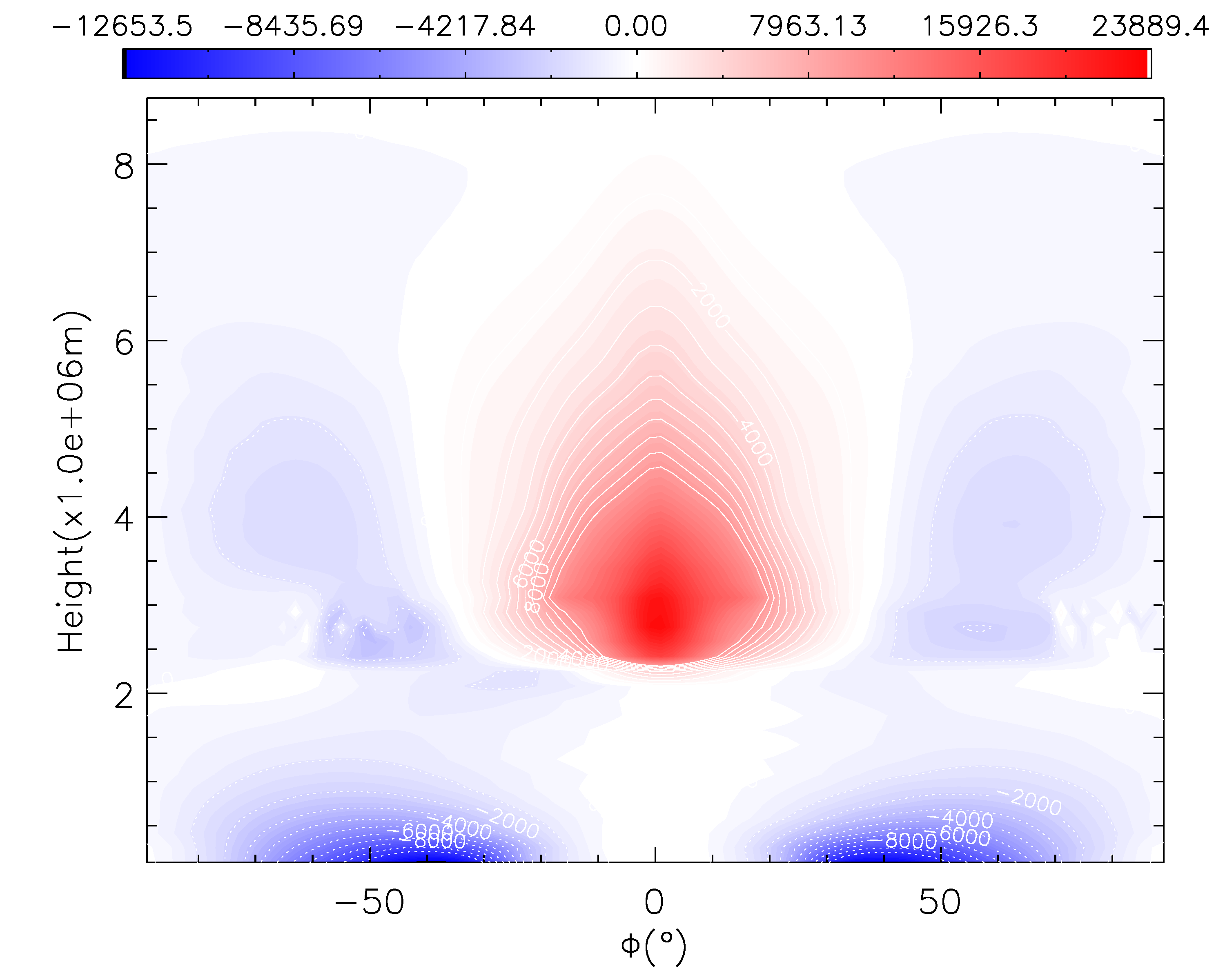}\label{std_full_lin_mmtm_slice_ylim_10000}}
 \end{center}
 \caption{Figure showing the same data as Figure \ref{lin_mmtm_comp},
   but for the Std Full simulation (see Table \ref{model_names} for
   explanation of simulation names) only, and after
   10\,000\,days. \label{lin_mmtm_long}}
\end{figure}

Figure \ref{lin_mmtm_comp} shows a broadly similar evolution in the
zonal momentum of the Std Full and Std RT simulations, when accounting
for the vertical offset. Each model covers a slightly different
pressure range, meaning the height axis will represent slightly
different pressure regimes. The initial acceleration occurs very
rapidly, as suggested by \citet{showman_2011}, with a jet structure
appearing in the first day. After around 10\,days the acceleration of
the flows in the low pressure, radiatively forced regions is largely
complete for the Std TF simulation, with slow evolution out to
1\,000\,days. After 1\,day the acceleration is broader and more
localised to low pressures, for the Std RT simulation, however this
may well simply be representative of transient evolution from slightly
different initial conditions. The Std RT simulation also undergoes
slower acceleration over the first 10\,days. The later stages are
broadly similar, but the Std RT presents a smoother jet profile. For
the Std Full simulation, Figure \ref{lin_mmtm_long} shows the same
data as Figure \ref{lin_mmtm_comp} but after 10\,000\,days, near the
end of our simulation. The overall structure of the zonal momentum has
not dramatically changed, from 1\,000\,days, however the zonal
momentum both in the jet (prograde), and at very high pressures
(retrograde), has significantly increased.

\subsubsection{Pseudo--Steady State Eddy Momentum Transport}
\label{subsub_section:steady_eddy}

For the rest of this section we focus on the eddy, and mean flow
interaction in our Std Full and Std RT simulations.  The work of
\citet{showman_2011} built on the analytical solutions of
\citet{matsuno_1966} and \citet{gill_1980}. \citet{matsuno_1966} and
\citet{gill_1980} studied the linear response (steady state) of an
atmosphere at rest, in 2D, to non-axisymmetric heating as present in
tidally-locked planets. \citet{matsuno_1966}, however, included a
positive and negative heating, on opposing `hemispheres', whereas
\citet{gill_1980} considered only the positive forcing, analogous to
the ``day side'' heating. These differences yield a subtly different
solution.  In order to compare the structure of the perturbations in
our eddies with the previous work of \citet{showman_2011},
\citet{matsuno_1966} and \citet{gill_1980}, we decompose the flow into
a zonally--averaged mean flow and a perturbation from this mean. For
example, $u\rightarrow \overline{u}+u^{\prime}$, where $\overline{u}$
is the zonally averaged zonal velocity. Note, these perturbations are
departures from a zonal mean, not from a \textit{temporal mean} flow.

Firstly, we explore the linear response of the atmosphere, from its
initial rest state, to the forcing, after the first day. The work of
\citet{showman_2011} incorporates simulations on a vertical pressure
grid, whereas our model adopts height as the vertical coordinate
\citep[see discussion in ][]{hardiman_2010}. Therefore, we interpolate
our prognostic variables onto pressure surfaces to aid comparison of
the eddy structures. However, for the divergence of the eddy momentum
fluxes, discussed earlier in this section, such a transformation is
much more complex so we retain our native height--based coordinate
system.

\begin{figure*}
\begin{center}
   \subfigure[Std Full: T, u, v at 3\,000\,Pa, 1\,day]{\includegraphics[width=9.0cm,angle=0.0,origin=c]{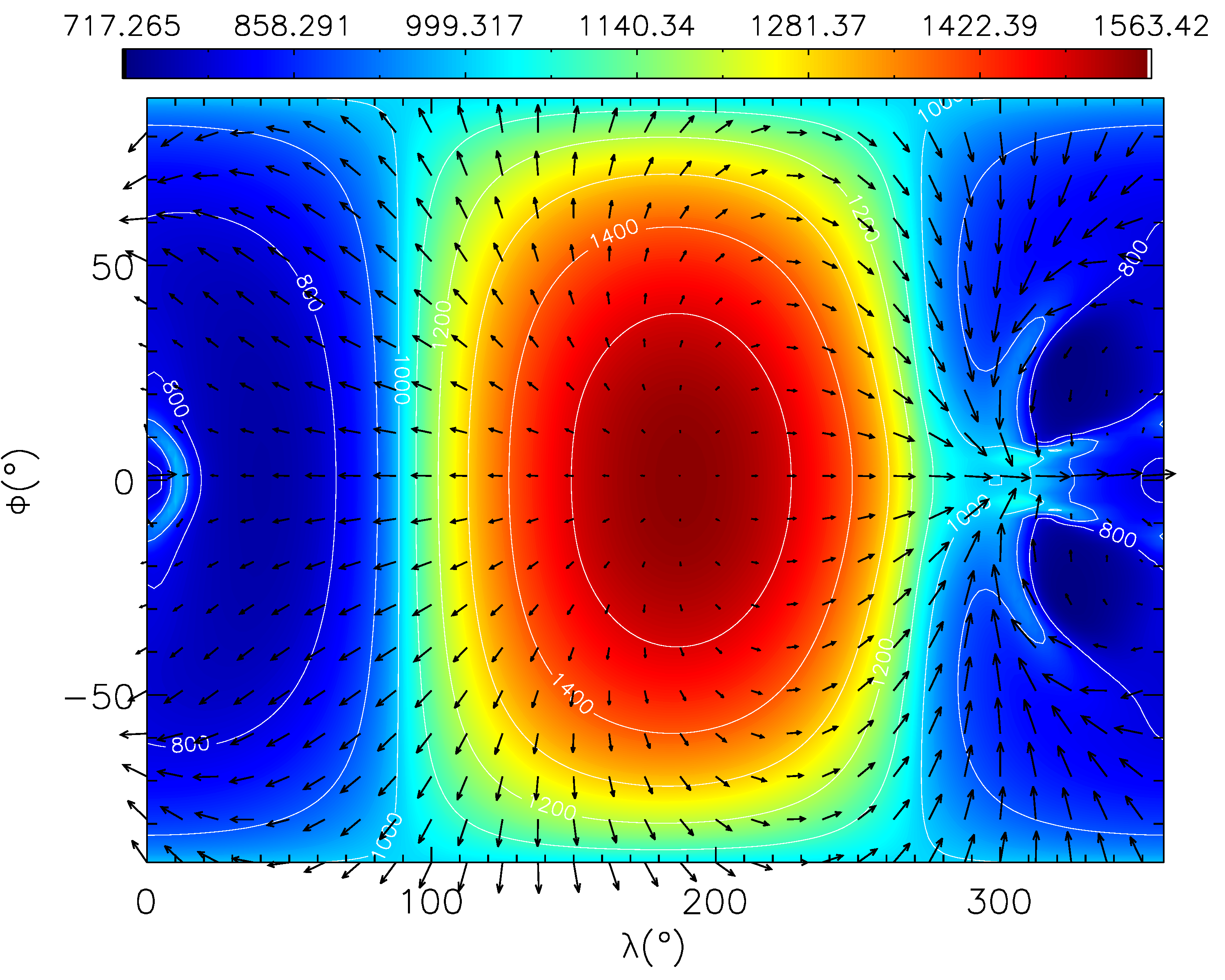}\label{std_full_slice_30mbar_1}}
  \subfigure[Std RT: T, u, v at 3\,000\,Pa, 1\,day]{\includegraphics[width=9.0cm,angle=0.0,origin=c]{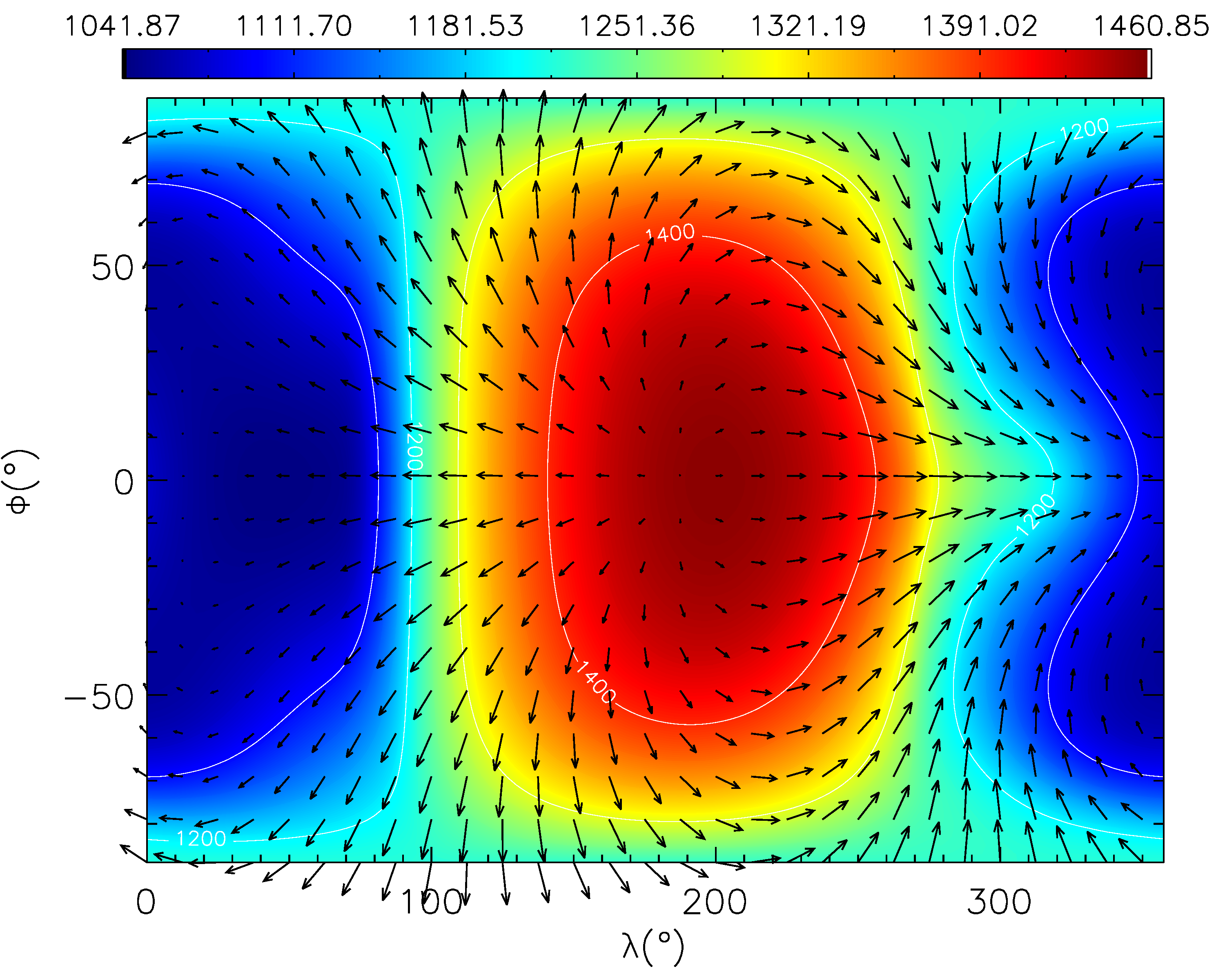}\label{rt_notiovo_full_slice_30mbar_1}}
   \subfigure[Std Full: T$^{\prime}$, u$^{\prime}$, v$^{\prime}$ at 3\,000\,Pa, 1\,day]{\includegraphics[width=9.0cm,angle=0.0,origin=c]{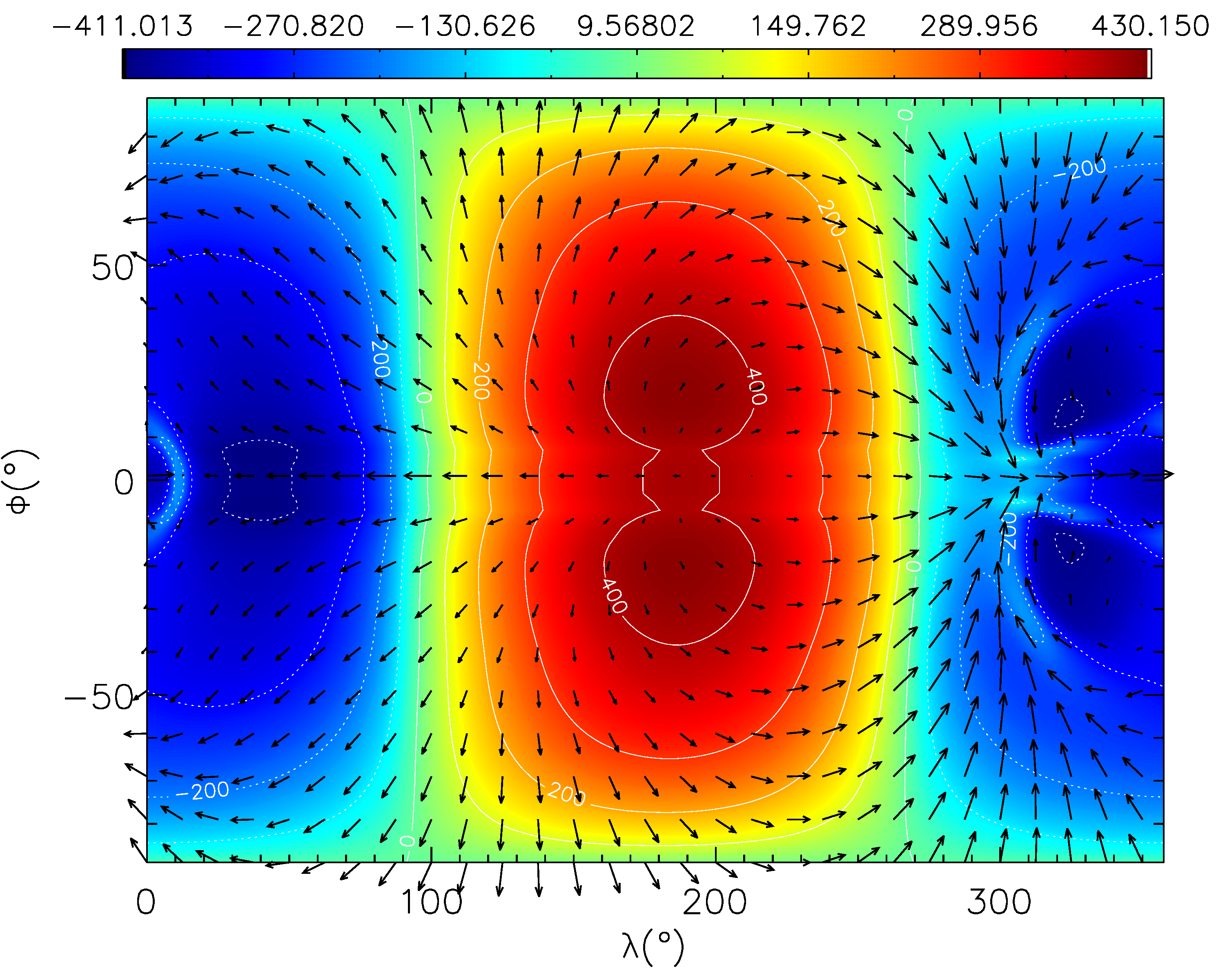}\label{std_full_eddy_slice_30mbar_1}}
   \subfigure[Std RT: T$^{\prime}$, u$^{\prime}$, v$^{\prime}$ at 3\,000\,Pa, 1\,day]{\includegraphics[width=9.0cm,angle=0.0,origin=c]{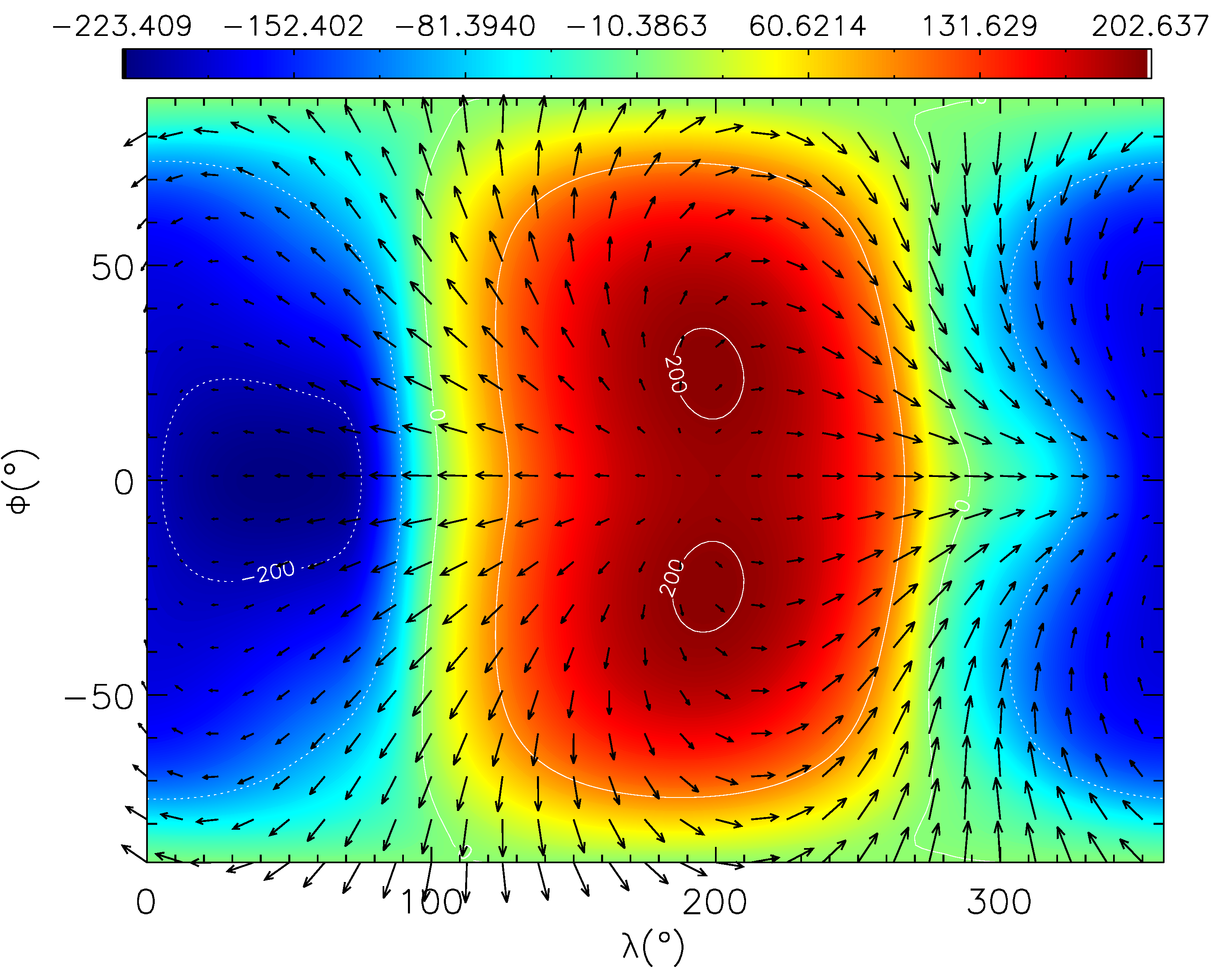}\label{rt_notiovo_full_eddy_slice_30mbar_1}}
\end{center}
\caption{Figure showing isobaric slices at 30\,mbar, 3\,000\,Pa, of
  temperature (K, colour scale) and horizontal wind (ms$^{-1}$,
  vectors), and their eddy components as the \textit{top} and
  \textit{bottom rows}, respectively. The \textit{left} and
  \textit{right columns} show results from the Std Full and Std RT
  simulation (see Table \ref{model_names} for explanation of
  simulation names), respectively. Features similar to that of
  \citet{matsuno_1966} and \citet{gill_1980} are apparent. The maximum
  magnitudes of the horizontal velocities (\& eddy components) are
  $\sim$4\,300, $\sim$1\,400, $\sim$4\,200 \& 1\,400\,ms$^{-1}$ for
  the \textit{top left}, \textit{top right}, \textit{bottom left} \&
  \textit{bottom right panels}, respectively. \label{30mbar_slice}}
\end{figure*}

Figure \ref{30mbar_slice} shows the temperature and horizontal wind
velocity (\textit{top row}) alongside their eddy components
(\textit{bottom row}), as the contours and vectors, respectively, for
the Std Full and Std RT simulation after 1\,day (\textit{left} and
\textit{right columns}, respectively), on the 30\,mbar 3\,000\,Pa
isobaric surface \citep[approximately the infrared
photosphere,][]{showman_2011}. Figure \ref{30mbar_slice} shows a
similar picture for both the Std Full and Std RT
simulations. Vortices, as presented in \citet{showman_2011} are seen
either side of the equator, both to the west and east of the hot spot,
with perturbations tilting westward, and thereby driving momentum to
the equator \citep{vallis_2006,showman_2011}. The pattern is apparent
even after 1\,day which is much faster than the advective timescale
for a $\sim$kms$^{-1}$ wind traversing a planet with a radius of
$\sim$10$^8$m. This indicates, as discussed in \citet{showman_2011}
that fast waves rapidly setup the perturbation pattern which in turn
accelerates the zonal flow. However, as we are starting from an
unphysical, rest state, it is not clear that this activity is relevant
to the study of hot Jupiters. The perturbations also show similar
features around the equator, where peaks and troughs in the zonal
velocity can be seen, corresponding to the equatorially trapped Kelvin
or Rossby waves previously identified
\citep{matsuno_1966,gill_1980,showman_2011}\footnote{Note
  \citet{matsuno_1966,gill_1980} plot pressure perturbations, whereas
  we plot temperature perturbations. However, we have checked density
  perturbations from our simulations, and the patterns match those in
  the temperature structure.}.

Figure \ref{30mbar_slice} shows that the linear response of our
simulations, under the radiative forcing, is similar to previous
results \citep[e.g.][]{showman_2011}. However, given that our
simulations are accelerating from a non-physical initial condition, it
is not clear what relevance this has to real hot Jupiter
atmospheres. Therefore, we now move to diagnosing the divergence of
the eddy momentum fluxes, which tracks the transport of momentum by
eddies in the simulated atmospheres. As discussed in Section
\ref{sub_section:deep_atmosphere} the deeper layers of our atmosphere
are still evolving, therefore, we restrict this analysis to upper, low
pressures regions. Diagnosing the momentum fluxes in the atmosphere is
done using the eddy-mean flow interaction equation \citep[see for
example][]{hardiman_2010}. \citet{showman_2011} explore the meridional
and vertical eddy terms of this equation in its primitive hydrostatic
form. For our purposes, \citet{hardiman_2010} derive the same equation
in height--based coordinates for our ``Full'' equation system. The
derivation, and different forms of this equation are included in
Appendix \ref{app_section:eddy_mean_derive}, but for the simulations
we analyse here we adopt (see Equation \ref{hardiman_eqn})
\begin{align}
(\overline{\rho}\,\overline{u})_{, t}&=&-\frac{(\overline{\rho v}\,\overline{u}\cos^2\phi)_{, \phi}}{r\cos^2\phi}-\frac{(\overline{\rho w}\,\overline{u}r^3)_{, r}}{r^3}\\\nonumber
&&+2\Omega\overline{\rho v}\sin\phi-2\Omega\overline{\rho w}\cos\phi\\\nonumber
&&-(\overline{\rho^{\prime}u^{\prime}})_{, t}-\frac{\left[ \overline{(\rho v)^{\prime}u^{\prime}}\cos^2\phi\right]_{, \phi}}{r\cos^2\phi}-\frac{\left[ \overline{(\rho w)^{\prime}u^{\prime}}r^3\right]_{, r}}{r^3}+\\\nonumber
&&\overline{\rho G_{\lambda}},\nonumber
\end{align}
where $r$ is radial position and $G_\lambda$ represents body forces
acting in the zonal direction. The subscript preceded by a comma
denotes a partial derivative, e.g,. $(X)_{,r}\rightarrow\frac{\partial
  X}{\partial r}$. As mentioned previously in this section we present
results in our height--based system, as conversion to a
pressure--based framework results in much more complex terms in the
eddy--mean flow interaction equation.

In the mechanisms proposed by \citet{showman_2011} the main balance,
over the equator, is between the meridional ($-\frac{\left[
    \overline{(\rho v)^{\prime}u^{\prime}}\cos^2\phi\right]_{,
    \phi}}{r\cos^2\phi}$) and vertical ($-\frac{\left[ \overline{(\rho
      w)^{\prime}u^{\prime}}r^3\right]_{, r}}{r^3}$) eddy momentum
transport terms i.e. those containing $v^{\prime}$ and $w^{\prime}$
(and gradients in the latitudinal, $\phi$ and the vertical, $r$,
directions, respectively). The remaining important terms can be
classified as mean flow terms involving the zonal velocity paired with
either the meridional or vertical flow ($-\frac{(\overline{\rho
    v}\,\overline{u}\cos^2\phi)_{, \phi}}{r\cos^2\phi}$,
$-\frac{(\overline{\rho w}\,\overline{u}r^3)_{, r}}{r^3}$), or the
mean Coriolis terms derived from the contributions of the Coriolis
terms of the momentum equations ($+2\Omega\overline{\rho v}\sin\phi$,
$-2\Omega\overline{\rho w}\cos\phi$). \citet{tsai_2014} test the
presence in their simulations, of the proposed pumping mechanism of
\citet{showman_2011} in the context of the $\beta$-plane where the
Coriolis parameter is assumed to vary linearly with latitude (as
opposed to $\propto\cos\phi$). \citet{tsai_2014} present figures of
the meridional and vertical eddy momentum transport (strictly the
divergence of the eddy momentum fluxes), supporting an eddy--driven
jet, although they do not present contributions from mean flow terms.

\begin{figure*}
\begin{center}
\subfigure[Std Full:$\Sigma_{\lambda}-\left(\frac{\left[ \overline{(\rho v)^{\prime}u^{\prime}}\cos^2\phi\right]_{, \phi}}{r\cos^2\phi}\right)$, 1\,000\,days]{\includegraphics[width=8.5cm,angle=0.0,origin=c]{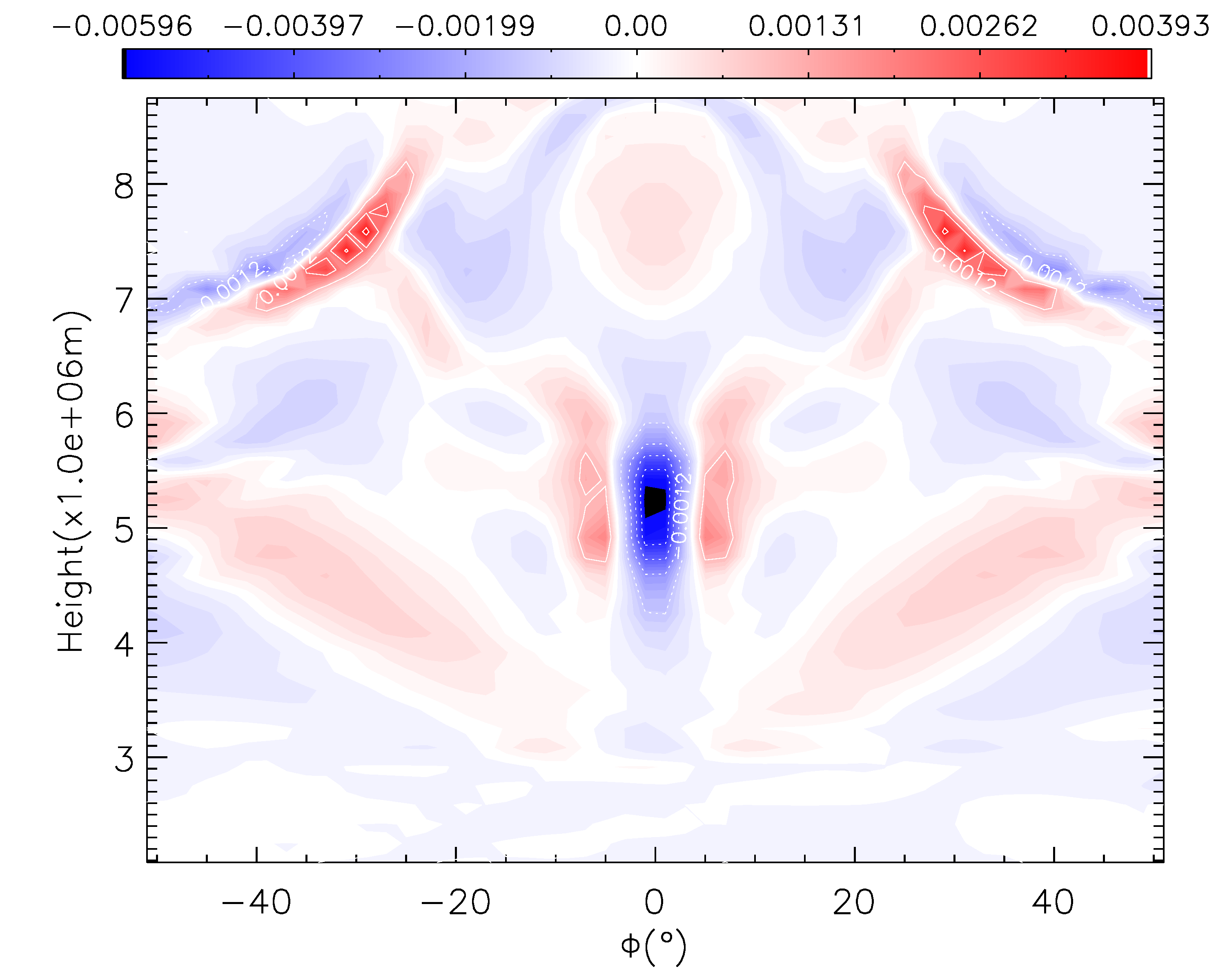}\label{std_full_meri_eddy_mmtm_slice_xylim_1000}}
\subfigure[Std RT: $\Sigma_{\lambda}-\left(\frac{\left[ \overline{(\rho v)^{\prime}u^{\prime}}\cos^2\phi\right]_{, \phi}}{r\cos^2\phi}\right)$, 1\,000\,days]{\includegraphics[width=8.5cm,angle=0.0,origin=c]{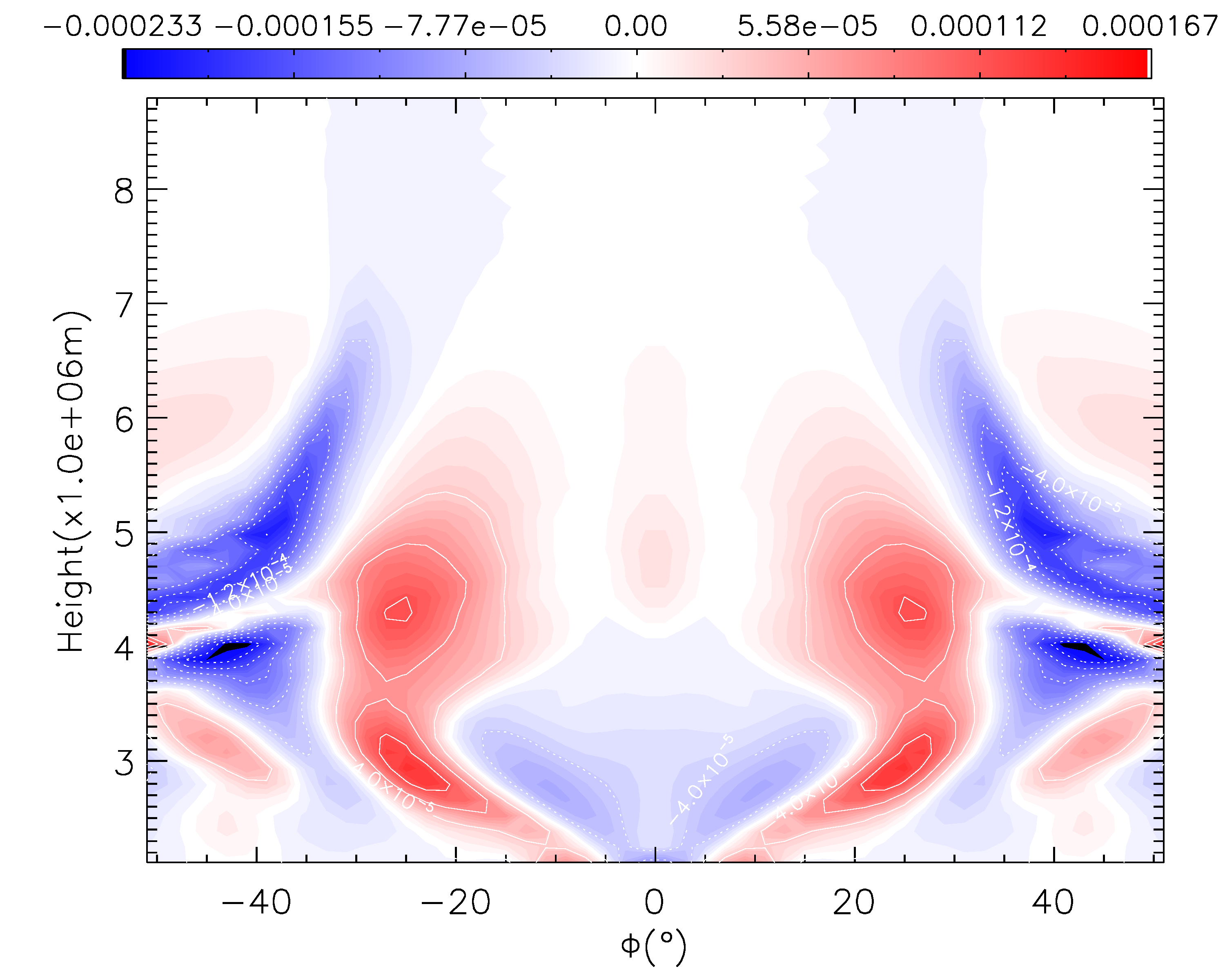}\label{rt_notiovo_full_meri_eddy_mmtm_slice_xylim_1000}}
\subfigure[Std Full:$\Sigma_{\lambda}-\left(\frac{\left[ \overline{(\rho w)^{\prime}u^{\prime}}r^3\right]_{, r}}{r^3}\right)$, 1\,000\,days]{\includegraphics[width=8.5cm,angle=0.0,origin=c]{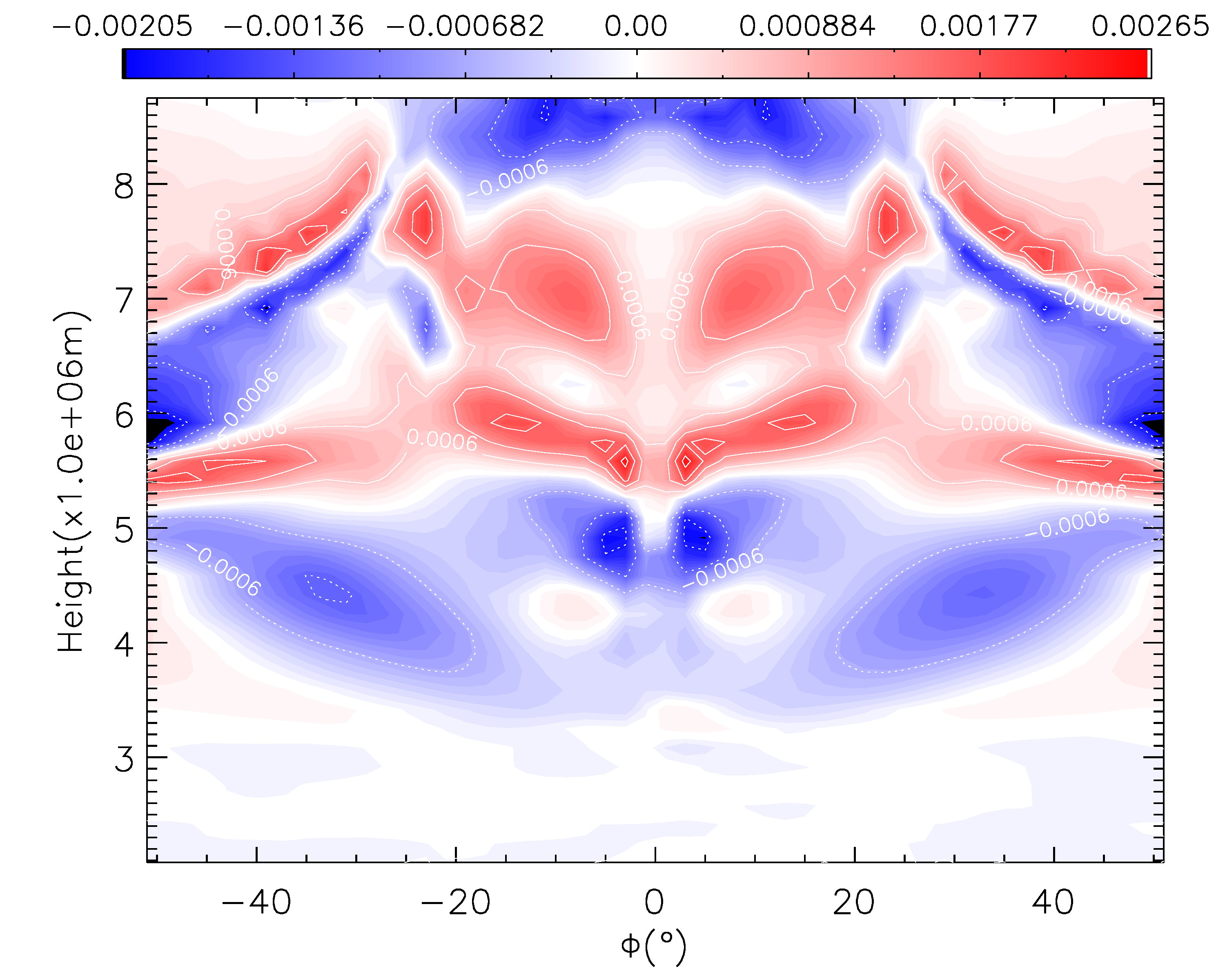}\label{std_full_vert_eddy_mmtm_slice_xylim_1000}}
\subfigure[Std RT: $\Sigma_{\lambda-}\left(\frac{\left[ \overline{(\rho w)^{\prime}u^{\prime}}r^3\right]_{, r}}{r^3}\right)$, 1\,000\,days]{\includegraphics[width=8.5cm,angle=0.0,origin=c]{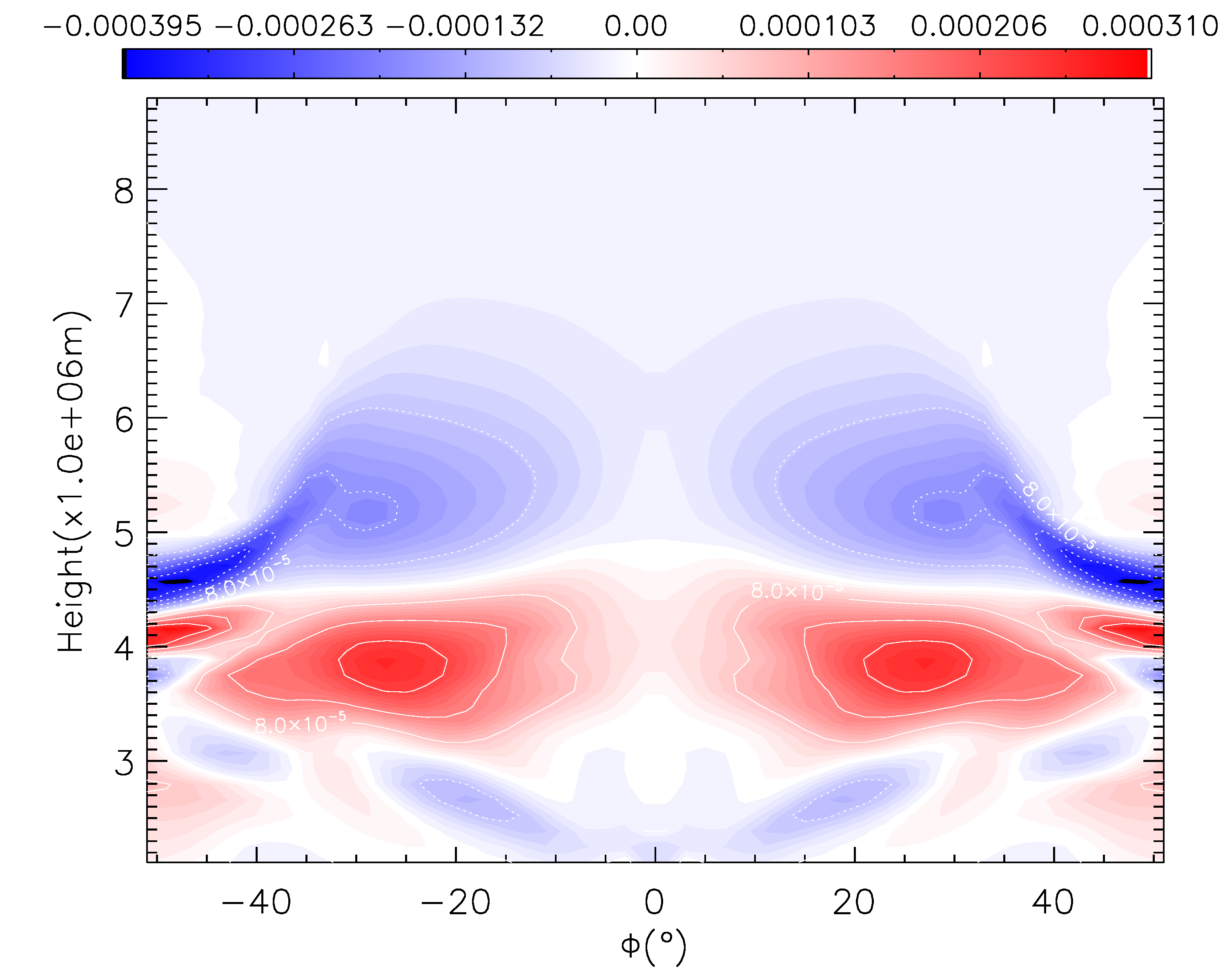}\label{rt_notiovo_full_vert_eddy_mmtm_slice_xylim_1000}}
\subfigure[Std Full: $\Sigma_{\lambda}-\left(\frac{\left[ \overline{(\rho v)^{\prime}u^{\prime}}\cos^2\phi\right]_{, \phi}}{r\cos^2\phi}+\frac{\left[ \overline{(\rho w)^{\prime}u^{\prime}}r^3\right]_{, r}}{r^3}\right)$, 1\,000\,days]{\includegraphics[width=8.5cm,angle=0.0,origin=c]{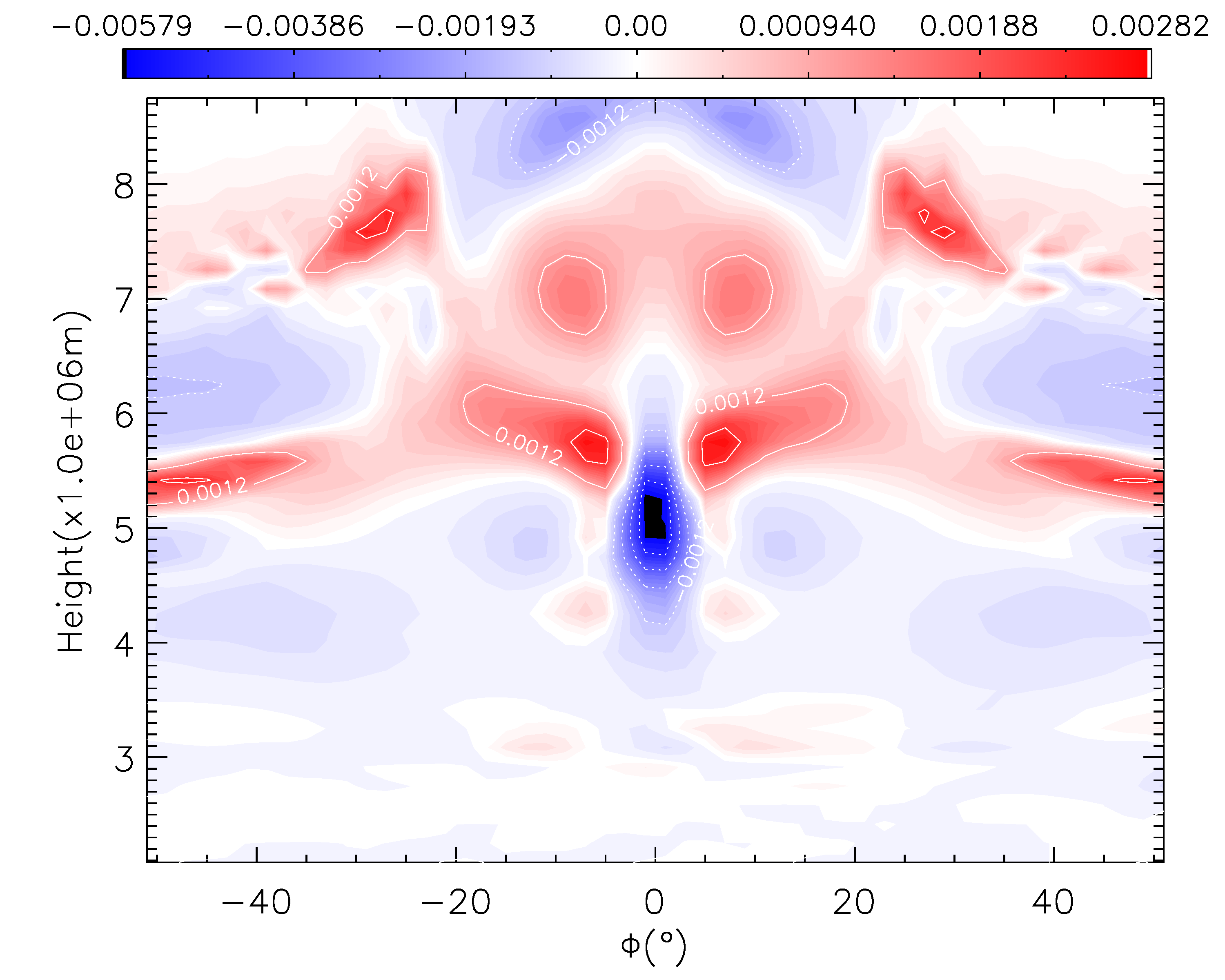}\label{std_full_sum_eddy_mmtm_slice_xylim_1000}}
\subfigure[Std RT: $\Sigma_{\lambda}-\left(\frac{\left[ \overline{(\rho v)^{\prime}u^{\prime}}\cos^2\phi\right]_{, \phi}}{r\cos^2\phi}+\frac{\left[ \overline{(\rho w)^{\prime}u^{\prime}}r^3\right]_{, r}}{r^3}\right)$, 1\,000\,days]{\includegraphics[width=8.5cm,angle=0.0,origin=c]{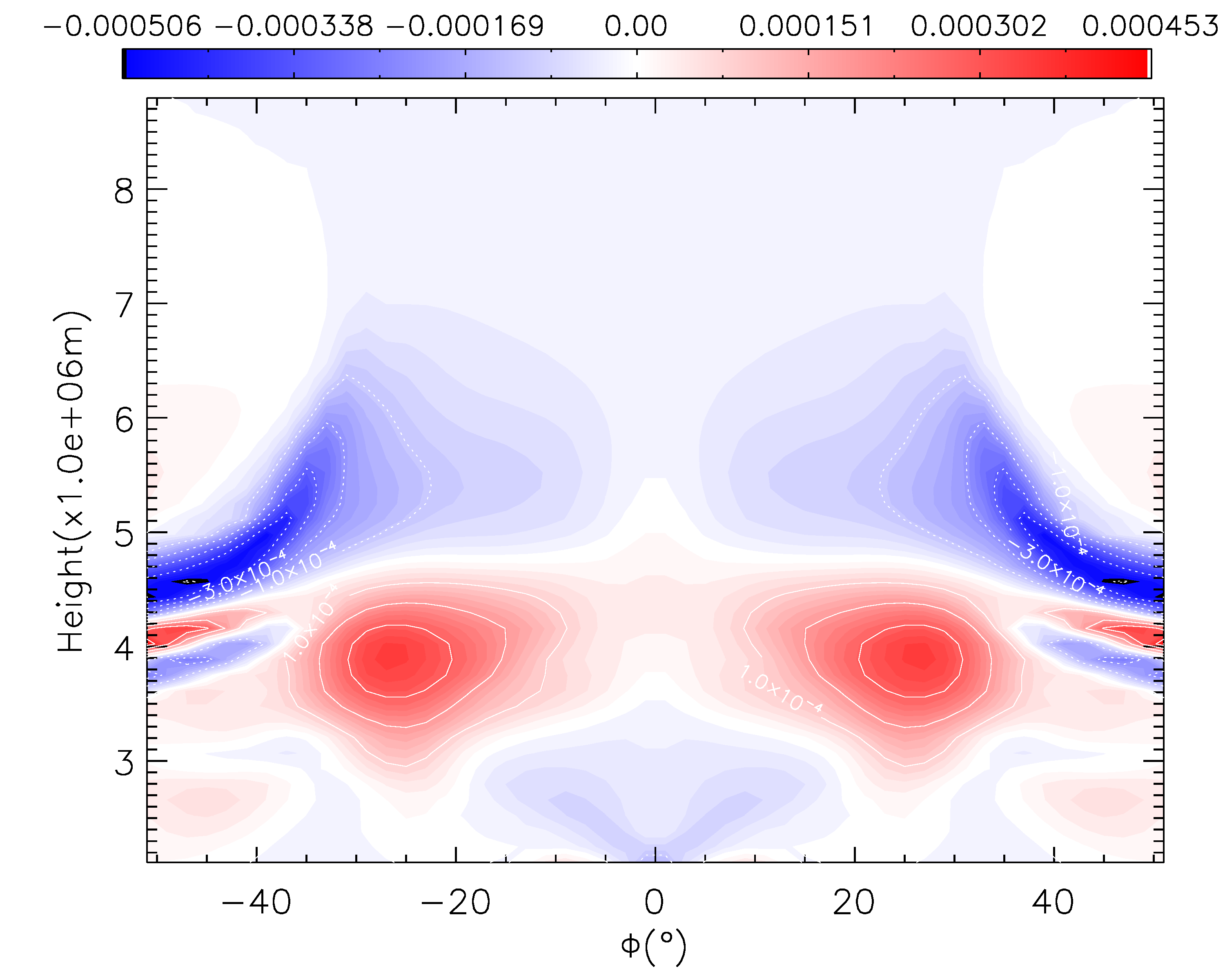}\label{rt_notiovo_full_sum_eddy_mmtm_slice_xylim_1000}}
 \end{center}
\vspace{-0.6cm}
 \caption{Figure showing latitudinal gradient in the meridional eddy
   momentum flux, the vertical gradient in the vertical eddy momentum
   flux, and their sum (as a function of latitude, $\phi$ and height,
   $z$), are shown as the \textit{top}, \textit{middle} and
   \textit{bottom rows}, respectively (see text \& Appendix
   \ref{app_section:eddy_mean_derive} for details), after
   1\,000\,days. The \textit{left} and \textit{right columns} then
   show data for the Std Full and Std RT simulation, respectively (see
   Table \ref{model_names} for explanation of simulation
   names). \label{eddy_mmtm_1000}}
\end{figure*}

Figure \ref{eddy_mmtm_1000} shows the divergence of the eddy momentum
fluxes for the Std Full and Std RT simulations, as the \textit{left}
and \textit{right columns}, respectively, after 1\,000\,days. The
latitudinal gradient in the meridional eddy momentum flux, the
vertical gradient in the vertical eddy momentum flux, and their sum,
are shown as the \textit{top}, \textit{middle} and \textit{bottom
  rows}, respectively. This figure, and subsequently similar figures
are truncated to the upper atmosphere i.e. $z>2\times 10^6$\,m to aid
interpretation\footnote{Contributions from transient perturbations,
  within these instantaneous plots, from the deeper regions (which
  have much higher densities, and therefore, momentum fluxes) can
  significantly alter the range of the plotted data. As discussed in
  Section \ref{sub_section:deep_atmosphere}, the deep atmosphere is
  still evolving there is not yet a mean flow, meaning the analysis we
  are performing is not meaningful in this region.}.

Figure \ref{eddy_mmtm_1000} reveals some similarities in the momentum
transport due to eddies, in our Std Full simulation, to those
presented in \citet{tsai_2014}. These structures are broadly
consistent with the mechanism proposed by \citet{showman_2011}, as the
eddy momentum transport largely acts to bring prograde momentum into
the jet region, however at around $4\times 10^6$\,m, over the equator,
a significant transport of retrograde momentum is present. The
vertical term (\textit{middle left panel}), shows predominantly
prograde momentum being transport from lower altitudes (from below
$5\times 10^6$\,m) to the upper atmosphere and retrograde momentum at
lower altitudes. The sum of these two terms, of the Std Full
simulation, after 1\,000\,days, does not demonstrate clear balance
between them (\textit{bottom left panel}) and residual eddy momentum
transport is present of the same order of the terms themselves. These
patterns are steady in time, albeit gradually decreasing in magnitude
(see Figure \ref{eddy_mmtm_10000} in Appendix
\ref{app_section:eddy_terms}). Interestingly, eddy momentum transport
after 1\,000\,days is very different in the Std RT case (\textit{right
  column} of Figure \ref{eddy_mmtm_1000}), compared to the Std Full
simulation. The balance between the meridional and vertical component
is closer, i.e. the magnitude of the summed momentum transports
(\textit{bottom right}) is much lower than the Std Full
counterpart. The vertical term is the opposite to that shown for the
Std Full simulation, as prograde momentum is transported into regions
below an altitude of $4\times 10^6$\,m. Additionally, the meridional
component is transporting prograde eddy momentum in the flanks of the
jet ($\phi=\pm 20^{\degree}$).

\begin{figure*}
\begin{center}
\subfigure[Std Full:$\Sigma_{\lambda}-\frac{(\overline{\rho v}\,\overline{u}\cos^2\phi)_{,\phi}}{r\cos^2\phi}$, 1\,000\,days]{\includegraphics[width=8.6cm,angle=0.0,origin=c]{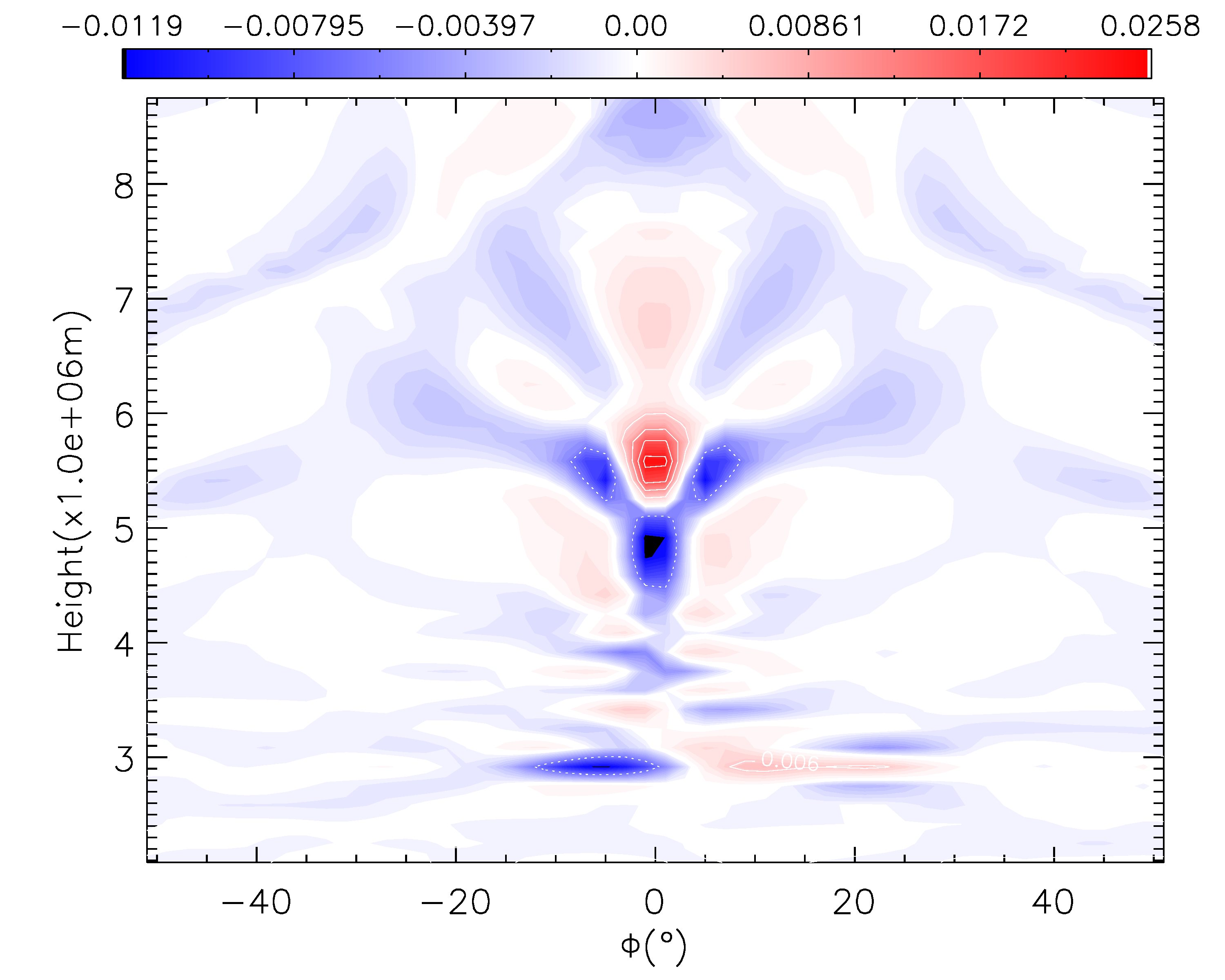}\label{std_full_meri_mean_mmtm_slice_xylim_1000}}
\subfigure[Std RT:$\Sigma_{\lambda}-\frac{(\overline{\rho v}\,\overline{u}\cos^2\phi)_{,\phi}}{r\cos^2\phi}$, 1\,000\,days]{\includegraphics[width=8.6cm,angle=0.0,origin=c]{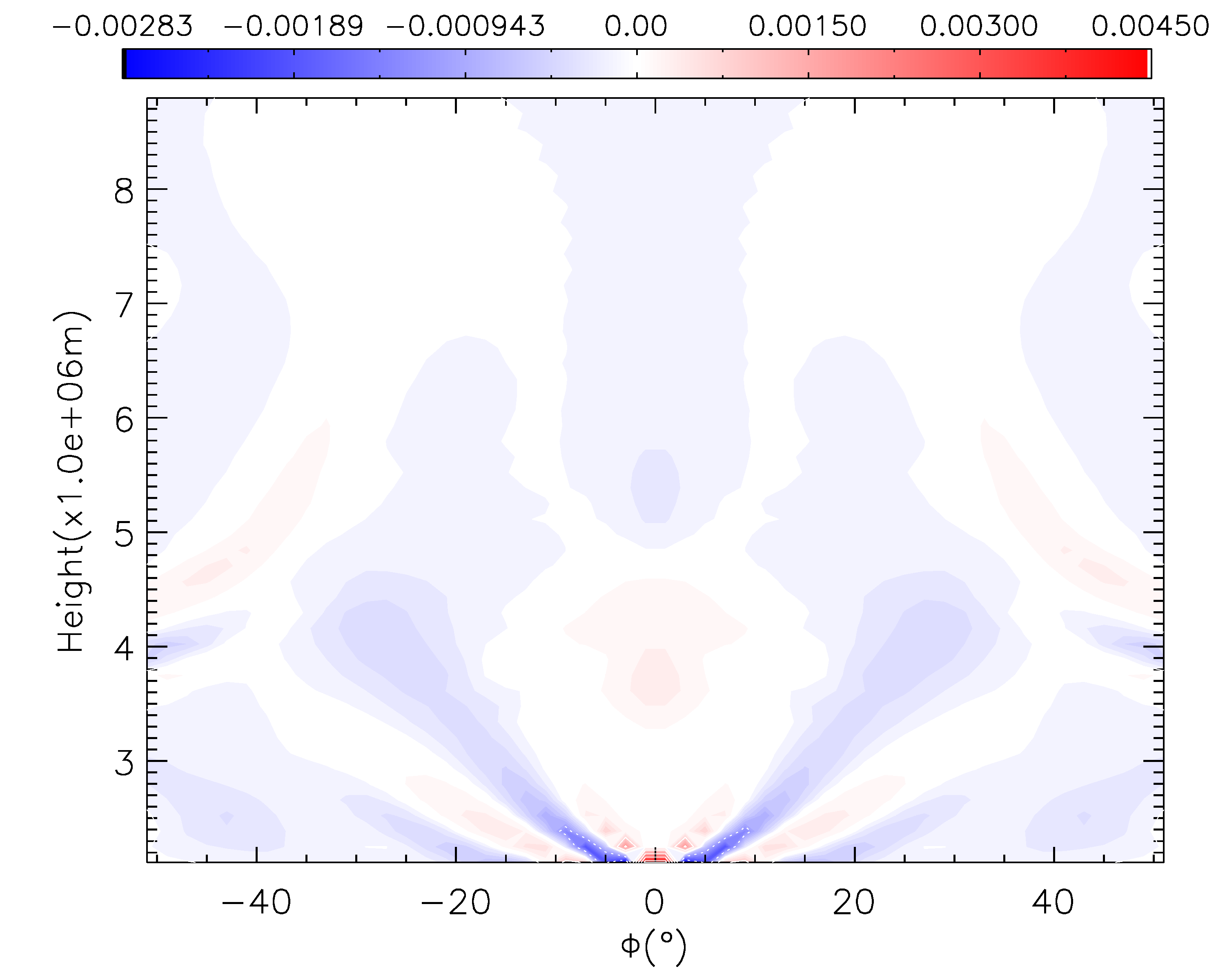}\label{rt_notiovo_full_meri_mean_mmtm_slice_xylim_1000}}
\subfigure[Std Full:$\Sigma_{\lambda}-\frac{(\overline{\rho w}\,\overline{u}r^3)_{, r}}{r^3}$, 1\,000\,days]{\includegraphics[width=8.6cm,angle=0.0,origin=c]{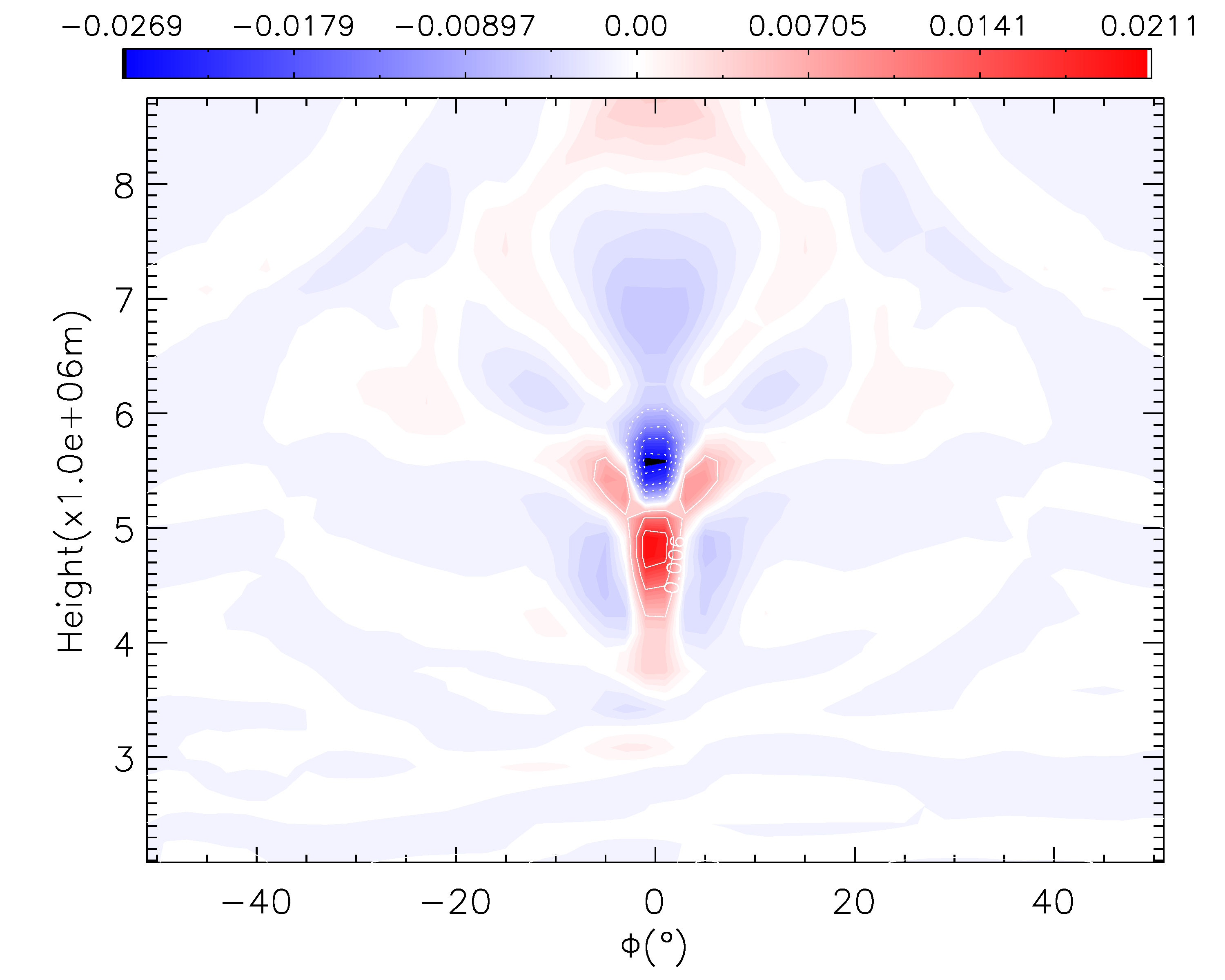}\label{std_full_vert_mean_mmtm_slice_xylim_1000}}
\subfigure[Std RT:$\Sigma_{\lambda}-\frac{(\overline{\rho w}\,\overline{u}r^3)_{, r}}{r^3}$, 1\,000\,days]{\includegraphics[width=8.6cm,angle=0.0,origin=c]{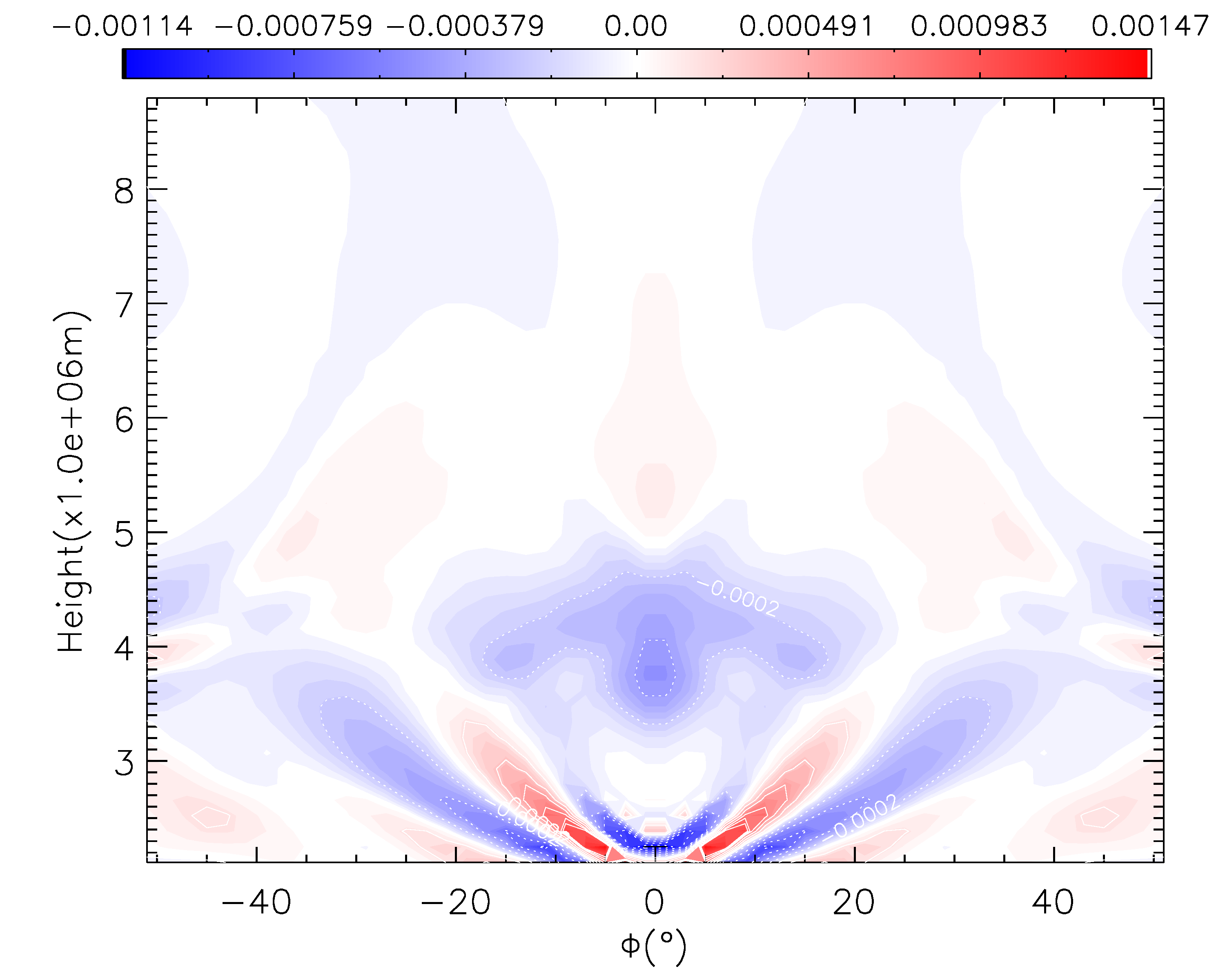}\label{rt_notiovo_full_vert_mean_mmtm_slice_xylim_1000}}
\subfigure[Std Full:$\Sigma_{\lambda}-\left(\frac{(\overline{\rho v}\,\overline{u}\cos^2\phi)_{,\phi}}{r\cos^2\phi}+\frac{(\overline{\rho w}\,\overline{u}r^3)_{, r}}{r^3}\right)$, 1\,000\,days]{\includegraphics[width=8.6cm,angle=0.0,origin=c]{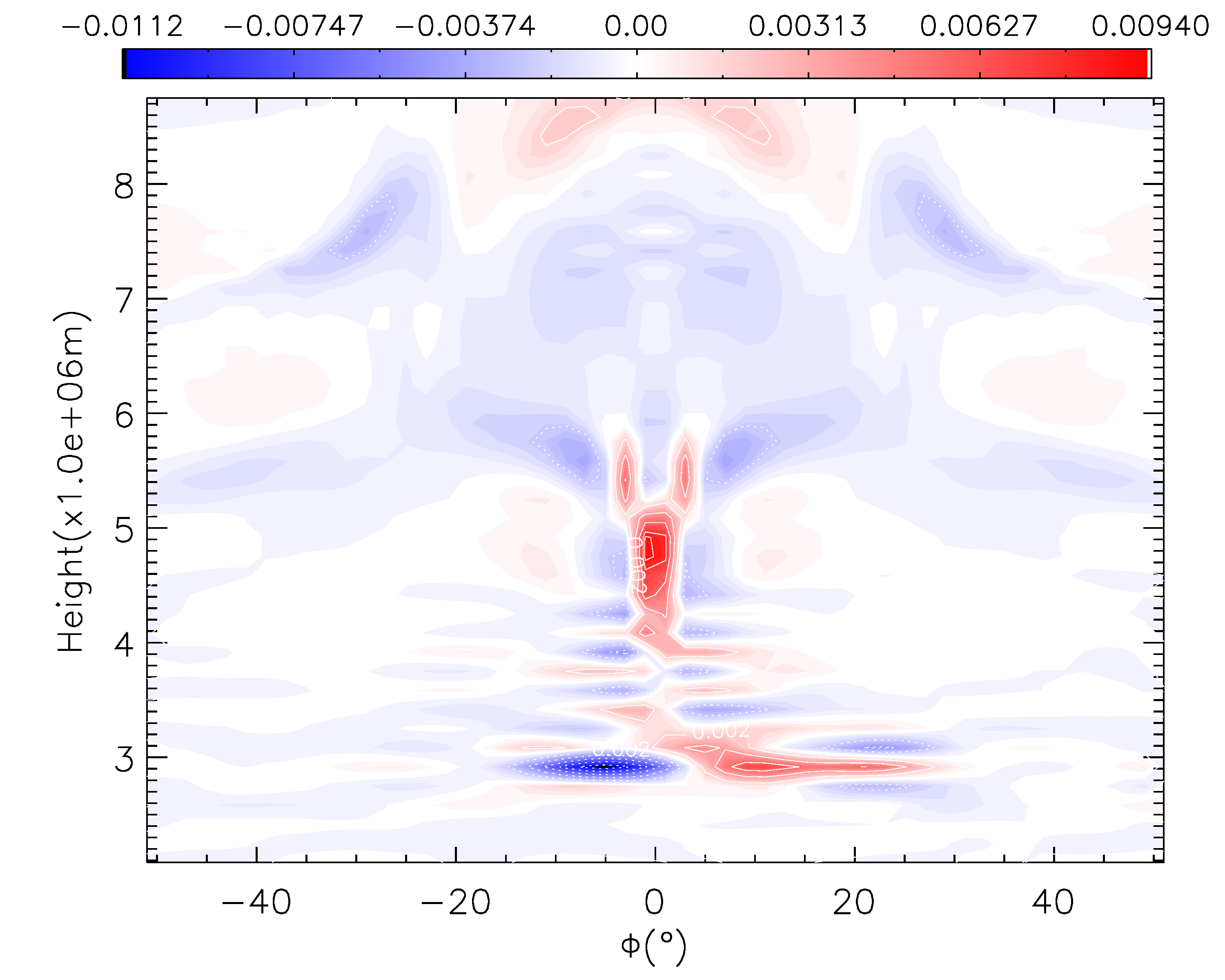}\label{std_full_sum_mean_mmtm_slice_xylim_1000}}
\subfigure[Std RT:$\Sigma_{\lambda}-\left(\frac{(\overline{\rho v}\,\overline{u}\cos^2\phi)_{,\phi}}{r\cos^2\phi}+\frac{(\overline{\rho w}\,\overline{u}r^3)_{, r}}{r^3}\right)$, 1\,000\,days]{\includegraphics[width=8.6cm,angle=0.0,origin=c]{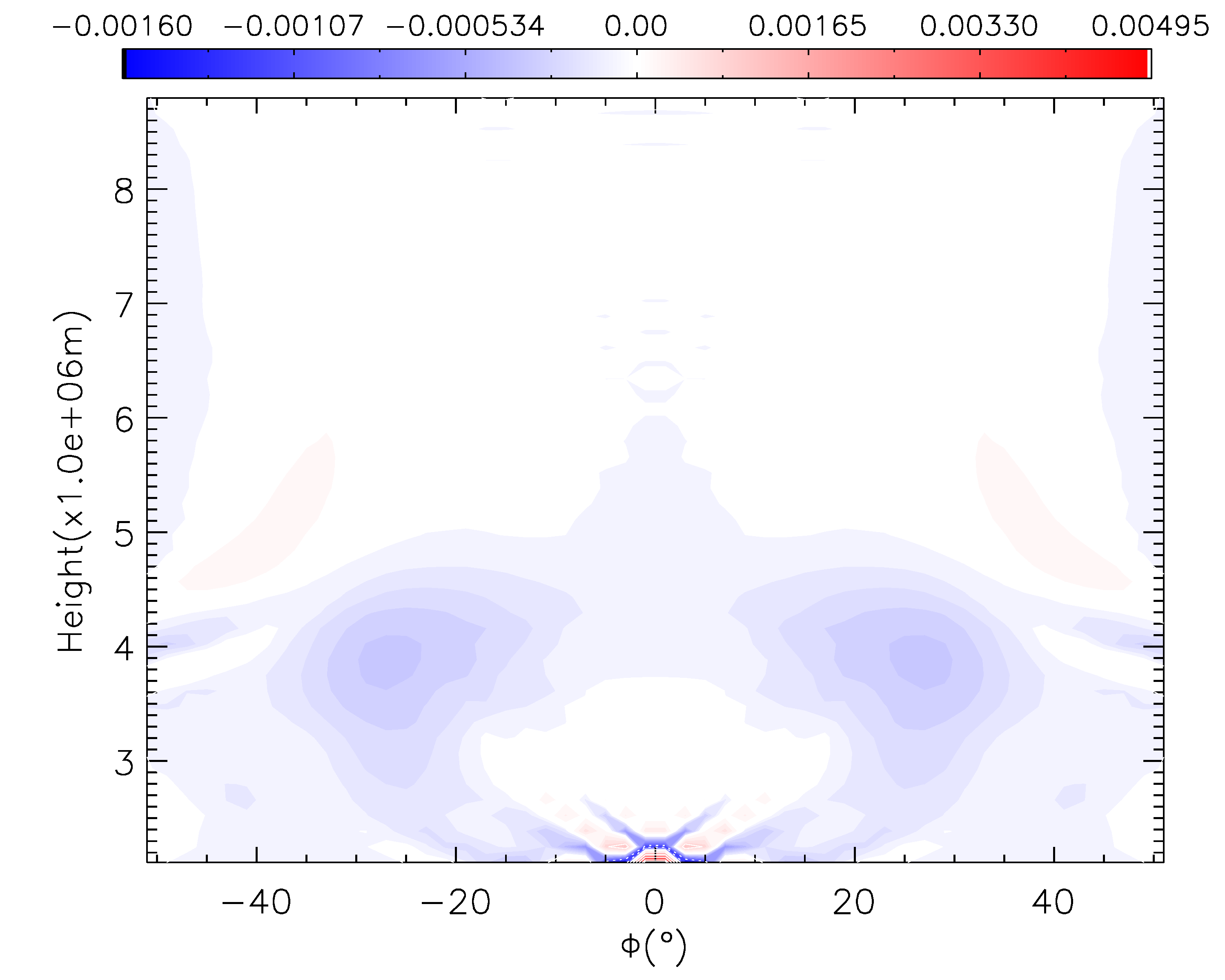}\label{rt_notiovo_full_sum_mean_mmtm_slice_xylim_1000}}
 \end{center}
\vspace{-0.6cm}
 \caption{Figure showing the latitudinal gradient of the meridional
   mean flow momentum flux, the vertical gradient of the vertical mean
   flow momentum flux and their sum, as the \textit{top},
   \textit{middle} and \textit{bottom panels}, respectively. The data
   are in the same format as Figure \ref{eddy_mmtm_1000}, showing the
   Std Full and Std RT simulations, as the \textit{left} and
   \textit{right columns} respectively, after
   1\,000\,days. \label{mean_mmtm_1000}}
\end{figure*}

Figure \ref{mean_mmtm_1000} presents the mean flow contributions for
the Std Full and Std RT simulations, as the \textit{left} and
\textit{right columns}, respectively, after 1\,000\,days. The
\textit{top}, \textit{middle} and \textit{bottom rows} show the mean
meridional ($-\frac{(\overline{\rho v}\,\overline{u}\cos^2\phi)_{,
    \phi}}{r\cos^2\phi}$) term, mean vertical ($-\frac{(\overline{\rho
    w}\,\overline{u}r^3)_{, r}}{r^3}$) term, and their sum,
respectively.

Figure \ref{mean_mmtm_1000} shows that the mean flow contributions are
somewhat balanced throughout much of the atmosphere, but show a
localised peak contribution in the sum over the equator at
$4\times 10^6$\,m, at $\sim 2\times 10^6$\,m for the Std Full and Std
RT simulations, respectively. These contributions are larger in
magnitude than those provided by the divergence of the eddy momentum
fluxes (compare with Figure \ref{eddy_mmtm_1000}), indicating a jet
maintained by a mix of eddy momentum transport and the mean flow. The
contributions from the remaining terms, deriving from the Coriolis
terms of the momentum equations ($2\Omega\overline{\rho v}\sin\phi$
and $-2\Omega\overline{\rho w}\cos\phi$) contribute an order of
magnitude lower momentum transport within the atmosphere (see Appendix
\ref{app_section:eddy_terms}). Therefore, our results suggest that the
jet driving mechanism, in our simulations, may actually be a
combination of eddy and mean flow momentum transport.

\section{Conclusions}
\label{section:conclusions}

In this study we have presented a set of simulations, nominally based
on HD~209458b, aimed at investigating the robustness of the
atmospheric dynamical morphology. These simulations have been designed
to explore the dependence of the atmospheric flows on various
assumptions made in standard models, primarily the state of the
deeper, higher pressure atmosphere which is inaccessible to
observations. We have presented the atmospheric dynamical structure
found in this simulation set and for a subset, explored the eddy-mean
interaction revealing momentum transport in the atmospheres.

Our results have shown that the super-rotating equatorial jet, found
as a solution for hot Jupiter atmospheres from several models
\citep[see for
example][]{cooper_2005,menou_2009,rauscher_2010,heng_2011,dobbs_dixon_2013,parmentier_2013,showman_2015,helling_2016,kataria_2016,lee_2016},
is robust in our setup. Our model does not include a drag or friction
at the bottom boundary potentially responsible for removing some of
the dependence on initial conditions
\citep[see][]{liu_2013,cho_2015}. Only when artificially forcing the
deep atmosphere over long timescales do we find the jet is
significantly slowed, and reduced in latitudinal breadth. However,
caution must still be exercised when interpreting results of
simulations of hot Jupiters as, although the quantitative dynamical
structures across our simulation set are consistent (apart from
extreme cases where we strongly force the deep atmosphere over long
timescales), the quantitative results are not. This variation in the
dynamical structure, between our simulations (and indeed over time
during the simulation) will in turn alter the thermodynamic profile of
the atmosphere (this latter point warrants a more detailed follow-up
study which we are engaged in, comparing results from the SPARC/MITgcm
and the UM).

For simulations adopting a simplified radiative transfer scheme, our
results show attributes of the proposed mechanism of
\citet{showman_2011}, however, the structure of the eddy momentum
transport is complex, and there is a non-negligible contribution from
the mean flow terms. Although the broad jet structure is similar
between the Full and the RT simulations, we find very different
patterns for the eddy momentum transport between these two
simulations, suggesting indeed that the eddy momentum transport may
not be the unique (or even the main) driver for the jet. Essentially,
we find evidence for a jet driven by both contributions from eddies
and mean flow interaction. A full investigation of the nature of the
jet acceleration is beyond the scope of this work. However, we are
performing a follow up study aimed at exploring the full, 3D solution
to the standing linear problem studied by \citet{showman_2011} and
\citet{tsai_2014} including an accurate representation of the heating
rate (Debras et al., in prep).

\begin{acknowledgements}
  We thank the anonymous referee for comments which improved this
  manuscript. NM acknowledges funding from the Leverhulme Trust via a
  Research Project Grant. JM and CS acknowledge the support of a Met
  Office Academic Partnership secondment. DSA acknowledges support
  from the NASA Astrobiology Program through the Nexus for Exoplanet
  System Science. This work is partly supported by the European
  Research Council under the European Community's Seventh Framework
  Programme (FP7/2007-2013 Grant Agreement No. 247060-PEPS and grant
  No. 320478-TOFU). The calculations for this paper were performed on
  the University of Exeter Supercomputer, a DiRAC Facility jointly
  funded by STFC, the Large Facilities Capital Fund of BIS, and the
  University of Exeter.
\end{acknowledgements}

\begin{appendix}

\section{Atmospheric flow structure}
\label{app_section:flow_details}

\begin{figure*}
\begin{center}
  \subfigure[Std Prim: 213\,Pa, 1\,200\,days]{\includegraphics[width=9.0cm,angle=0.0,origin=c]{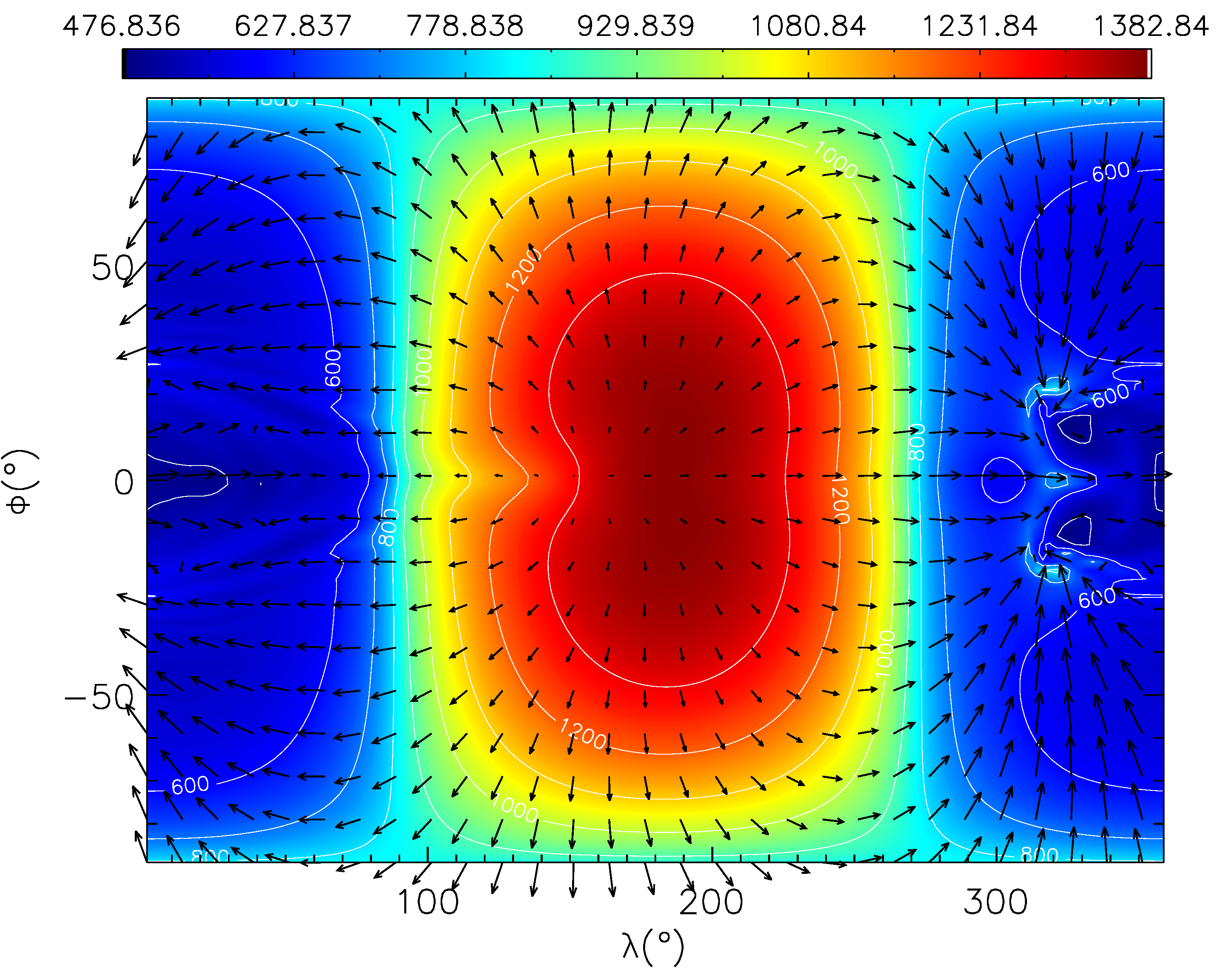}\label{std_prim_213_1200_slice}}
  \subfigure[Std Full: 213\,Pa, 1\,200\,days]{\includegraphics[width=9.0cm,angle=0.0,origin=c]{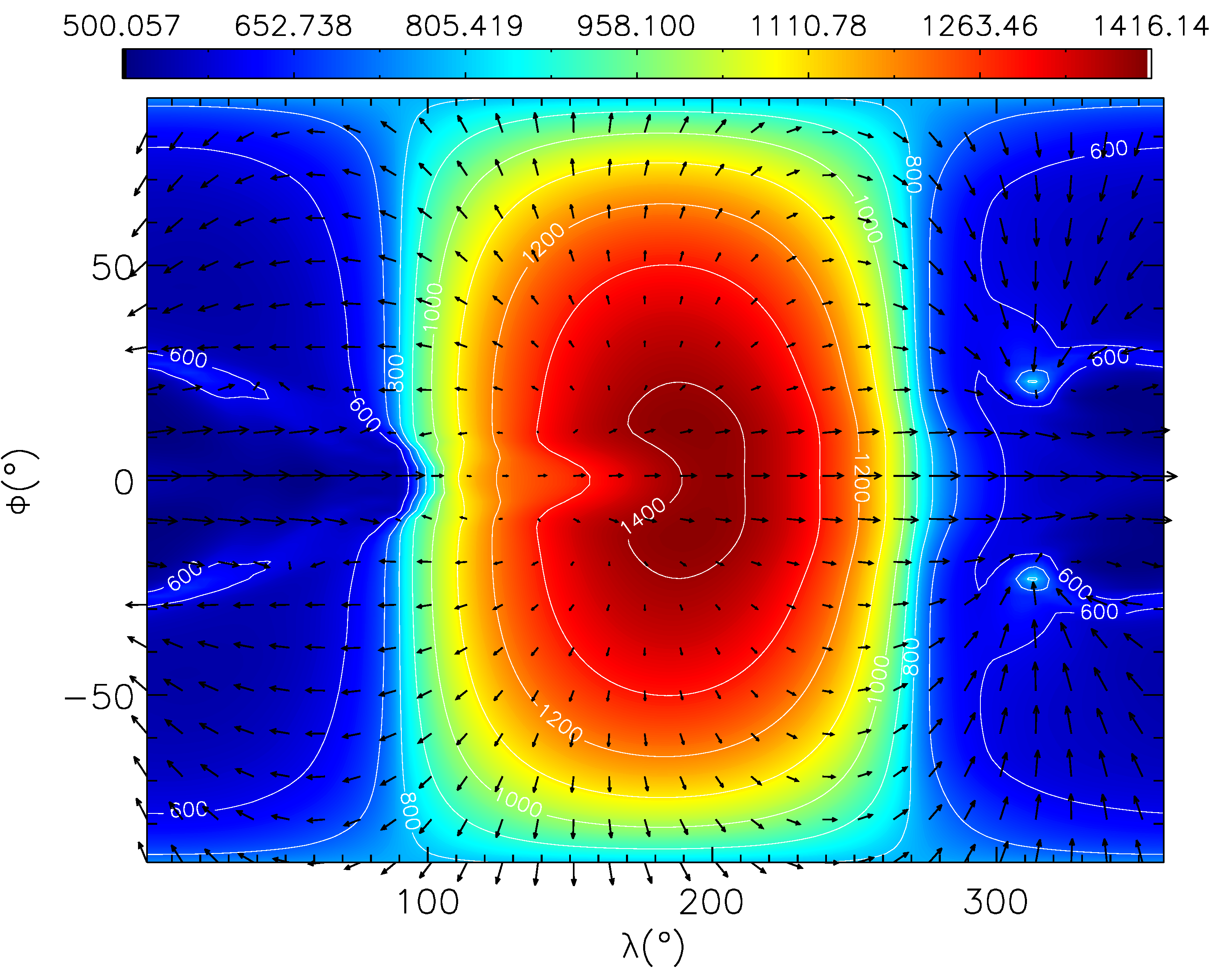}\label{std_full_213_1200_slice}}
  \subfigure[Std Prim: 21\,600\,Pa, 1\,200\,days]{\includegraphics[width=9.0cm,angle=0.0,origin=c]{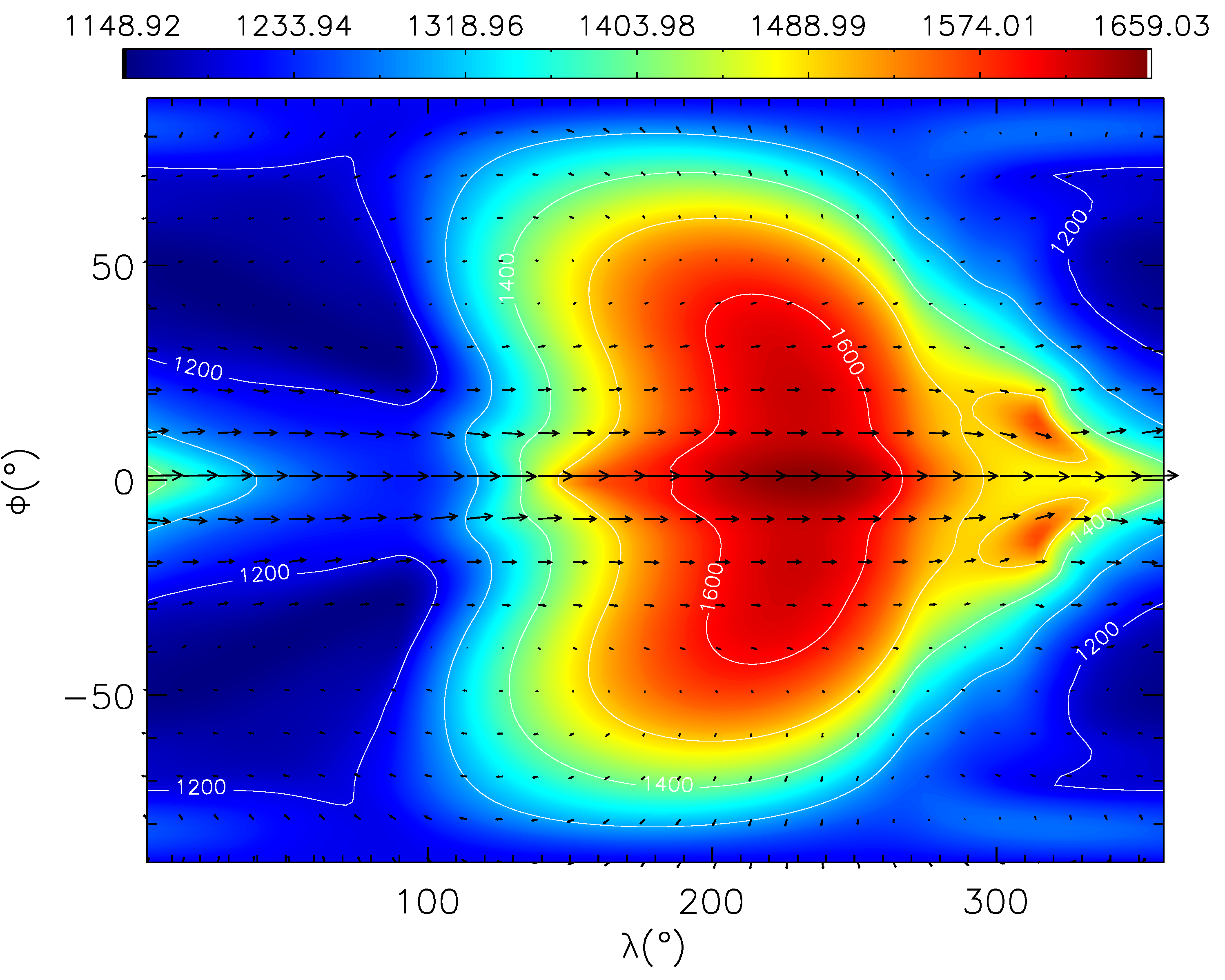}\label{std_prim_21600_1200_slice}}
  \subfigure[Std Full: 21\,600\,Pa, 1\,200\,days]{\includegraphics[width=9.0cm,angle=0.0,origin=c]{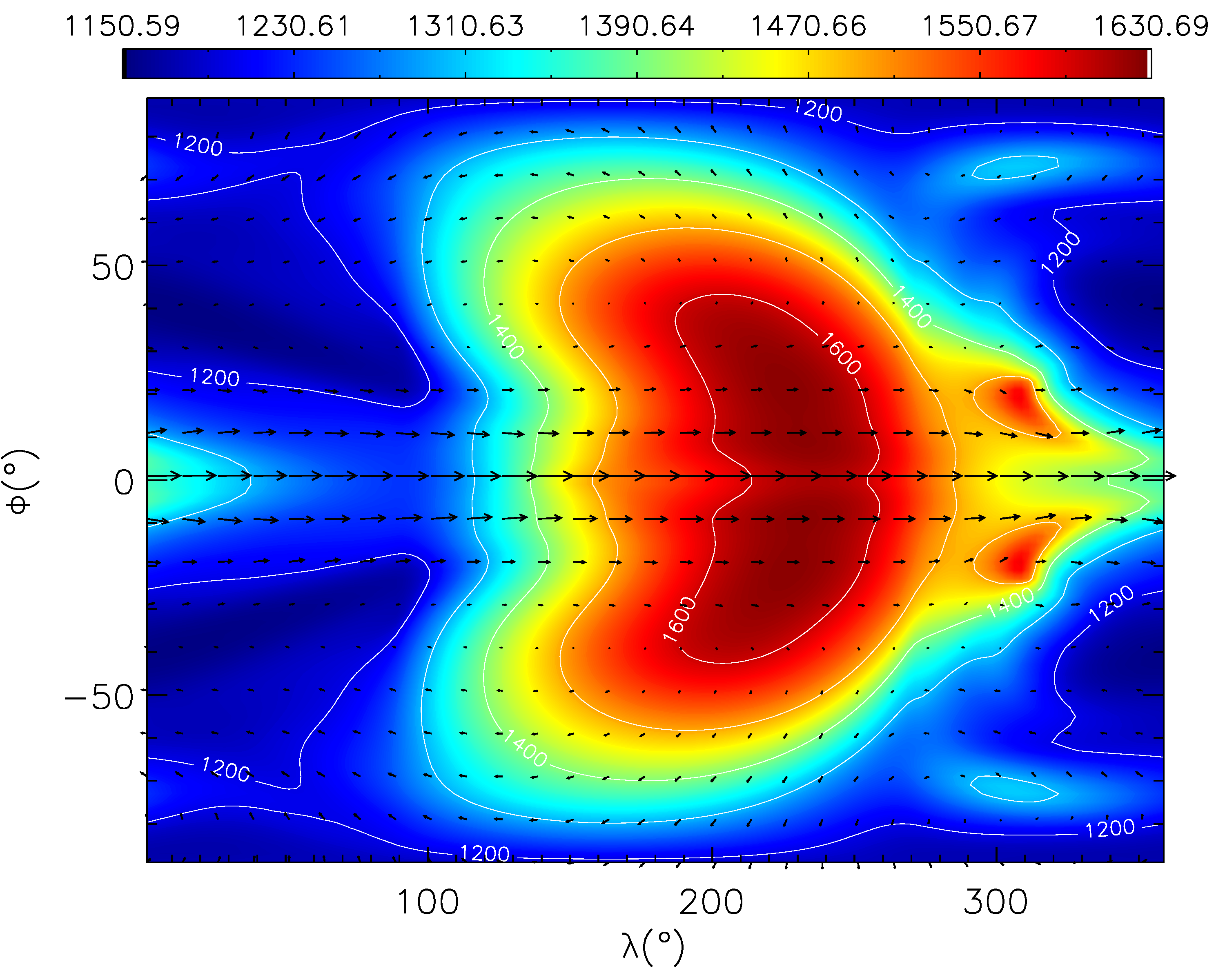}\label{std_full_21600_1200_slice}}
\end{center}
\caption{Figure showing the same informations as Figure
  \ref{slice_std_1200_bot} but at pressures of 213 \& 21\,600\,Pa
  (\textit{top} and \textit{bottom rows}, respectively) after
  1\,200\,days for the Std Prim and Std Full simulations as the
  \textit{left} and \textit{right columns}, respectively (see Table
  \ref{model_names} for explanation of simulation
  names). The maximum magnitudes of the horizontal velocities are
  $\sim$7\,000, $\sim$6\,500, $\sim$1\,000 \& 1\,100\,ms$^{-1}$ for
  the \textit{top left}, \textit{top right}, \textit{bottom left} \&
  \textit{bottom right panels}, respectively. \label{slice_std_1200_top}}
\end{figure*}

\begin{figure*}
\begin{center}
  \subfigure[Std Prim: 213\,Pa, 10\,000\,days]{\includegraphics[width=9.0cm,angle=0.0,origin=c]{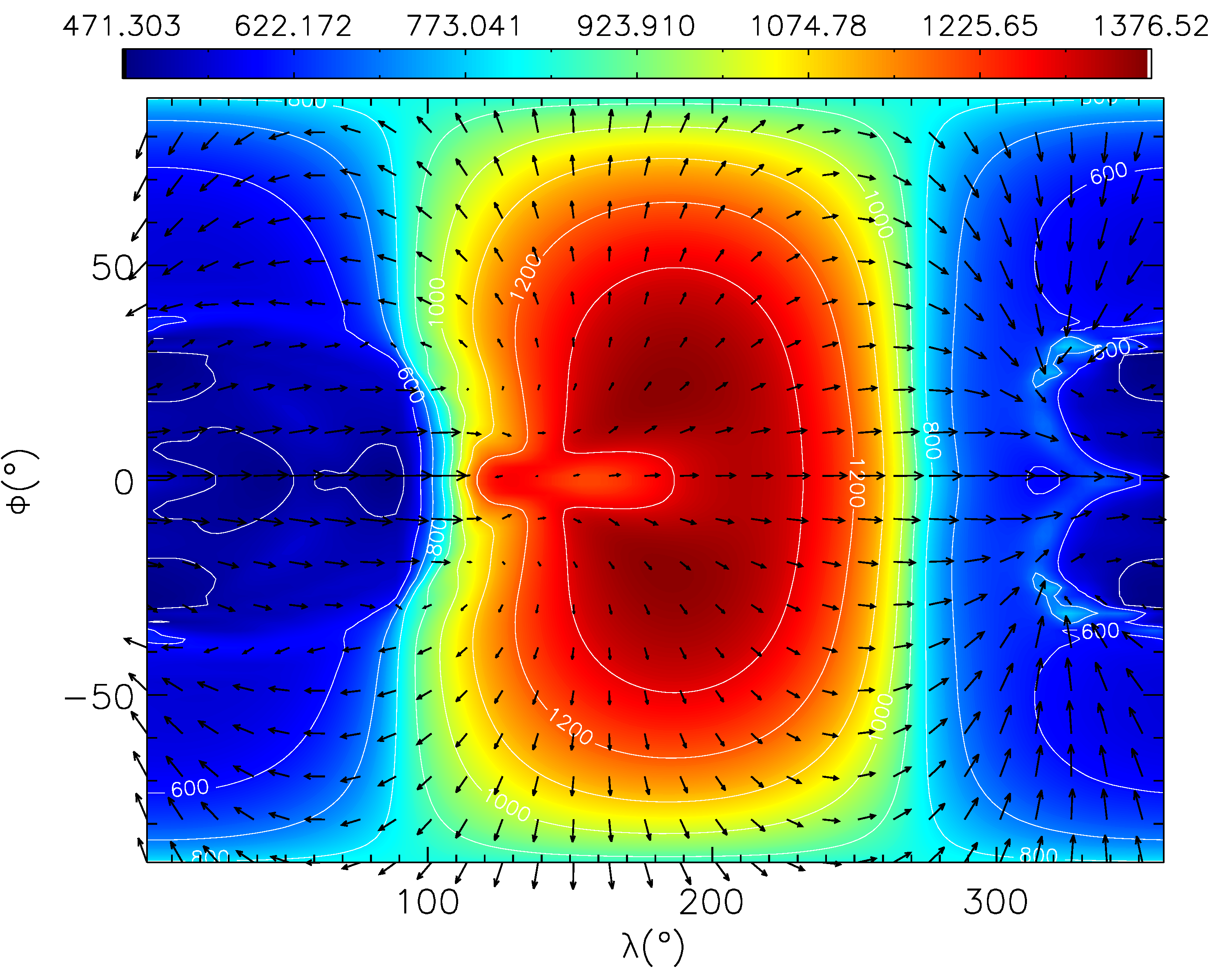}\label{std_prim_213_10000_slice}}
\subfigure[Std Full: 213\,Pa, 10\,000\,days]{\includegraphics[width=9.0cm,angle=0.0,origin=c]{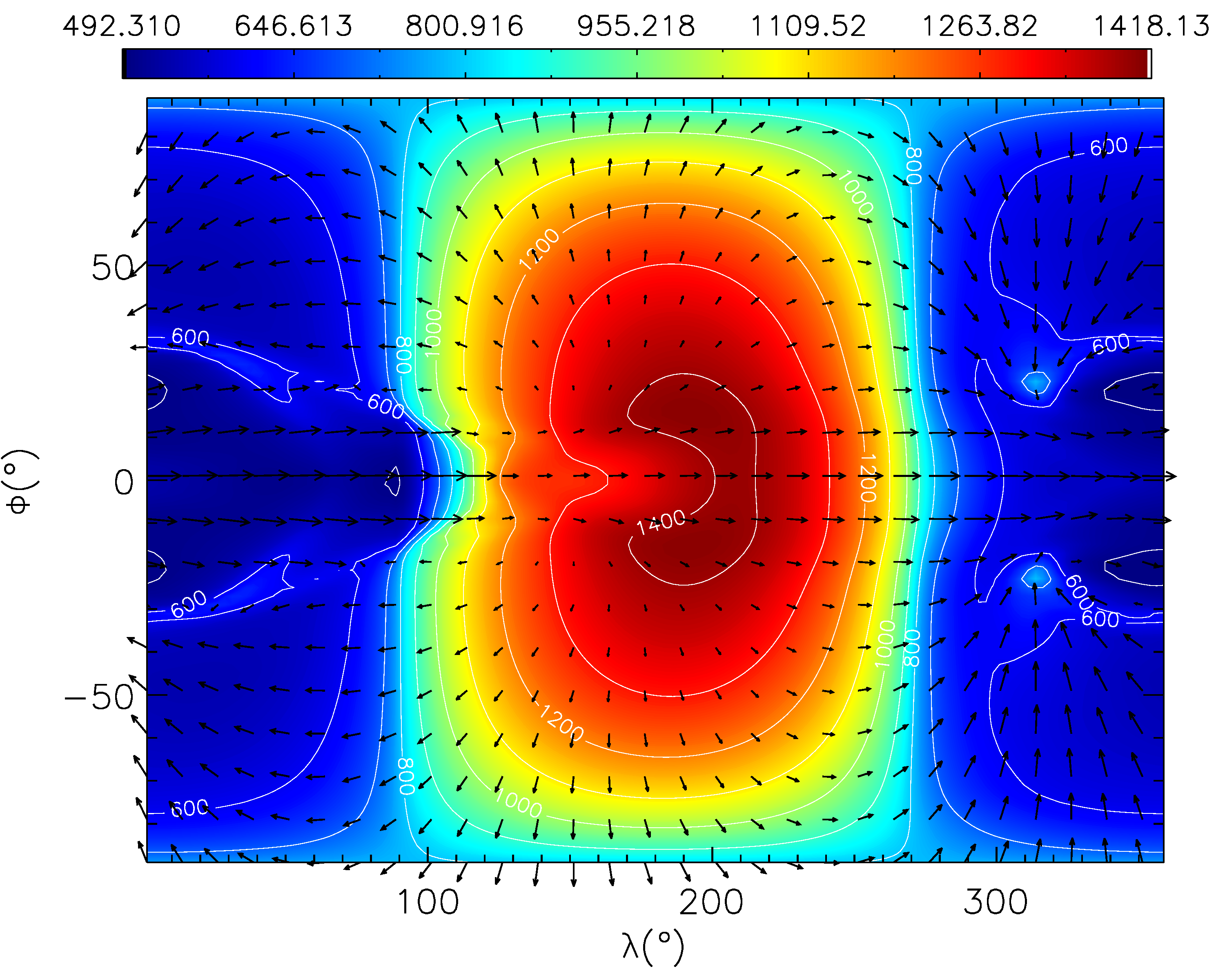}\label{std_full_213_10000_slice}}
  \subfigure[Std Prim: 21\,600\,Pa, 10\,000\,days]{\includegraphics[width=9.0cm,angle=0.0,origin=c]{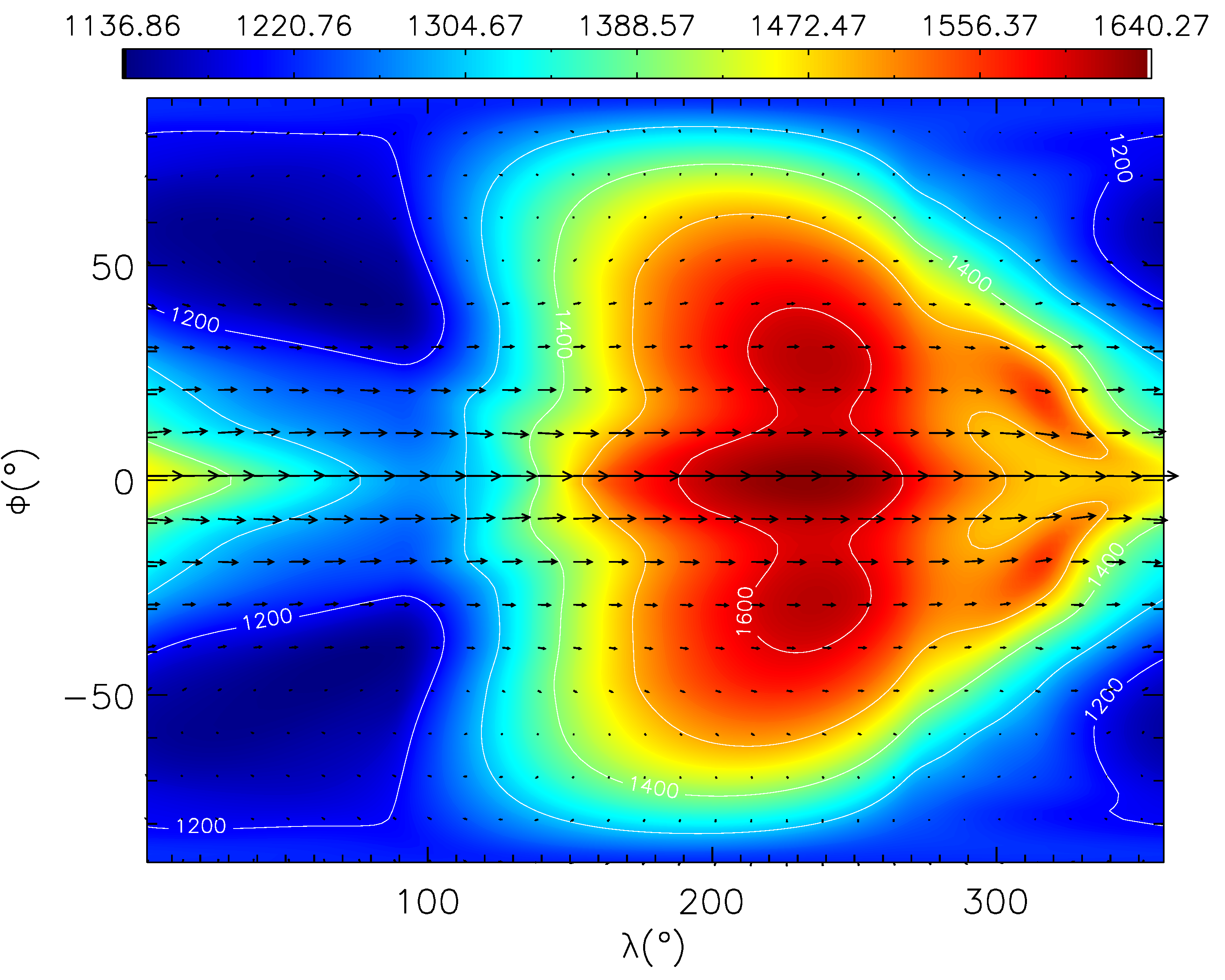}\label{std_prim_21600_10000_slice}}
  \subfigure[Std Full: 21\,600\,Pa, 10\,000\,days]{\includegraphics[width=9.0cm,angle=0.0,origin=c]{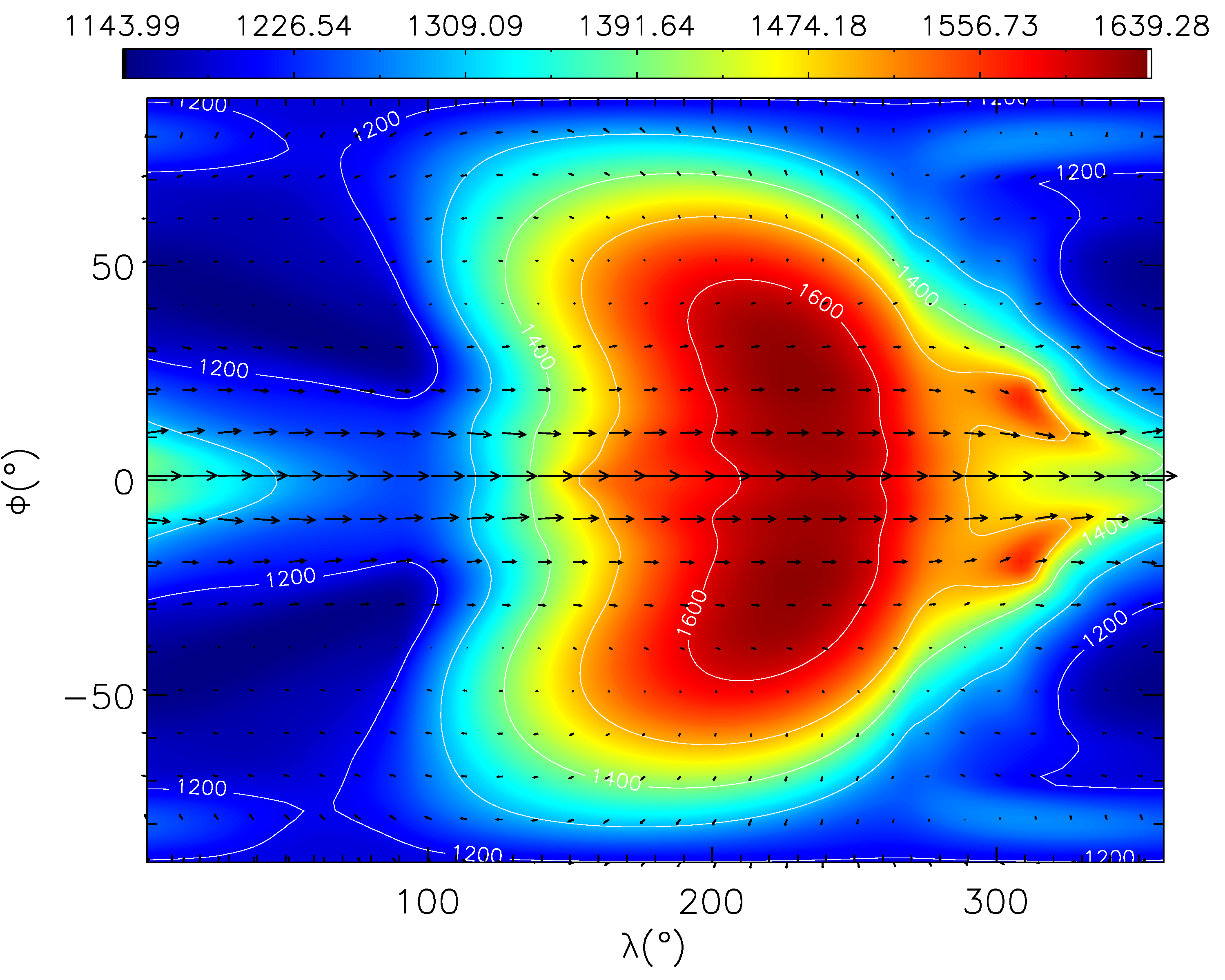}\label{std_full_21600_10000_slice}}
\end{center}
\caption{As Figure \ref{slice_std_1200_top} but after 10,000
  \,days. The maximum magnitudes of the horizontal velocities are
  $\sim$7\,000, $\sim$6\,600, $\sim$600 \& 1\,000\,ms$^{-1}$ for the
  \textit{top left}, \textit{top right}, \textit{bottom left} \&
  \textit{bottom right panels},
  respectively. \label{slice_std_10000_top}}
\end{figure*}

Figures \ref{slice_std_1200_top} and \ref{slice_std_10000_top} show
the same information as Figures \ref{slice_std_1200_bot} and
\ref{slice_std_10000_bot}, shown in Section
\ref{sub_section:jet_robust} but for the lower pressure surfaces at
213 and 21\,600\,Pa (as the \textit{top} and \textit{bottom rows},
respectively), after 1\,200 and 10\,000\,days, respectively. These
Figures reveal a broadly similar flow structure between the two times,
and regardless of the completeness of the dynamical equations. This is
to be expected given the strong radiative forcing at these pressures.

\begin{figure*}
\begin{center}
  \subfigure[Std RT: 213\,Pa, 1\,600\,days]{\includegraphics[width=9.0cm,angle=0.0,origin=c]{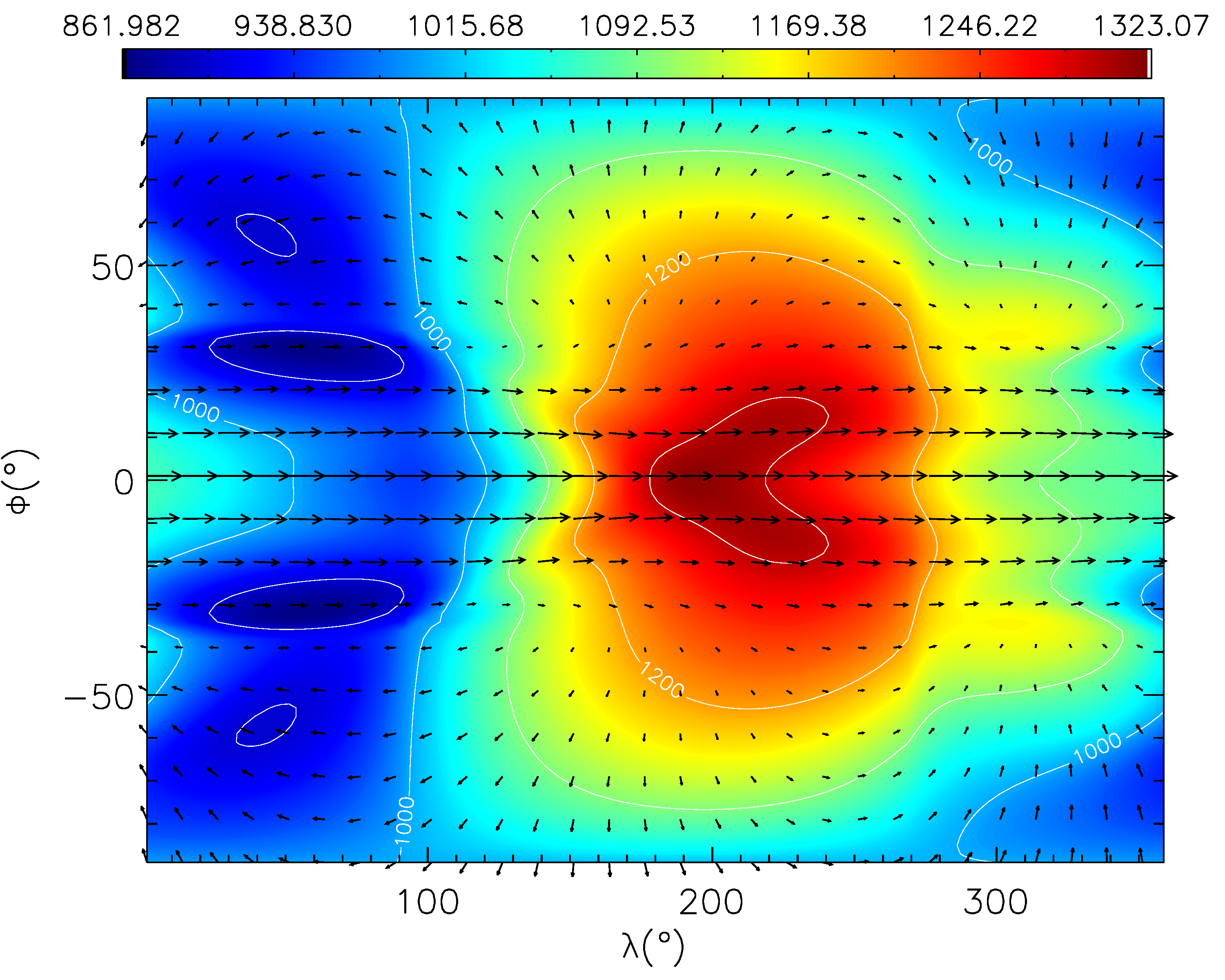}\label{rt_notiovo_full_213_1600_slice}}
  \subfigure[Std RT: 21\,600\,Pa, 1\,600\,days]{\includegraphics[width=9.0cm,angle=0.0,origin=c]{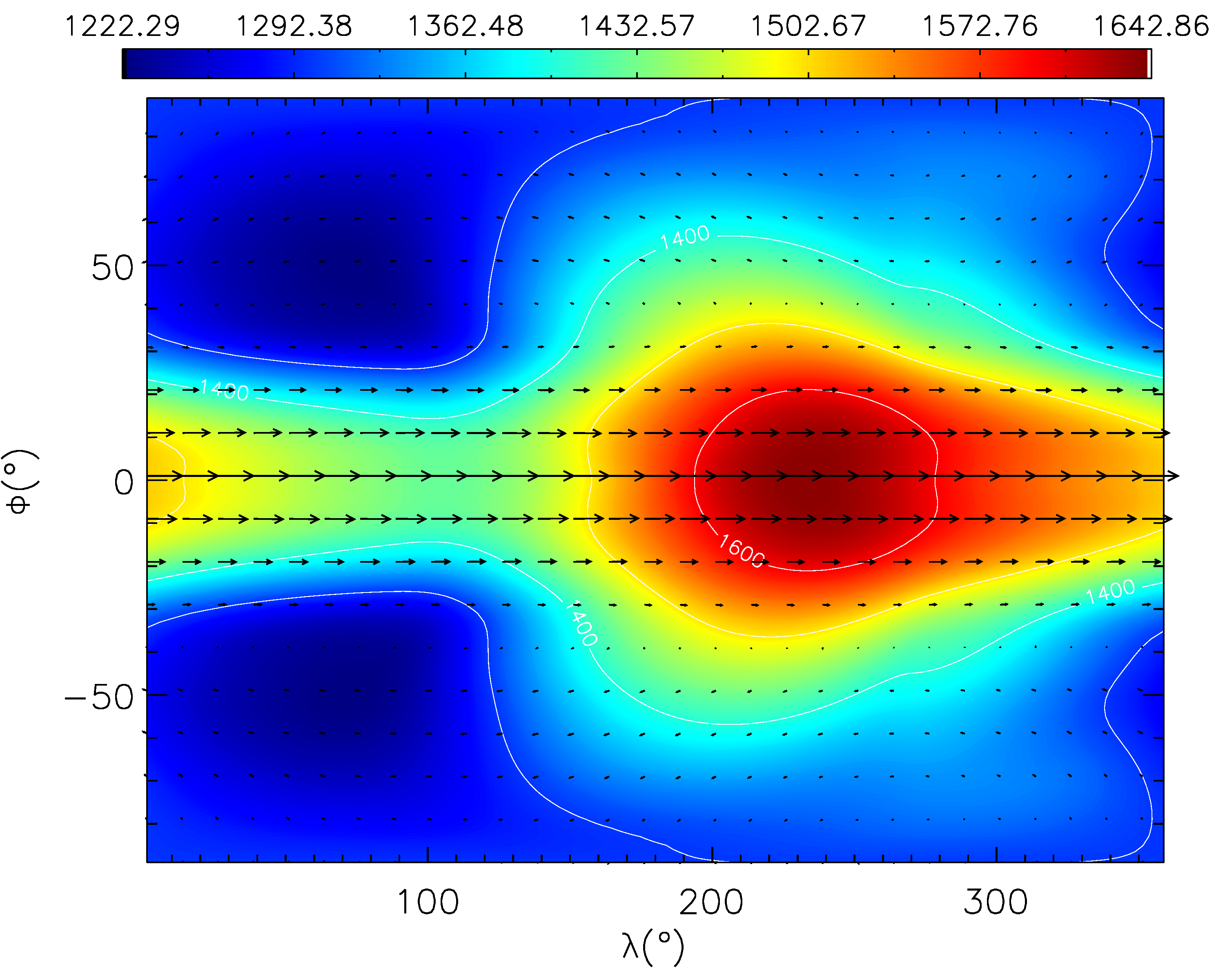}\label{rt_notiovo_full_21600_1600_slice}}
\end{center}
\caption{As for Figure \ref{slice_std_1200_top} but for the Std RT
  simulation (see Table \ref{model_names} for explanation of
  simulation names). The time sampled is close to the end of the
  simulations, 1\,600\,days. The maximum magnitudes of the horizontal
  velocities are $\sim$2\,800, \& 300\,ms$^{-1}$ for the \textit{left}
  \& \textit{right panels}, respectively. \label{slice_rt_top}}
\end{figure*}

Figure \ref{slice_rt_top} presents the results for the Std RT
simulation, matching those presented in Figure
\ref{slice_std_1200_top}, but after 1\,600\,days (the end of this
simulation). Revealing differences similar to those previously
discussed in \citet{showman_2009} and \citet{amundsen_2016}.

\begin{figure*}
\begin{center}
  \subfigure[Deep $\Delta T_{\rm eq\rightarrow pole}$ Full: 213\,Pa, 10\,000\,days]{\includegraphics[width=9.0cm,angle=0.0,origin=c]{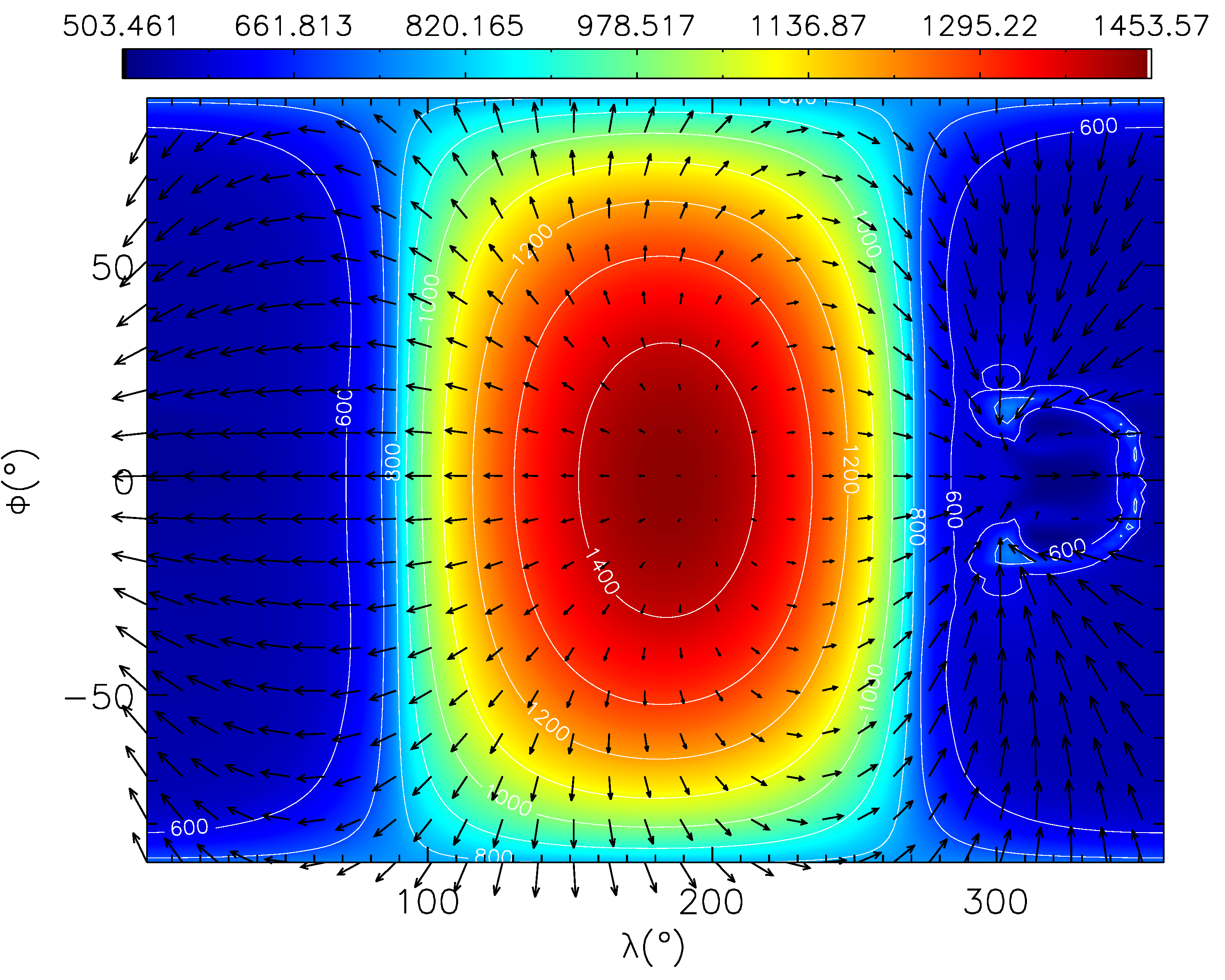}\label{teq_pole_deep_atm_full_213_10000_slice}}
  \subfigure[Deep $\Delta T_{\rm eq\rightarrow pole}$ Full: 21\,600\,Pa, 10\,000\,days]{\includegraphics[width=9.0cm,angle=0.0,origin=c]{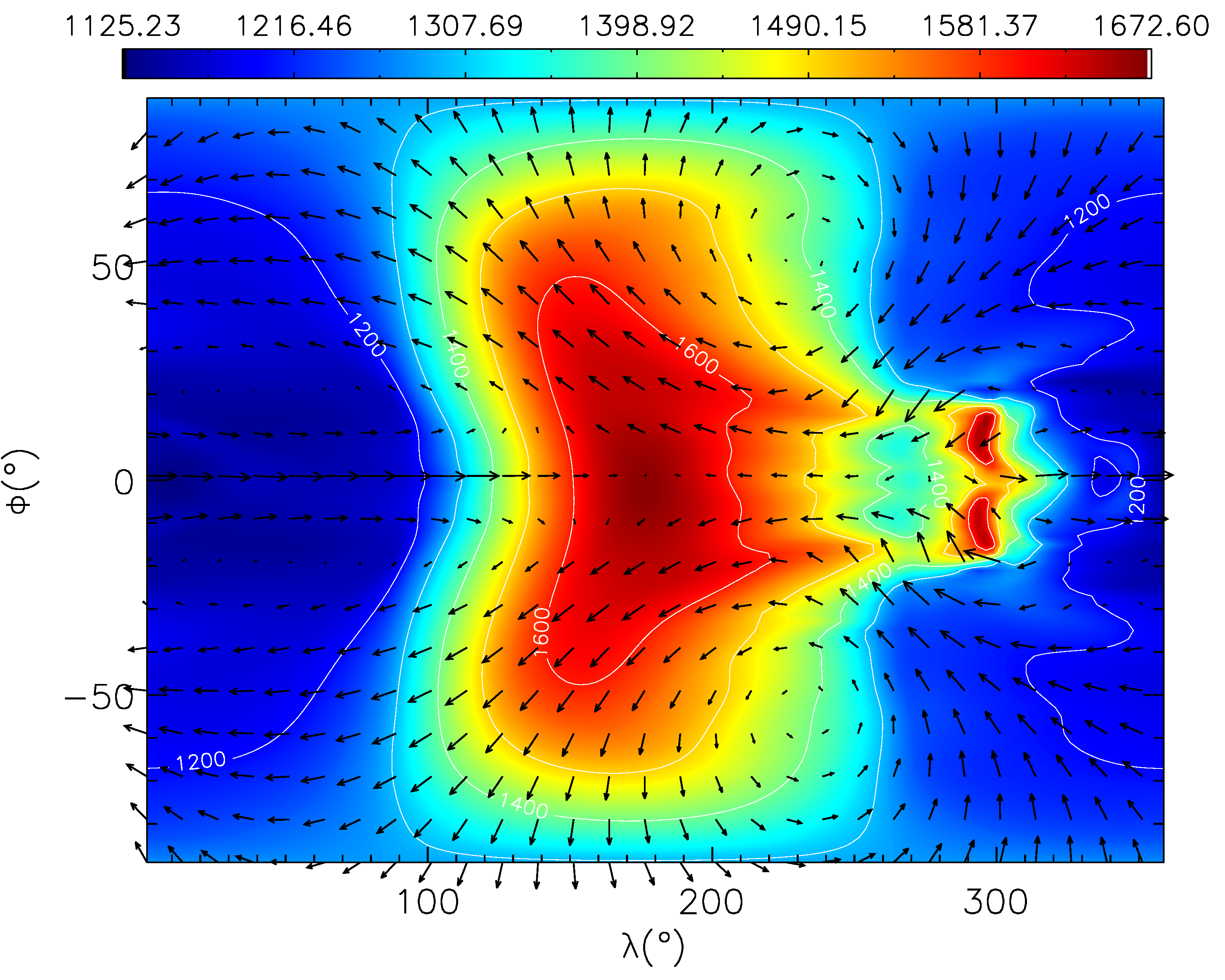}\label{teq_pole_deep_atm_full_21600_10000_slice}}
\end{center}
\caption{As for Figure \ref{slice_std_10000_top} but for the Deep
  $\Delta T_{\rm eq\rightarrow pole}$ simulation (see Table
  \ref{model_names} for explanation of simulation names). The maximum
  magnitudes of the horizontal velocities are $\sim$6\,400 \&
  2\,000\,ms$^{-1}$ for the \textit{left} \& \textit{right panels},
  respectively. \label{slice_deep_top}}
\end{figure*}

Figure \ref{slice_deep_top} presents the results for the
$\Delta T_{\rm eq\rightarrow pole}$ simulation, matching those
presented in Figure \ref{slice_std_10000_top}. These slices reveal
several differences when compared with those presented in Figure
\ref{slice_std_10000_top}. The lowest pressure slices, 213\,Pa, show
the diverging flow centred closer to $180^{\circ}$, and a sharper
day--night temperature structure in the
$\Delta T_{\rm eq\rightarrow pole}$, as opposed to the Std Full
simulation. For the 21\,600\,Pa surface, the equatorial zonal flow is
not uniformly prograde for the $\Delta T_{\rm eq\rightarrow pole}$
simulation, as also seen in the zonal and temporal mean of the zonal
wind, Figure \ref{deep_uvel_bar}.

\section{Shallow-hot Jupiter}
\label{app_section:shj}

\begin{figure*}
\begin{center}
  \subfigure[SHJ, $K_\lambda\sim0.158$, $K^\prime_\lambda \lesssim 1.2\times 10^{-12}$: 200-1\,200\,days]{\includegraphics[width=9.0cm,angle=0.0,origin=c]{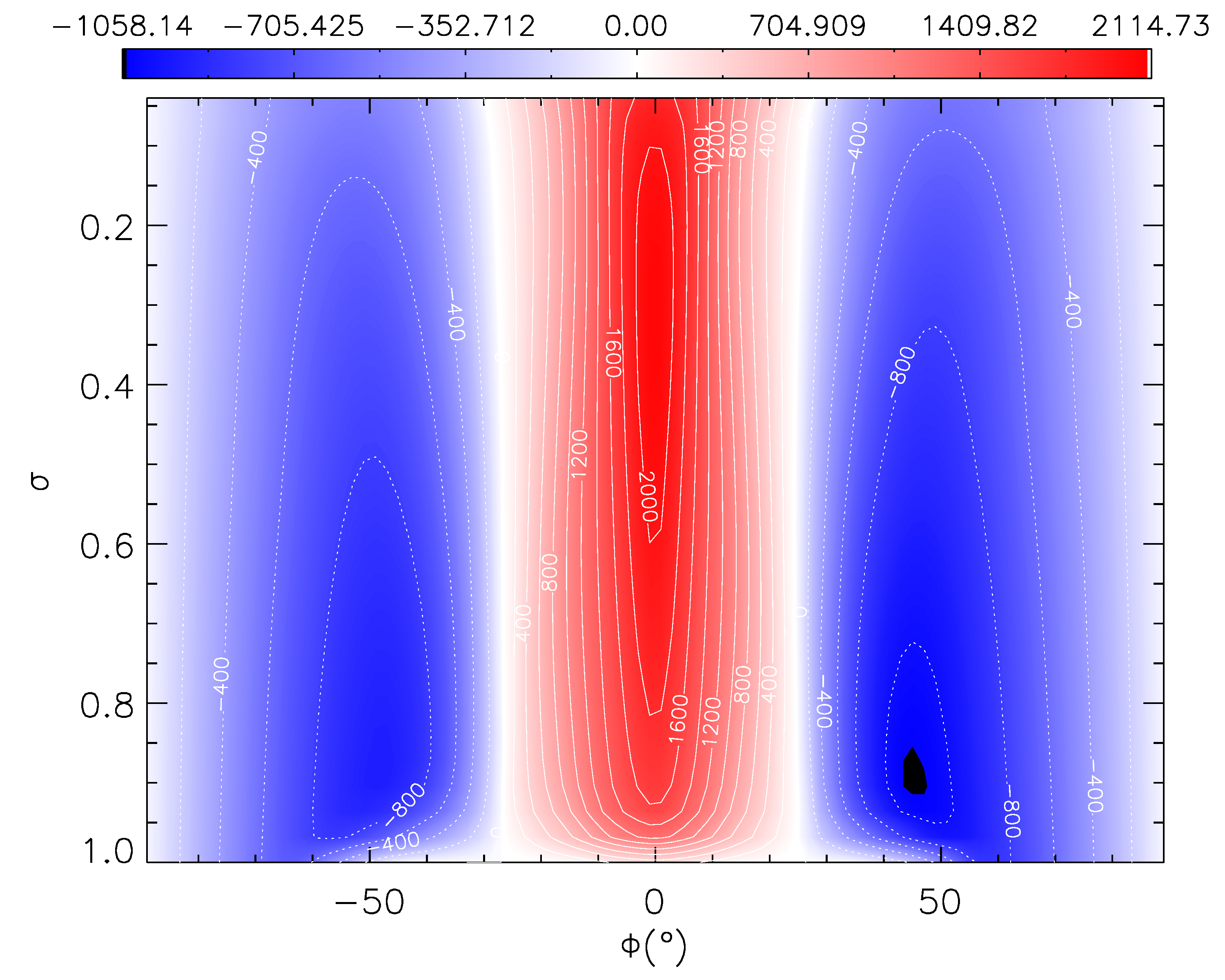}\label{shj_diff_t_1_200_1200_uvel_bar}}
  \subfigure[SHJ, $K_\lambda\sim2.6\times 10^{-3}$, $K^\prime_\lambda \lesssim 2.0\times 10^{-14}$: 200-1\,200\,days]{\includegraphics[width=9.0cm,angle=0.0,origin=c]{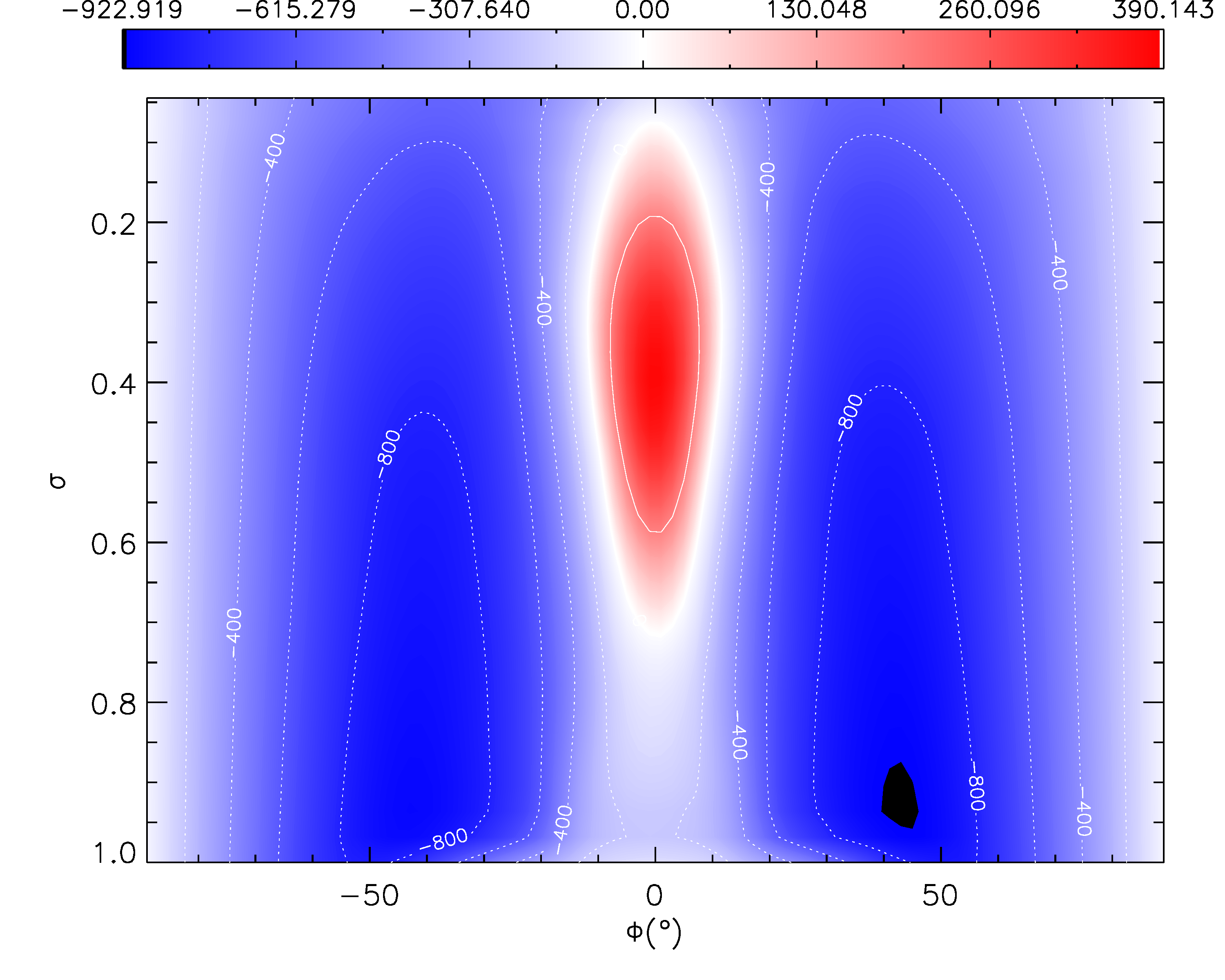}\label{shj_diff_t_96_200_1200_uvel_bar}}
\end{center}
\caption{Figure showing the zonal and temporal (200--1\,200 days) mean
  of the zonal wind (ms$^{-1}$) for the SHJ simulations in the same
  format as Figure \ref{std_uvel_bar}, but against a vertical axis of
  $\sigma$ (where $\sigma=p/p_0$ and $p_0=10^5$\,Pa in this case). The
  \textit{left} and \textit{right panels} show simulations using
  larger, and smaller diffusion coefficients, respectively (see text
  and Section \ref{sub_section:model_damp} for
  explanation). \label{shj_uvel_bar}}
\end{figure*}

\citet{mayne_2014} performed the Shallow-hot Jupiter (SHJ) test of
\citet{menou_2009}, and compared to the results of \citet{menou_2009}
and \citet{heng_2011}. Our simulations largely matched previous works
except that the super rotating equatorial jet did not intersect the
bottom, or inner, boundary. We assumed this to be caused by the
difference in using a pressure-based \citep[as is the case
for][]{menou_2009,heng_2011}, or height-based model (the UM). However,
the SHJ simulation has now been performed by \citet{mendonca_2016}
using a height-based approach who were able to match the previous
simulations, and obtain a jet intersecting the high pressure
boundary. Therefore, we have revisited the SHJ test, and investigated
this discrepancy further. Figure \ref{shj_uvel_bar} shows the zonally
and temporally averaged, from 200-1\,200\,days, zonal wind, in the
same format as Figure \ref{std_uvel_bar} but for the SHJ case
\citep[see][for details of the setup etc.]{mayne_2014}. The
\textit{left} and \textit{right panels} of Figure \ref{shj_uvel_bar}
show simulations applying `stronger' ($K_\lambda\sim 0.158$,
$K^\prime_\lambda \lesssim 1.2\times 10^{-12}$) and `weaker'
($K_\lambda\sim2.6\times 10^{-03}$,
$K^\prime_\lambda \lesssim 2.0\times 10^{-14}$) diffusion,
respectively. See Section \ref{sub_section:model_damp} for explanation
of how the diffusion coefficient is set ($K_\lambda$) and subsequently
applied ($K^\prime_\lambda$). Figure \ref{shj_uvel_bar} shows that the
super rotating equatorial jet intersects the bottom boundary when
using the larger diffusion coefficient, and therefore `stronger'
diffusion. The jet speeds are significantly slower when applying
`weaker' diffusion. This is counter--intuitive, and demonstrates
clearly that not only the maximum wind/jet velocity, but also the
shape of the atmospheric flow can be altered by the choice of
diffusion settings \citep[see also discussion
in][]{heng_2011,li_2010b}. We are currently exploring this effect
further, which is likely caused by the diffusion altering the eddy
transport in the atmosphere.

\section{Derivation of the eddy mean interaction equation}
\label{app_section:eddy_mean_derive}

Starting with the `full' or `deep' equations as defined in
\citep[see][]{mayne_2014}. The derivation for the eddy--mean
interaction equation, in height coordinates, is as follows: the zonal
momentum equation is
\begin{align}
\frac{Du}{Dt}&= \nonumber\\
&-\frac{uw}{r}+\frac{uv}{r}\tan\phi+2\Omega v\sin\phi-2\Omega w\cos\phi \nonumber\\
&-\frac{1}{r\cos\phi}\frac{1}{\rho}\frac{\partial p}{\partial\lambda}+G_{\lambda}\label{momentum_full}\\
\frac{Du}{Dt}&=\frac{\partial u}{\partial t}+\textbf{u}.\nabla{u}, \label{momentum_full_comp}
\end{align}
where $\frac{D}{Dt}$ is the material derivative. The mass continuity
equation is
\begin{align}
\frac{\partial\rho}{\partial t}+\nabla.\left(\rho\textbf{u}\right)=0. \label{mass}
\end{align}
Multiplying the momentum equation (Eqn \ref{momentum_full}) by $\rho$,
and adding to the mass continuity equation (Eqn \ref{mass}) multiplied
by $u$, to the right hand side (which is equal to zero), gives
\begin{align}
\rho\frac{Du}{Dt}&=\rho\frac{\partial u}{\partial t}+\rho\textbf{u}.\nabla{u}+u\left[\frac{\partial\rho}{\partial t}+\nabla.\left(\rho\textbf{u}\right)\right]. \label{stage_one}
\end{align}
Using the product rule,
\begin{align}
\frac{\partial\left(\rho u\right)}{\partial t}&=\rho\frac{\partial u}{\partial t}+u\frac{\partial \rho}{\partial t}\label{product}\\\nonumber
\nabla.\left(\rho u\textbf{u}\right)&=\rho\textbf{u}.\nabla u+u\nabla.\left(\rho\textbf{u}\right).
\end{align}
Therefore, Eqn \ref{stage_one} becomes,
\begin{align}
\rho\frac{Du}{Dt}&=\frac{\partial\left(\rho u\right)}{\partial t}+\nabla.\left(\rho u\textbf{u}\right).\label{stage_two}
\end{align}
As, 
\begin{align}
&\nabla.\left(X\textbf{Y}\right)=\nonumber \\
&\frac{1}{r\cos\phi}\frac{\partial\left(XY_\lambda\right)}{\partial \lambda}+\frac{1}{r\cos\phi}\frac{\partial\left(XY_\phi\cos\phi\right)}{\partial\phi}+\frac{1}{r^2}\frac{\partial\left(r^2XY_r\right)}{\partial r}, \label{div_full}
\end{align}
Eqn \ref{stage_two} can be expanded as,
\begin{align}
\rho\frac{Du}{Dt}&=\nonumber \\
&\frac{\partial\left(\rho u\right)}{\partial t}+\frac{1}{r\cos\phi}\frac{\partial\left(\rho u^2\right)}{\partial \lambda}\nonumber\\
&+\frac{1}{r\cos\phi}\frac{\partial\left(\rho u v\cos\phi\right)}{\partial\phi}+\frac{1}{r^2}\frac{\partial\left(r^2\rho u w\right)}{\partial r} \label{stage_three}
\end{align}
i.e. $X=\rho u$, and $\textbf{Y}=\textbf{u}=\left(u,v,w\right)$.
Using the momentum equation (Eqn \ref{momentum_full}),
\begin{align}
\rho\frac{Du}{Dt}&= \nonumber\\
&-\frac{\rho uw}{r}+\frac{\rho uv}{r}\tan\phi+2\Omega\rho v\sin\phi-2\Omega\rho w\cos\phi \nonumber\\
&-\frac{1}{r\cos\phi}\frac{\partial p}{\partial\lambda}+\rho G_{\lambda}, \label{stage_four}
\end{align}
i.e. multiply Eqn \ref{momentum_full} by $\rho$.  Therefore, combining
Eqns \ref{stage_three} and \ref{stage_four}, to eliminate
$\rho\frac{Du}{Dt}$ yields,
\begin{align}
&\frac{\partial\left(\rho u\right)}{\partial t}+\frac{1}{r\cos\phi}\frac{\partial\left(\rho u^2\right)}{\partial \lambda}+\nonumber\\
&\frac{1}{r\cos\phi}\frac{\partial\left(\rho u v\cos\phi\right)}{\partial\phi}+\frac{1}{r^2}\frac{\partial\left(r^2\rho u w\right)}{\partial r}= \nonumber\\
&-\frac{\rho uw}{r}+\frac{\rho uv}{r}\tan\phi+2\Omega\rho v\sin\phi-2\Omega\rho w\cos\phi \nonumber\\
&-\frac{1}{r\cos\phi}\frac{\partial p}{\partial\lambda}+\rho G_{\lambda}. \label{stage_five}
\end{align}
Rearranging by collecting all terms involving $v$ and $w$ on one side yields,
\begin{align}
&\frac{\partial\left(\rho u\right)}{\partial t}+\frac{1}{r\cos\phi}\frac{\partial\left(\rho u^2\right)}{\partial \lambda}+\frac{1}{r\cos\phi}\frac{\partial p}{\partial\lambda}-\rho G_{\lambda}= \nonumber \\
&-\frac{1}{r\cos\phi}\frac{\partial\left(\rho u v\cos\phi\right)}{\partial\phi}+\frac{\rho uv}{r}\tan\phi+2\Omega\rho v\sin\phi \nonumber\\
&-\frac{1}{r^2}\frac{\partial\left(r^2\rho u w\right)}{\partial r}-\frac{\rho uw}{r}-2\Omega\rho w\cos\phi. \label{before_rearrange}
\end{align}
Now we combine two of the $uw$ and $uv$ terms. Starting with the $uv$, expand the $\tan\phi$,
\begin{align}
&-\frac{1}{r\cos\phi}\frac{\partial\left(\rho u v\cos\phi\right)}{\partial\phi}+\frac{\rho uv}{r}\tan\phi=\nonumber \\
&-\frac{1}{r\cos\phi}\frac{\partial\left(\rho u v \cos\phi\right)}{\partial\phi}+\frac{\rho uv\sin\phi}{r\cos\phi}, \label{uv_one}
\end{align}
rearranging and then multiplying the fractions by
$\frac{\cos\phi}{\cos\phi}$, we can transform the $uw$ terms from Eqn
\ref{uv_one} to
\begin{align}
&-\frac{1}{r\cos\phi}\frac{\partial\left(\rho u v\cos\phi\right)}{\partial\phi}+\frac{\rho uv\sin\phi}{r\cos\phi}=\nonumber \\
&-\frac{1}{r\cos\phi}\frac{\partial\left(\rho u v\cos\phi\right)}{\partial\phi}-\frac{\rho uv}{r\cos\phi}\frac{\partial\left(\cos\phi\right)}{\partial\phi}= \nonumber\\
&-\frac{\cos\phi}{r\cos^2\phi}\frac{\partial\left(\rho u v\cos\phi\right)}{\partial\phi}-\frac{\rho uv\cos\phi}{r\cos^2\phi}\frac{\partial\left(\cos\phi\right)}{\partial\phi}.
\end{align}
This can then be compressed using the inverse product rule,
i.e. $\partial\left(\left(\rho
    uv\cos\phi\right)\left(\cos\phi\right)\right)$ to yield,
\begin{align}
&-\frac{\cos\phi}{r\cos^2\phi}\frac{\partial\left(\rho u v\cos\phi\right)}{\partial\phi}-\frac{\rho uv\cos\phi}{r\cos^2\phi}\frac{\partial\left(\cos\phi\right)}{\partial\phi}= \nonumber\\
&-\frac{1}{r\cos^2\phi}\frac{\partial\left(\rho uv\cos^2\phi\right)}{\partial\phi}. \label{compress_uv}
\end{align}
For the $uw$ terms from Eqn \ref{before_rearrange}, multiplying the
first by $\frac{r}{r}$ and the second by
$\frac{r^2}{r^2}\frac{\partial r}{\partial r}$ yields
\begin{align}
 &-\frac{1}{r^2}\frac{\partial\left(r^2\rho u w\right)}{\partial r}-\frac{\rho uw}{r}=\nonumber \\
 &-\frac{r}{rr^2}\frac{\partial\left(r^2\rho u w\right)}{\partial r}-\frac{r^2\rho uw}{r^2r}\frac{\partial r}{\partial r},
\end{align}
which again can be compressed using the inverse product rule,
$\partial\left(\left(r^2\rho wu\right)\left(r\right)\right)$,
\begin{align}
&-\frac{r}{rr^2}\frac{\partial\left(r^2\rho u w\right)}{\partial r}-\frac{r^2\rho uw}{r^2r}\frac{\partial r}{\partial r}=\nonumber \\
&-\frac{1}{r^3}\frac{\partial\left(\rho uwr^3\right)}{\partial r}. \label{compress_uw}
\end{align}
Now, Eqns \ref{compress_uv} and \ref{compress_uw} can be substituted
into Eqn \ref{before_rearrange} to yield,
\begin{align}
&\frac{\partial\left(\rho u\right)}{\partial t}+\frac{1}{r\cos\phi}\frac{\partial\left(\rho u^2\right)}{\partial \lambda}+\frac{1}{r\cos\phi}\frac{\partial p}{\partial\lambda}-\rho G_{\lambda}= \nonumber \\
&-\frac{1}{r\cos^2\phi}\frac{\partial\left(\rho uv\cos^2\phi\right)}{\partial\phi}+2\Omega\rho v\sin\phi \nonumber\\
&-\frac{1}{r^3}\frac{\partial\left(\rho uwr^3\right)}{\partial r}-2\Omega\rho w\cos\phi. \label{before_zon_avg}
\end{align}
We can now apply a zonal average to this equation, meaning that
$\frac{\partial}{\partial\lambda}$ terms become
zero. \emph{Critically, this only applies if a zonal average is
  performed along surfaces of equal $r$, as the derivatives are
  performed at constant $r$} (see discussion in Section
\ref{sub_section:jet_pumping}),
\begin{align}
&\frac{\partial\left(\overline{\rho u}\right)}{\partial t}-\overline{\rho G_{\lambda}}= \nonumber \\
&-\frac{1}{r\cos^2\phi}\frac{\partial\left(\overline{\rho uv}\cos^2\phi\right)}{\partial\phi}+2\Omega\overline{\rho v}\sin\phi \nonumber\\
&-\frac{1}{r^3}\frac{\partial\left(\overline{\rho uw}r^3\right)}{\partial r}-2\Omega\overline{\rho w}\cos\phi.
\end{align}
Then the wind terms involving the zonal wind, $u$ only, can be
decomposed into a mean and perturbation using
$u=\overline{u}+u^\prime$, $\overline{\rho
  u}=\overline{\rho}\,\overline{u}+\overline{\rho^\prime u^\prime}$,
$\overline{\rho uv}=\overline{\rho
  v}\,\overline{u}+\overline{\left(\rho v\right)^\prime u^\prime}$ and
$\overline{\rho uw}=\overline{\rho
  w}\,\overline{u}+\overline{\left(\rho w\right)^\prime u^\prime}$,
yielding,
\begin{align}
&\frac{\partial\left(\overline{\rho}\, \overline{u}\right)}{\partial t}+\frac{\partial\left(\overline{\rho^\prime u^\prime}\right)}{\partial t}-\overline{\rho G_{\lambda}}= \nonumber \\
&-\frac{1}{r\cos^2\phi}\frac{\partial\left(\overline{\rho v}\,\overline{u}\cos^2\phi\right)}{\partial\phi}-\frac{1}{r\cos^2\phi}\frac{\partial\left(\overline{\left(\rho v\right)^\prime u^\prime}\cos^2\phi\right)}{\partial\phi}+\nonumber\\
&2\Omega\overline{\rho v}\sin\phi \nonumber\\
&-\frac{1}{r^3}\frac{\partial\left(\overline{\rho w}\,\overline{u}r^3\right)}{\partial r}-\frac{1}{r^3}\frac{\partial\left(\overline{\left(\rho w\right)^\prime u^\prime}r^3\right)}{\partial r}-2\Omega\overline{\rho w}\cos\phi.
\end{align}
Finally, rearranging this equation so that the mean terms are on the
left and the perturbation terms on the right yields 
\begin{align}
&\frac{\partial\left(\overline{\rho}\, \overline{u}\right)}{\partial t}+\frac{1}{r\cos^2\phi}\frac{\partial\left(\overline{\rho v}\,\overline{u}\cos^2\phi\right)}{\partial\phi}+\frac{1}{r^3}\frac{\partial\left(\overline{\rho w}\,\overline{u}r^3\right)}{\partial r}\nonumber \\
&-2\Omega\overline{\rho v}\sin\phi+2\Omega\overline{\rho w}\cos\phi-\overline{\rho G_{\lambda}}= \nonumber \\
&-\frac{\partial\left(\overline{\rho^\prime u^\prime}\right)}{\partial t}-\frac{1}{r\cos^2\phi}\frac{\partial\left(\overline{\left(\rho v\right)^\prime u^\prime}\cos^2\phi\right)}{\partial\phi}-\frac{1}{r^3}\frac{\partial\left(\overline{\left(\rho w\right)^\prime u^\prime}r^3\right)}{\partial r}. \label{hardiman_eqn}
\end{align}
which matches that given as Eqn (6) in \citet{hardiman_2010}.

Next in the `shallow' or `primitive' cases \citep[see][]{mayne_2014}.

The zonal momentum equation is
\begin{align}
\frac{Du}{Dt}&=\frac{uv}{R_{\rm p}}\tan\phi+2\Omega v\sin\phi-\frac{1}{R_{\rm p}\cos\phi}\frac{1}{\rho}\frac{\partial P}{\partial\lambda}+G_{\lambda}\label{momentum_shal}
\end{align}
where $r\rightarrow R_{\rm p}$, $\frac{uw}{r}\rightarrow 0$ and
$w\cos\phi\rightarrow 0$. Similarly, Eqn \ref{div_full} becomes,
\begin{align}
&\nabla.\left(X\textbf{Y}\right)=\nonumber \\
&\frac{1}{R_{\rm p}\cos\phi}\frac{\partial\left(XY_\lambda\right)}{\partial \lambda}+\frac{1}{R_{\rm p}\cos\phi}\frac{\partial\left(XY_\phi\cos\phi\right)}{\partial\phi}+\frac{1}{R_{\rm p}^2}\frac{\partial\left(R_{\rm p}^2XY_r\right)}{\partial z}, \label{div_shal}
\end{align}
where $r\rightarrow z$ in the derivative.

Therefore, the key steps of combining the $uv$ and $uw$ terms, using
Eqns \ref{compress_uv} and \ref{compress_uw} are simplified
greatly. For the $uw$ there is only one term (as $\frac{uw}{r}$ has
been removed), and for the $uv$ terms the process is exactly the same,
just replacing $r$ with $R_{\rm p}$. Therefore, this yields,
\begin{align}
&\frac{\partial\left(\overline{\rho}\, \overline{u}\right)}{\partial t}+\frac{1}{R_{\rm p}\cos^2\phi}\frac{\partial\left(\overline{\rho v}\,\overline{u}\cos^2\phi\right)}{\partial\phi}+\frac{1}{R_{\rm p}^2}\frac{\partial\left(\overline{\rho w}\,\overline{u}R_{\rm p}^2\right)}{\partial z}\nonumber \\
&-2\Omega\overline{\rho v}\sin\phi= \nonumber \\
&-\frac{\partial\left(\overline{\rho^\prime u^\prime}\right)}{\partial t}-\frac{1}{R_{\rm p}\cos^2\phi}\frac{\partial\left(\overline{\left(\rho v\right)^\prime u^\prime}\cos^2\phi\right)}{\partial\phi}-\frac{1}{R_{\rm p}^2}\frac{\partial\left(\overline{\left(\rho w\right)^\prime u^\prime}R_{\rm p}^2\right)}{\partial z}\nonumber \\
&+\overline{\rho G_{\lambda}}, \label{hardiman_deep_eqn}
\end{align}
which can clearly be simplified further, but we leave in this form to
provide comparison with Eqn \ref{hardiman_eqn}.

\section{Eddy mean interaction terms}
\label{app_section:eddy_terms}

\begin{figure}
\begin{center}
\subfigure[Std Full:$\Sigma_{\lambda}-\left(\frac{\left[ \overline{(\rho v)^{\prime}u^{\prime}}\cos^2\phi\right]_{, \phi}}{r\cos^2\phi}\right)$, 10\,000\,days]{\includegraphics[width=8.6cm,angle=0.0,origin=c]{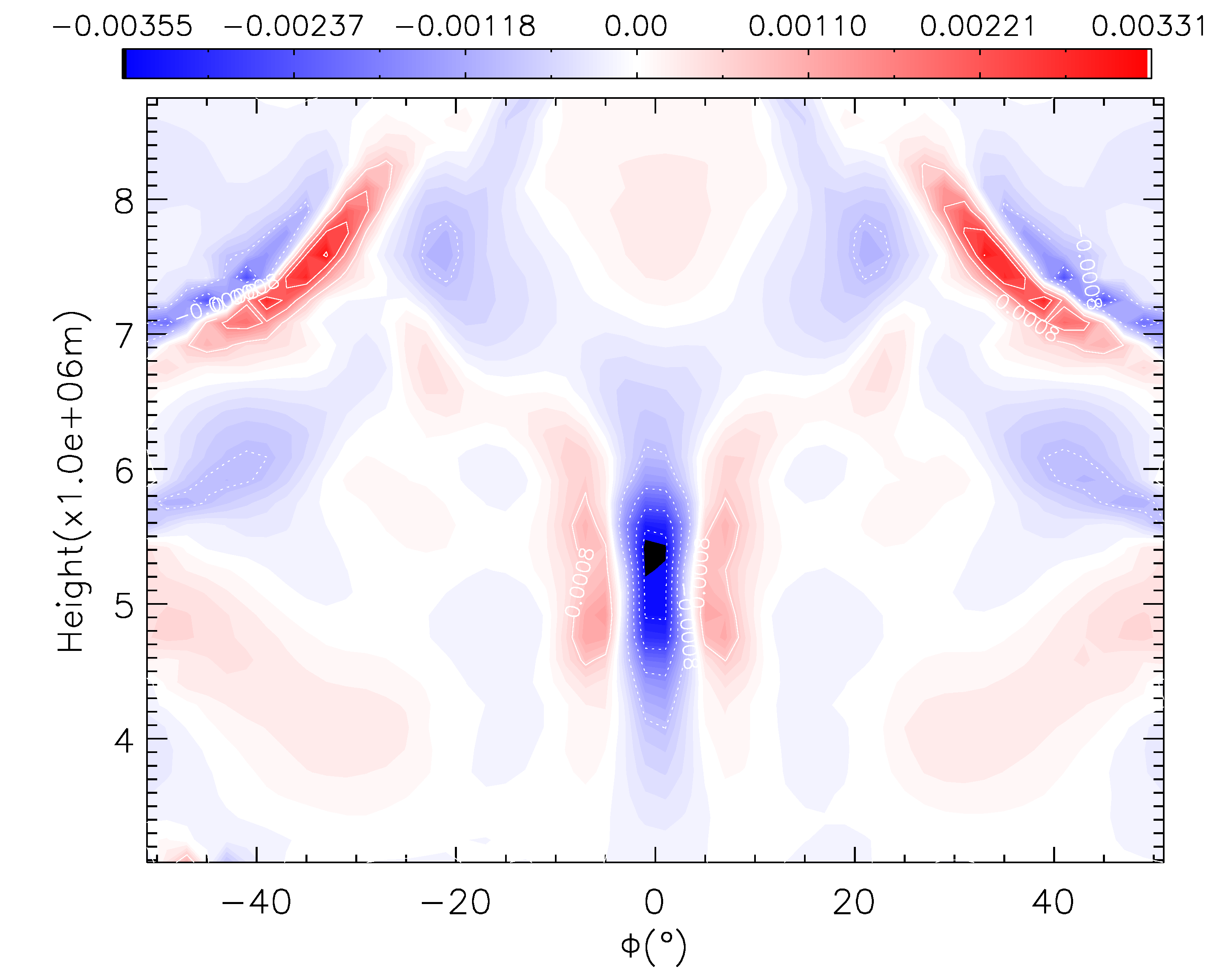}\label{std_full_meri_eddy_mmtm_slice_xylim_10000}}
\subfigure[Std Full:$\Sigma_{\lambda}-\left(\frac{\left[ \overline{(\rho w)^{\prime}u^{\prime}}r^3\right]_{, r}}{r^3}\right)$, 10\,000\,days]{\includegraphics[width=8.6cm,angle=0.0,origin=c]{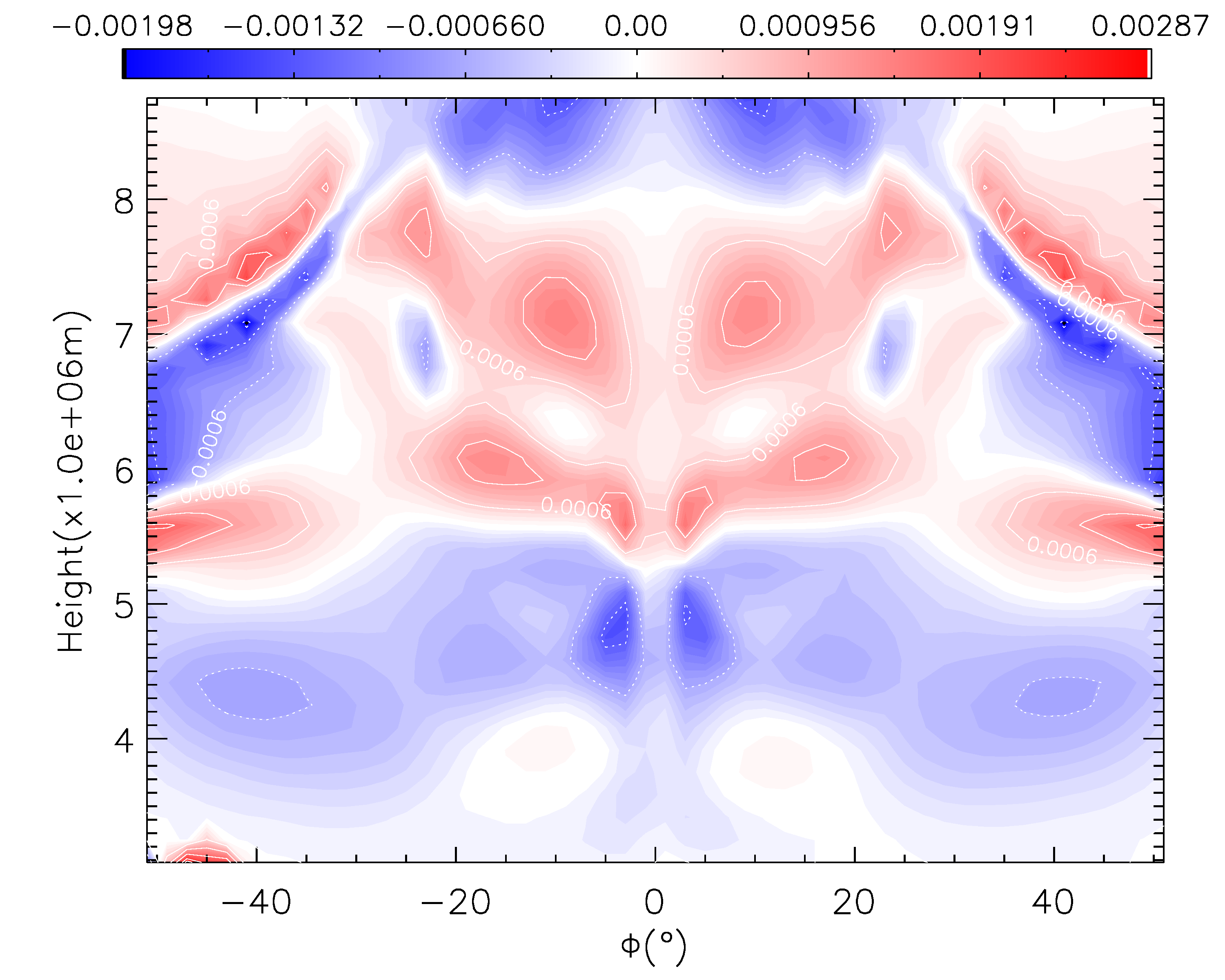}\label{std_full_vert_eddy_mmtm_slice_xylim_10000}}
\subfigure[Std Full: $\Sigma_{\lambda}-\left(\frac{\left[ \overline{(\rho v)^{\prime}u^{\prime}}\cos^2\phi\right]_{, \phi}}{r\cos^2\phi}+\frac{\left[ \overline{(\rho w)^{\prime}u^{\prime}}r^3\right]_{, r}}{r^3}\right)$, 10\,000\,days]{\includegraphics[width=8.6cm,angle=0.0,origin=c]{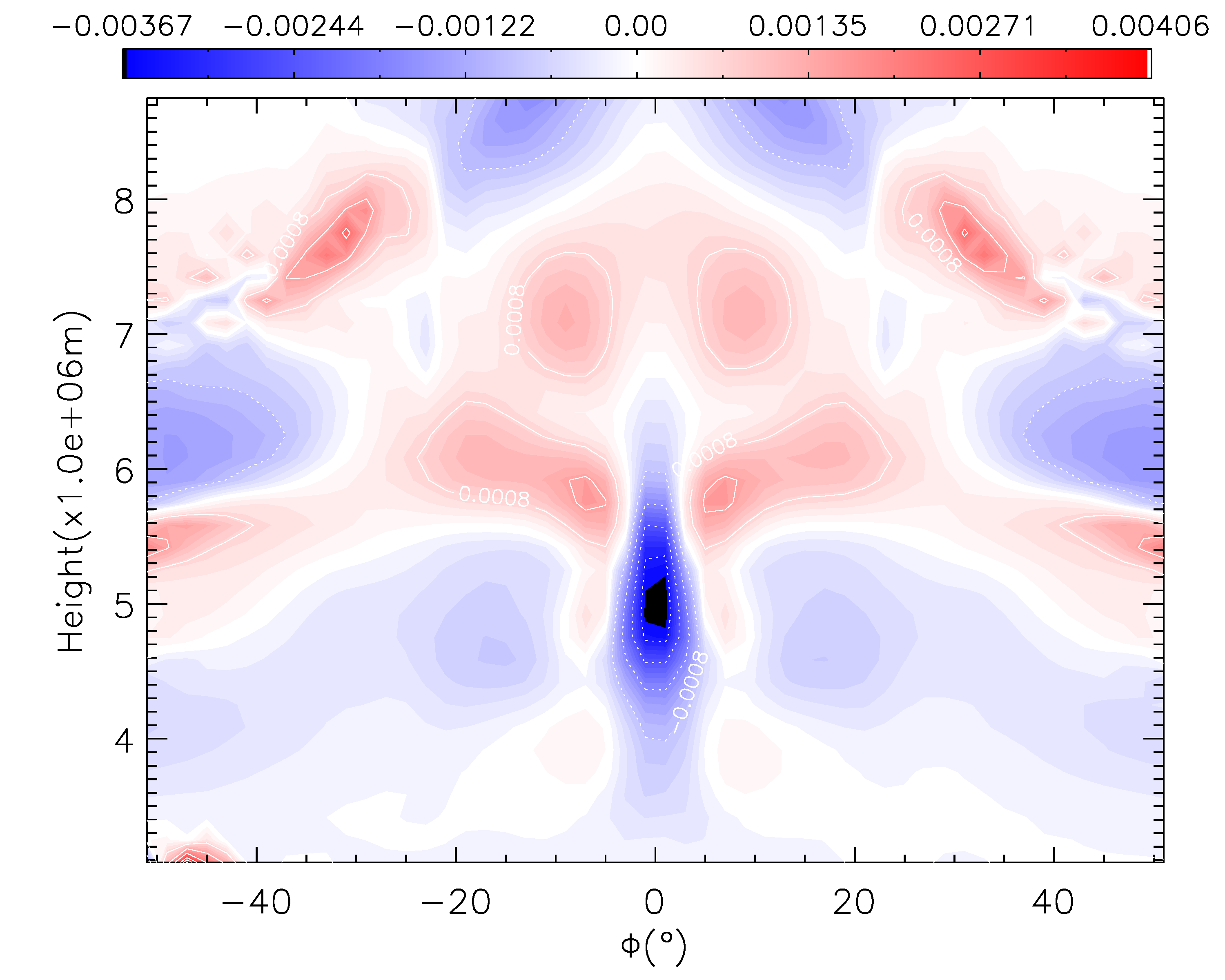}\label{std_full_sum_eddy_mmtm_slice_xylim_10000}}
 \end{center}
 \caption{Figure showing the same data (in the same format) as the
   \textit{left column} of Figure \ref{eddy_mmtm_1000} but after
   10\,000\,days. Note the reduced range of the height axis compared
   to Figure \ref{eddy_mmtm_1000}.\label{eddy_mmtm_10000}}
\end{figure}

In Section \ref{subsub_section:steady_eddy}, Figure
\ref{eddy_mmtm_1000} the latitudinal gradient of the meridional and
the vertical gradient of the vertical eddy momentum fluxes, are
presented after 1\,000\,days for the Std Full and Std RT
simulations. Figure \ref{eddy_mmtm_10000} presents the same data as
the \textit{left column} of Figure \ref{eddy_mmtm_1000}, i.e. the Std
Full simulation, but after 10\,000\,days. Here, the lowest plotted
height has been increased, to avoid the presence of fluctuations in
the deep atmosphere dominating the plot (as discussed in Section
\ref{subsub_section:steady_eddy}). The structures are very similar, to
the earlier time, albeit slight reduced in magnitude.

\begin{figure*}
\begin{center}
\subfigure[Std Full:$\Sigma_{\lambda}+2\Omega\overline{\rho v}\sin\phi$, 1\,000\,days]{\includegraphics[width=8.6cm,angle=0.0,origin=c]{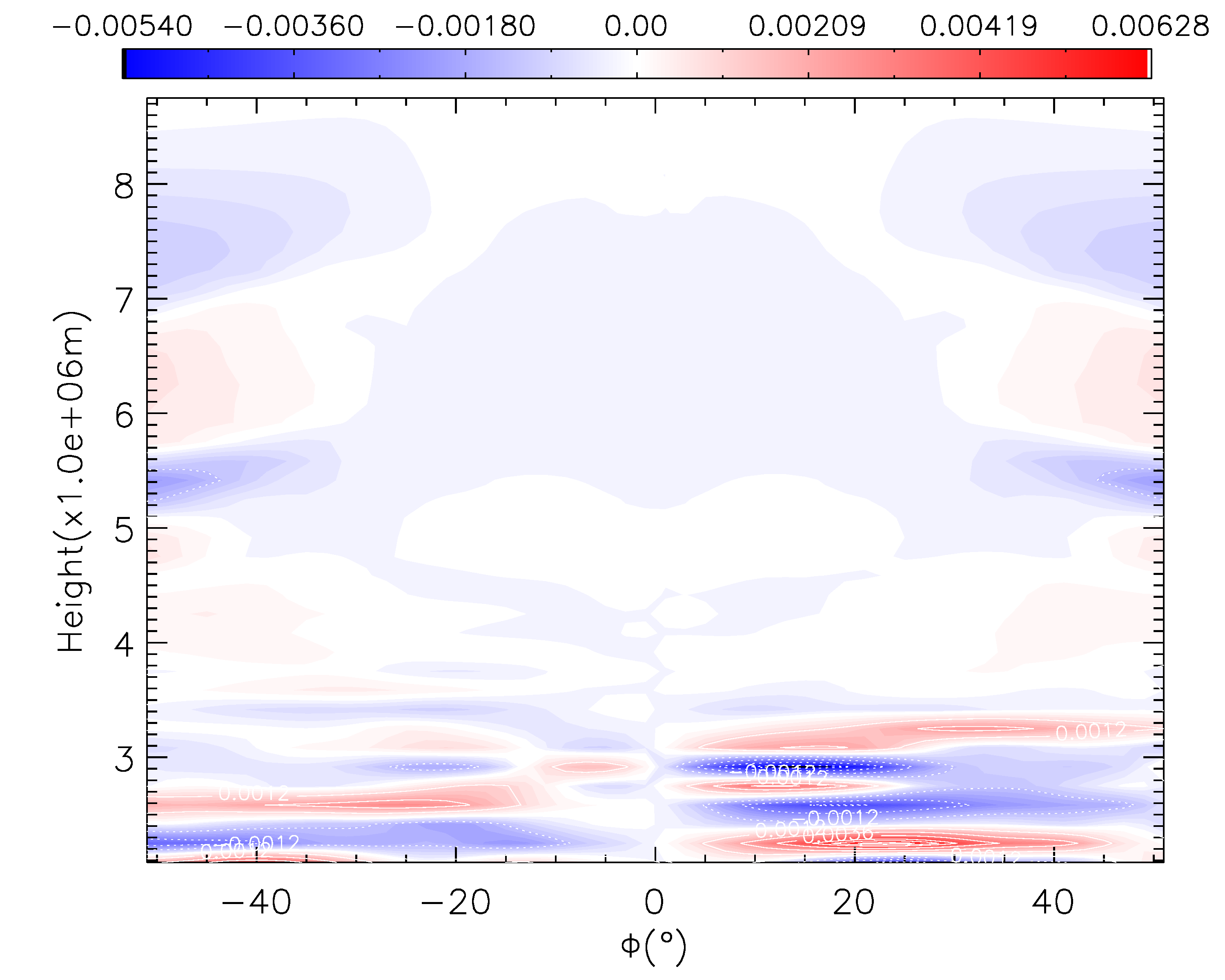}\label{std_full_meri_mean_sin_mmtm_slice_xylim_1000}}
\subfigure[Std RT:$\Sigma_{\lambda}+2\Omega\overline{\rho v}\sin\phi$, 1\,000\,days]{\includegraphics[width=8.6cm,angle=0.0,origin=c]{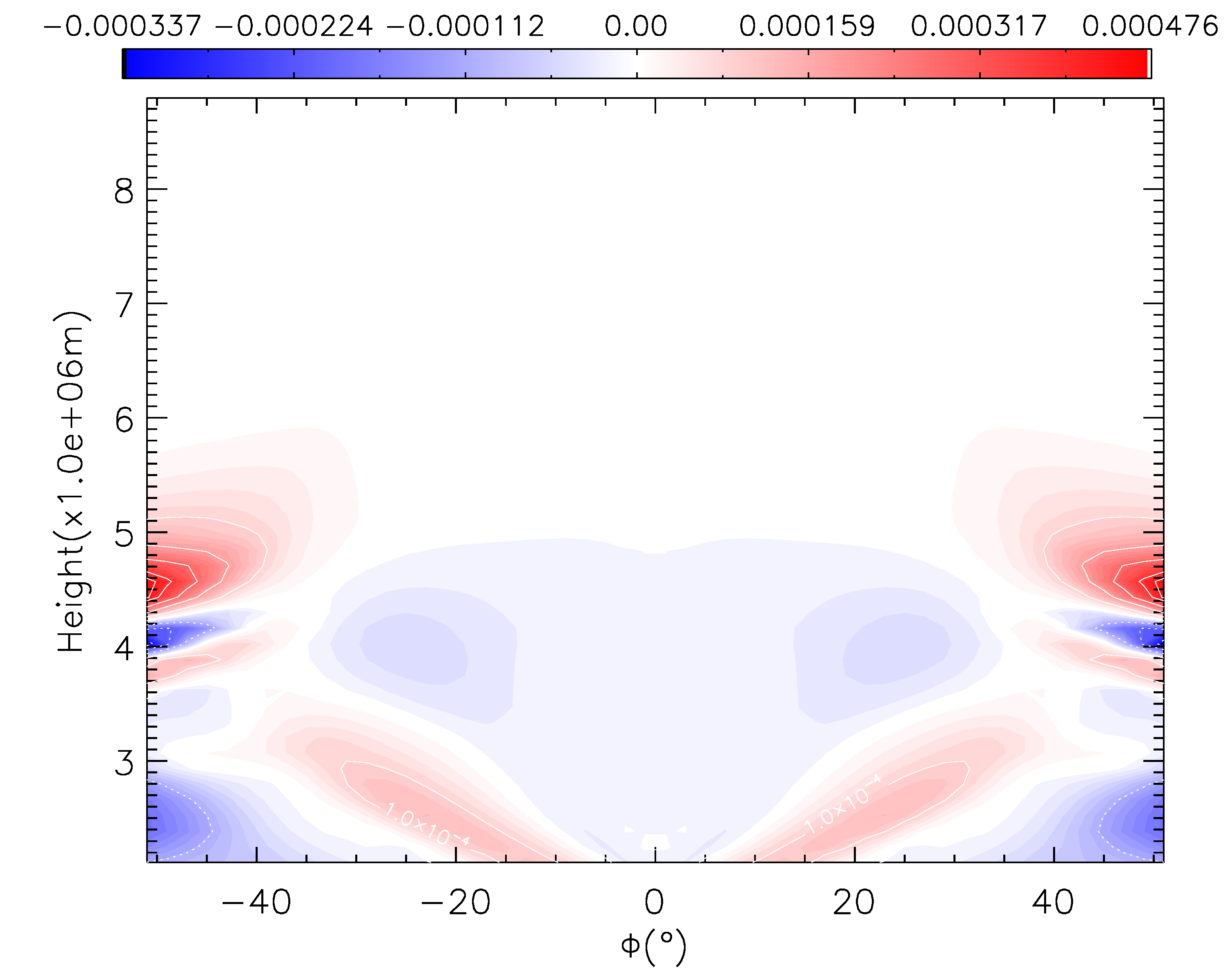}\label{rt_notiovo_full_meri_mean_sin_mmtm_slice_xylim_1000}}
\subfigure[Std Full:$\Sigma_{\lambda}-2\Omega\overline{\rho w}\cos\phi$, 1\,000\,days]{\includegraphics[width=8.6cm,angle=0.0,origin=c]{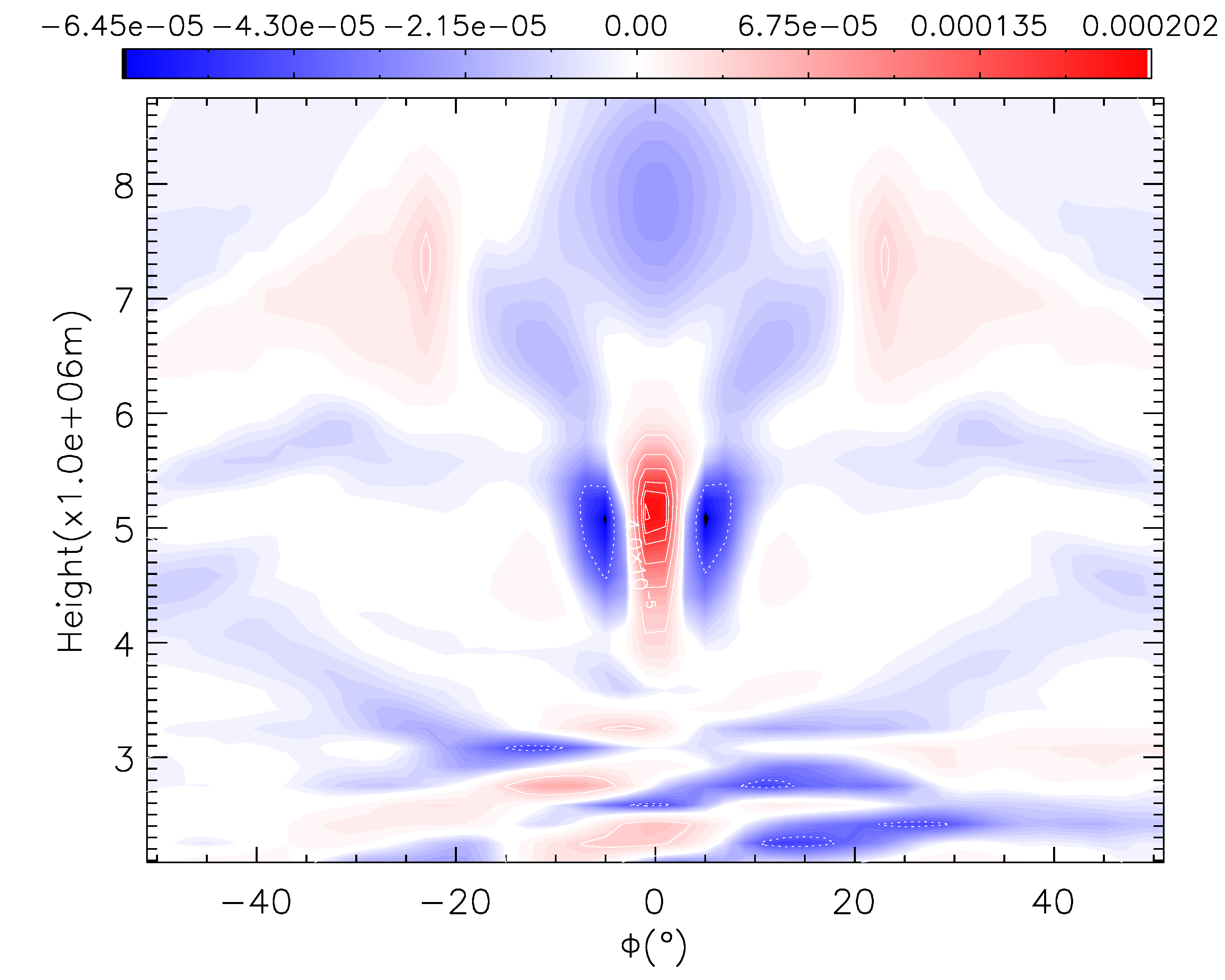}\label{std_full_vert_mean_cos_mmtm_slice_xylim_1000}}
\subfigure[Std RT:$\Sigma_{\lambda}-2\Omega\overline{\rho w}\cos\phi$, 1\,000\,days]{\includegraphics[width=8.6cm,angle=0.0,origin=c]{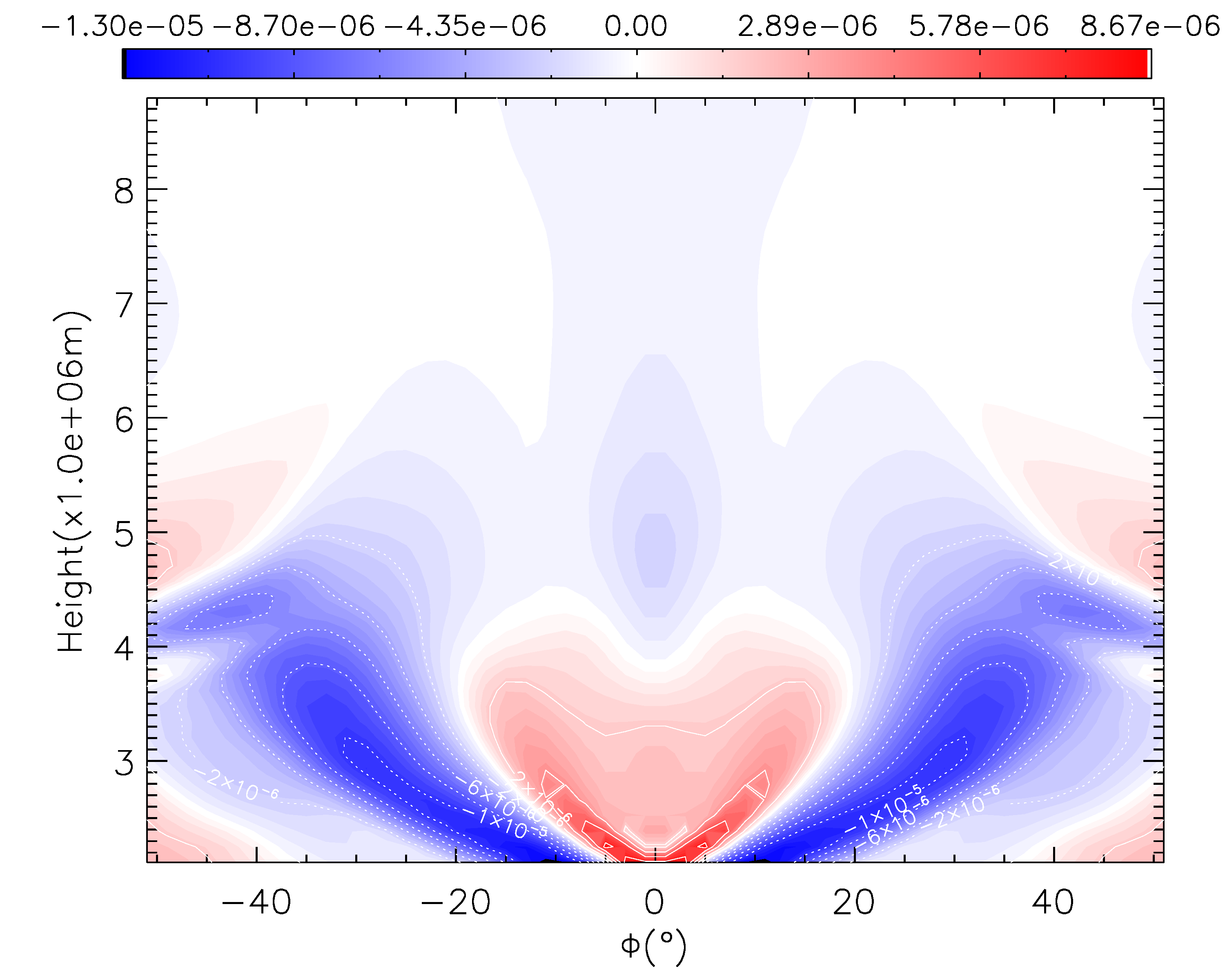}\label{rt_notiovo_full_vert_mean_cos_mmtm_slice_xylim_1000}}
\subfigure[Std Full:$\Sigma_{\lambda}+\left(2\Omega\overline{\rho v}\sin\phi-2\Omega\overline{\rho w}\cos\phi\right)$, 1\,000\,days]{\includegraphics[width=8.6cm,angle=0.0,origin=c]{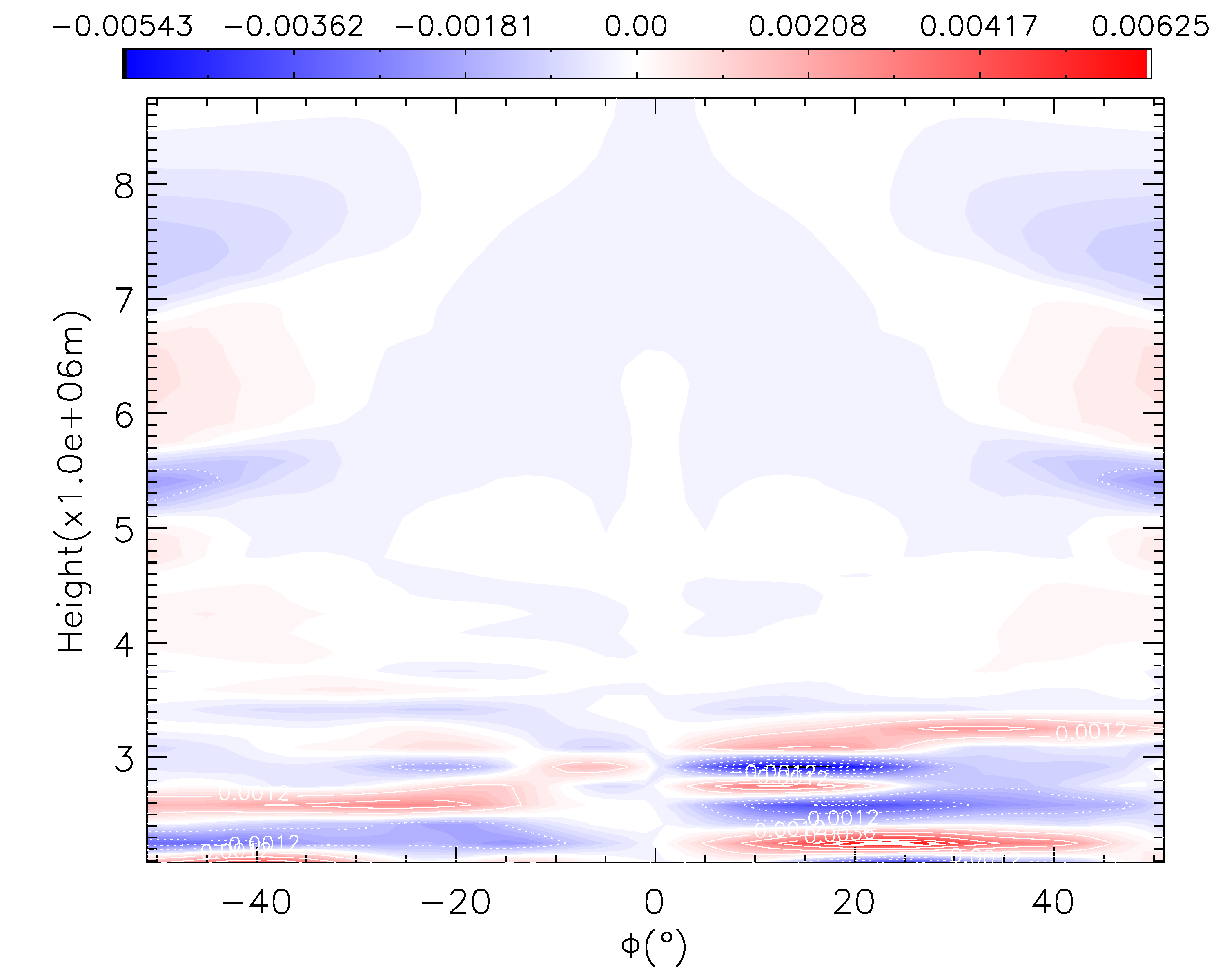}\label{std_full_sum_mean_trig_mmtm_slice_xylim_1000}}
\subfigure[Std RT:$\Sigma_{\lambda}+\left(2\Omega\overline{\rho v}\sin\phi-2\Omega\overline{\rho w}\cos\phi\right)$, 1\,000\,days]{\includegraphics[width=8.6cm,angle=0.0,origin=c]{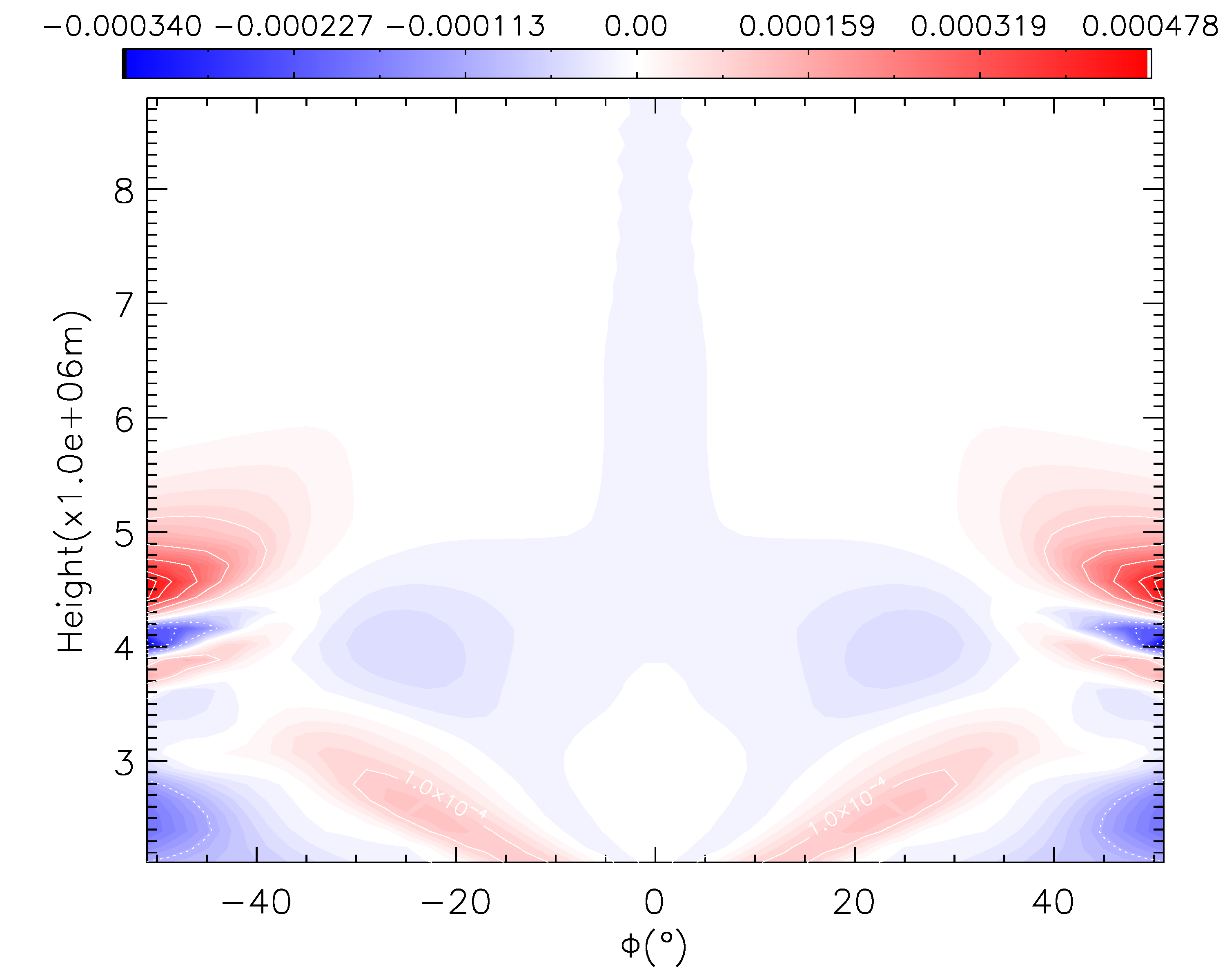}\label{rt_notiovo_full_sum_mean_trig_mmtm_slice_xylim_1000}}
 \end{center}
 \caption{Figure showing the meridional and vertical Coriolis terms
   (see Section \ref{subsub_section:steady_eddy}), and their sum, as
   the \textit{top}, \textit{middle} and \textit{bottom panels},
   respectively. The data are in the same format as Figure
   \ref{eddy_mmtm_1000}, showing the Std Full and Std RT simulations,
   as the \textit{left} and \textit{right columns} respectively, after
   1\,000\,days. \label{cor_mmtm_1000}}
\end{figure*}

The remaining terms in the eddy-mean interaction equation, aside from
those in Figures \ref{eddy_mmtm_1000} and \ref{mean_mmtm_1000} in
Section \ref{subsub_section:steady_eddy}, are the mean flow Coriolis
terms (see Section \ref{subsub_section:steady_eddy}). The mean
Coriolis terms, in the meridional ($+2\Omega\overline{\rho
  v}\sin\phi$) and vertical ($-2\Omega\overline{\rho w}\cos\phi$), as
well as their sum, are shown as the \textit{top}, \textit{middle} and
\textit{bottom rows} of Figure \ref{cor_mmtm_1000} in the same format
as Figure \ref{eddy_mmtm_1000}. The Std Full and Std RT simulations
are shown as the \textit{left} and \textit{right columns},
respectively, after 1\,000\,days.  Figure \ref{cor_mmtm_1000} shows
that the remaining Coriolis terms have a significantly reduced
contribution over the eddy and mean flow momentum flux transport, in
the region of the jet (i.e. aside from fluctuations in the low
altitude atmosphere).

\end{appendix}

%-------------------------------------------------------------------
\bibliographystyle{aa}
\bibliography{references}

\end{document}